\documentclass{emulateapj}
\usepackage{url}
\usepackage{natbib}
\usepackage{array}
\usepackage{rotating}

\newcommand{\beq}{\begin{equation}}
\newcommand{\eeq}{\end{equation}}
\newcommand{\bdi}{\begin{displaymath}}
\newcommand{\edi}{\end{displaymath}}

\newcommand{\spitzer}{\textit{Spitzer}}
\newcommand{\irac}{IRAC}

\newcommand{\degree}{$^{\circ}$}

\begin{document}
\shorttitle{Star Formation in the W3 GMC}
\shortauthors{Rivera-Ingraham, A.~et al.}

\title{Star Formation \& Young Stellar Content in the W3 GMC}

\author{Alana~Rivera-Ingraham,\altaffilmark{1}
        Peter~G.~Martin,\altaffilmark{2}
        Danae~Polychroni,\altaffilmark{3}
        Toby~J.~T.~Moore\altaffilmark{4}}

\altaffiltext{1}{Department of Astronomy \& Astrophysics, University of Toronto, 50 St. George Street, Toronto, ON  M5S~3H4, Canada}
\altaffiltext{2}{Canadian Institute for Theoretical Astrophysics, University of Toronto, 60 St. George Street, Toronto, ON M5S~3H8, Canada}
\altaffiltext{3}{INAF-IFSI, via Fosso del Cavaliere 100, 00133 Roma}
\altaffiltext{4}{Astrophysics Research Institute, Liverpool John Moores University, Twelve Quays House, Egerton Wharf, Birkenhead, CH41 1LD, UK}

\begin{abstract}
In this work we have carried out an in-depth analysis of the young stellar content in the W3 GMC. 
The YSO population was identified and classified in the IRAC/MIPS color-magnitude space according to the `Class' scheme and compared to other classifications based on intrinsic properties. Class 0/I and II candidates were also compared to low/intermediate-mass pre-main-sequence stars selected through their colors and magnitudes in 2MASS. We find that a reliable color/magnitude selection of low-mass PMS stars in the infrared requires prior knowledge of the protostar population, while intermediate mass objects can be more reliably identified.
By means of the MST algorithm and our YSO spatial distribution and age maps we investigated the YSO groups and the star formation history in W3. We find signatures of clustered and distributed star formation in both triggered and quiescent environments. The central/western parts of the GMC are dominated by large scale turbulence likely powered by isolated bursts of star formation that triggered secondary star formation events. 
Star formation in the eastern high density layer also shows signs of quiescent and triggered stellar activity, as well as extended periods of star formation. While our findings support triggering as a key factor for inducing and enhancing some of the major star forming activity in the HDL (e.g., W3 Main/W3(OH)), we argue that some degree of quiescent or spontaneous star formation is required to explain the observed YSO population. Our results also support previous studies claiming an spontaneous origin for the isolated massive star(s) powering KR 140. 
\end{abstract}


%%%%%%%%%%%%%%%%%%%%%%%%%%%%%%%%%%%%%%%%%%%%%%%%%%%%%%%%%%%%%%%%%%%%%%%%%%%%
\section{Introduction} \label{sec:intro}
The W3 giant molecular cloud (GMC) together with W4 and W5 form part of a massive star forming complex in the Perseus Arm. 
With an estimated mass of $\sim4\times10^5$\,M$_\odot$ (\citealp{moore2007}; \citealp{polychroni2010}) and a size of $\sim1.5$\,deg$^2$ this cloud is one of the most massive in the outer Galaxy. Located at a distance of $\sim2$\,kpc (e.g., \citealp{xu2006}; \citealp{navarete2011}), W3 comprises a wealth of \ion{H}{2} regions, clusters, and active star forming sites with clear signatures of both low-mass and high-mass star formation.

\begin{figure*}[ht]
\centering
\includegraphics[scale=0.75,angle=270]{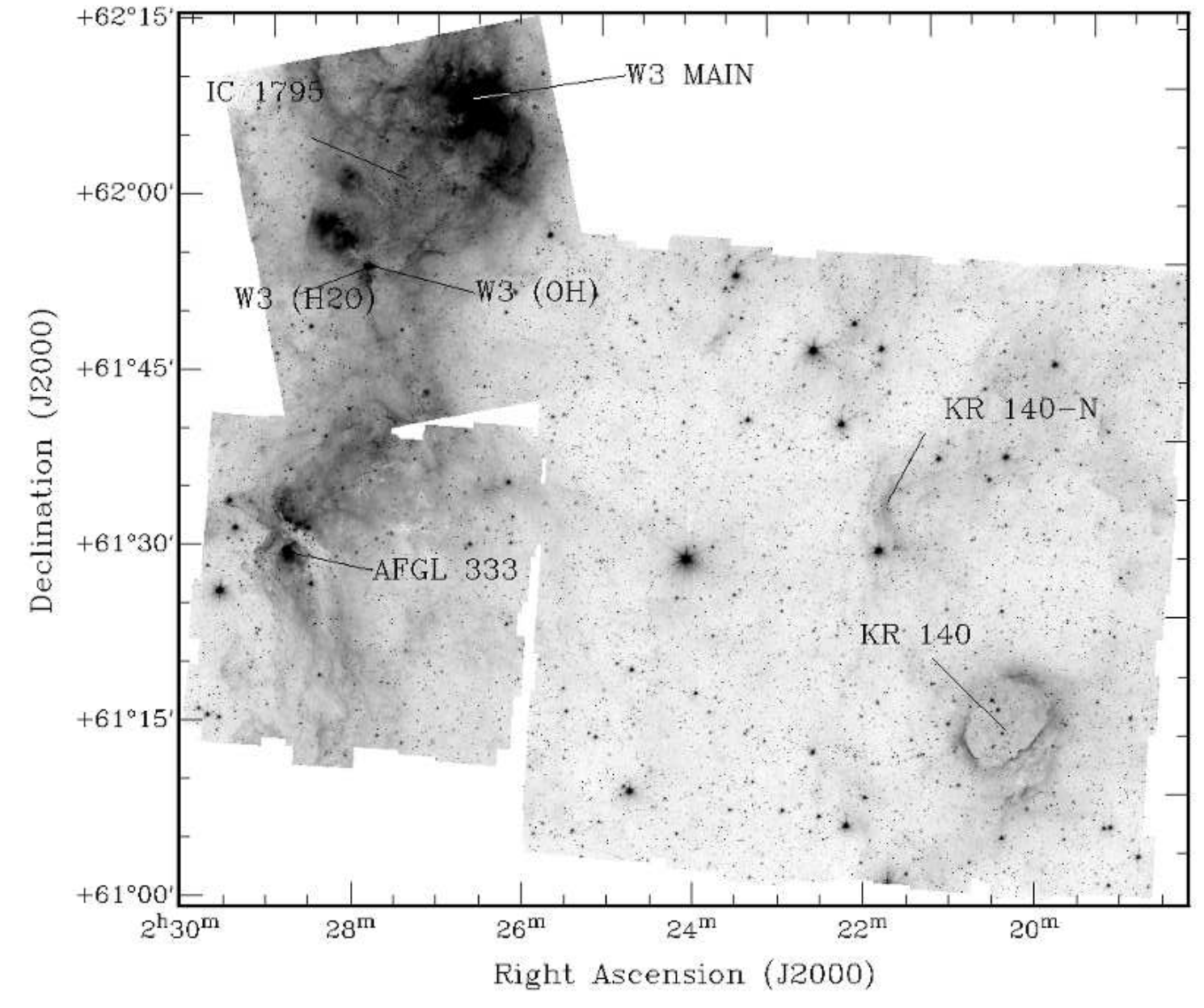}
\caption{Greyscale \spitzer\ channel 1 mosaic, with labels marking the regions and key features in the W3 GMC.}
\label{fig:intro}
\end{figure*}

\begin{figure*}[ht]
\centering
\includegraphics[scale=0.75,angle=0]{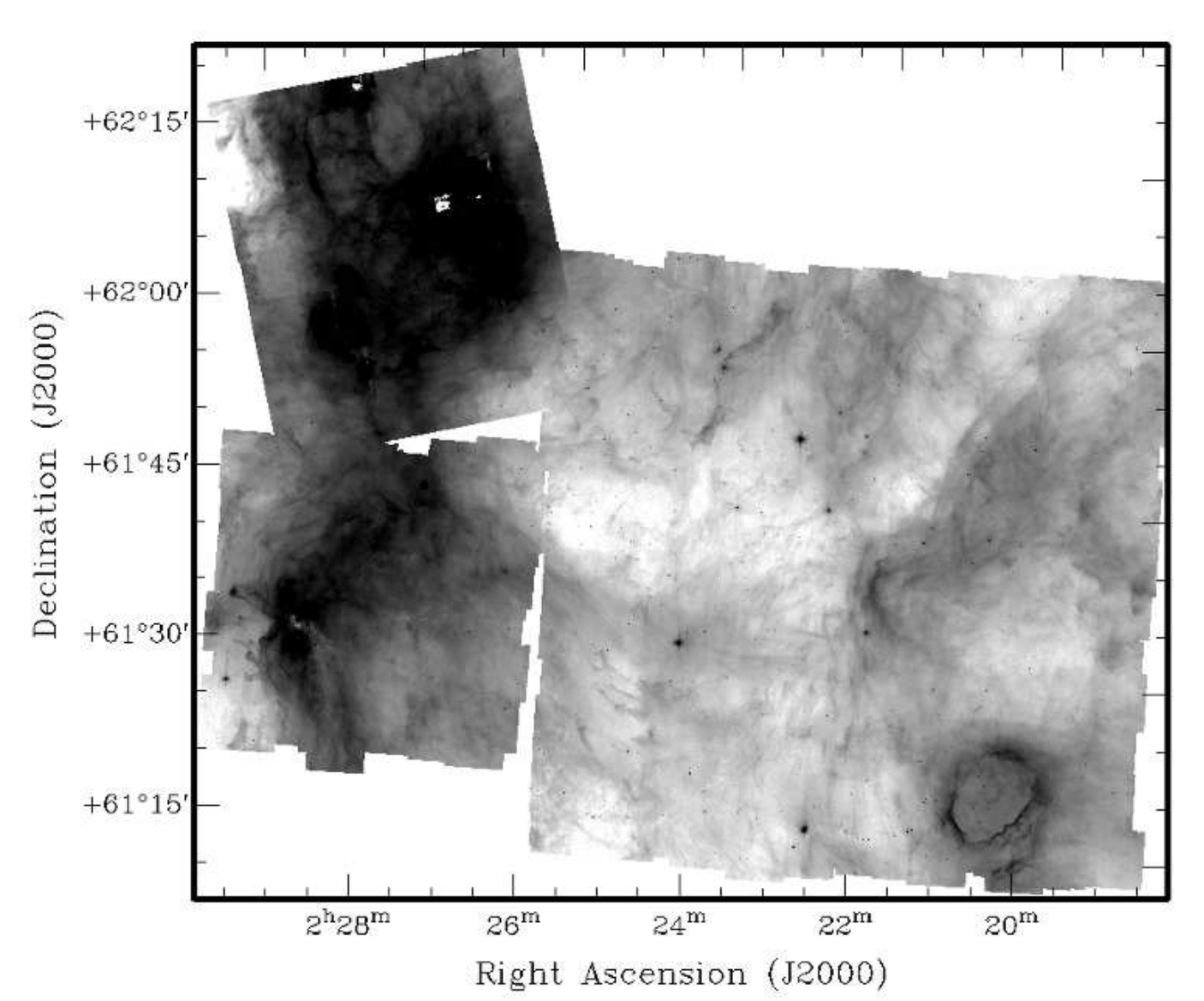}
\caption{Greyscale \spitzer\ channel 4 mosaic of W3. Intensity scale has been chosen to highlight the weaker features surrounding the main star forming regions labeled in Fig.\ref{fig:intro} (at the expense of the latter). Details in this image include several Infrared Dark Clouds (IRDCs) and filaments (e.g., ($\alpha$, $\delta$)=(02h 26m 57s, 61\degree\ 29\arcmin\ 45\arcsec)).}
\label{fig:intro_ch4}
\end{figure*}

Its relatively close location and its massive star content has made this cloud a prime target for the study of cluster and high-mass star formation (see \citealp{megeath2008} for a detailed review of the literature related to the study of W3), which contrary to the low-mass case is, overall, poorly understood. 
Massive OB (M$>8$\,M$_\odot$) stars are sources of UV photons, responsible for dissociating/ionizing molecules and atoms (HII regions), and of metals, which enrich the surrounding gas with heavier species. They also affect the local and global dynamical state of the interstellar medium (ISM) through ionization and radiation pressure, stellar winds, outflows and supernovae. These stars are therefore fundamental in determining the state and evolution of the ISM, as well as the formation, maintenance, and dissipation of structures ranging from the largest, Galactic scales, to giant molecular clouds, disks, and planetary systems.  
Theories such as the turbulent core model \citep{mckee2003} and competitive accretion (e.g., \citealp{bonnell2005}) have been proposed as possible theories explaining the formation of the rare and massive OB stars. However, the question as to whether massive star formation is simply a scaled-up version of low-mass star formation, or if it is the result of a completely different process, remains one of the main outstanding issues in star formation theory. Indeed, authors such as \citet{yorke1993} claim that a bimodality should arise from the shorter Kelvin-Helmholtz timescale for massive stars, their high $\mathrm{L}/\mathrm{M}$ ratio, and the dramatic effects their strong radiation field have on their evolution. The immediate effects of high-mass star activity can be observed, for instance, with masers, shocks, `hot spots', and the formation of hypercompact, ultracompact and classical HII regions (e.g., \citealp{zinnecker2007} and references therein). Furthermore, while low-mass stars are known to be able to form in isolation, most star formation is thought to occur in clusters embedded in their parent GMCs \citep{lada2003}. This is particularly true for massive stars, which have been theorized to form exclusively in clusters (e.g., \citealp{dewit2005}), making cluster studies crucial in order to understand and investigate the formation of massive stars.
W3 is believed to contain massive stars in various evolutionary stages (e.g., \citealp{tieftrunk1997}). The eastern high density layer (HDL) neighboring W4 contains the most active star forming sites W3 North, W3 Main, W3 (OH) and AFGL 333, with signatures of massive star formation in a triggered environment (e.g., \citealp{oey2005}).  A rare case of spontaneous massive star formation is also believed to have occurred in KR 140, west of the HDL (e.g., \citealp{kerton2008}).

\subsection{Proposed Analysis: The W3 GMC}
This work is the first of a series of papers focusing on W3 and which aim to shed some light on the still poorly understood massive star formation process. With the advent of high sensitivity instruments such as \spitzer\ and the Herschel Space Observatory, this GMC presents an ideal opportunity  to investigate the high-mass star/cluster star formation process and its relation to triggered/spontaneous star formation events, clustered, and distributed star formation.
In this analysis we follow up the study initiated by \citet{polychroni2010} to characterize the pre-stellar population in W3 by means of \spitzer\ data. This analysis will complement the Herschel data obtained under the Guaranteed Time Key Program HOBYS\footnote{http://starformation-herschel.iap.fr/hobys/} (Herschel imaging survey of OB young Stellar objects), as well as an extensive molecular analysis (Polychroni et al., in preparation), which focuses on the formation, evolution, and dynamics of the largely unknown progenitors of massive stars and clusters in this cloud.

In Section \ref{sec:data} we first describe the infrared data used to select and classify the young stellar objects (YSOs). Classification techniques and schemes, including a description of the final catalog of candidate YSOs have been included in Section \ref{sec:class}. The physical and observational properties of the YSO sample are presented in Section \ref{sec:discussion}. Their clustering and spatial distribution, as well as the techniques employed in our analysis, are described in Section \ref{sec:groupclass}. Section \ref{sec:history} investigates the cluster properties and ages across W3, and results are used to compare the star formation history and activity in the different subregions. We conclude with a summary of the star formation in W3.

%%%%%%%%%%%%%%%%%%%%%%%%%%%%%%%%%%%%%%%%%%%%%%%%%%%%%%%%%%%%%%%%%%%%%%%%%%%%%%%%%%%%%
\section{Photometry, Data Processing \& Datasets} \label{sec:data}
\subsection{\textit{Spitzer} IRAC and MIPS Infrared Observations}\label{irac}
The main active regions of W3 were observed by \textit{Spitzer} under two main Program Identification Number (PID) programs. The northern parts of the HDL comprising W3 Main and W3 (OH) (Figs.\ref{fig:intro} and \ref{fig:intro_ch4}) were first observed in 2004 under program P00127; AFGL 333 and central/western region of W3 (including KR 140 and the active region to its north: `KR 140-N') were subsequently observed in 2007 (program P30955). A preliminary analysis of the data obtained during the first observation has been presented in \citet{ruch2007}, while \citet{polychroni2010} combined data from both programs for their analysis of the stellar population associated with SCUBA cores \citep{moore2007}. For this work we also used data from both programs.

For the reduction process we downloaded the IRAC and MIPS 24\,\micron\ corrected Basic Calibrated Data (cBCD) from the \spitzer\ Archive. The \irac\ cBCDs were produced with the S18.7.0 pipeline. P00127 and P30955 MIPS data were processed with the S18.13.0 and S18.12.0 pipelines, respectively. These files already include all the pre-processing artifact mitigation corrections, including muxbleed, column pulldown/pullup and electronic banding. Tiles were reduced, background-matched and mosaicked with the MOPEX\footnote{http://ssc.spitzer.caltech.edu/dataanalysistools/tools/mopex/} tool, which with the exception of the MIPS observations from program P30955 (KR 140) produced maps of higher quality than the mosaicked pBCD data already provided by the Archive. The best available mosaics were chosen for our analysis. Photometry tests performed on both the pBCDs and our own reduced mosaics shows no systematic differences in channels 3 and 4, and negligible (0.04 mag) differences in channel 1 and channel 2, which is much smaller than any of our expected photometric uncertainties.

\subsection{\textit{Spitzer} Source Extraction and Photometry}\label{irac2} 
A preliminary list of sources was obtained using SExtractor \citep{bertin1996} and a mexican hat filter, which was observed to perform the best (with the highest detection rate of visually confirmed sources in the region) in the extremely crowded regions found in W3, especially W3 Main and W3 (OH). We note that this also introduced a significant number of artifacts and false detections which we eliminated through a series of cleaning steps, described below. SExtractor also produced noise and background maps, which we also checked visually to ensure optimal source extraction. This preliminary list was subsequently fed into the point source extraction package in MOPEX (APEX) for point response function (PRF) fitting, which provides a more accurate centroid calculation as well as additional statistics for each source resulting from the fitting (e.g., $\chi^2$, signal-to-noise ratio (SNR), etc). When bandmerging the IRAC list with the 2MASS Point Source Catalog (through the GATOR interface in the NASA/IPAC Infrared Science Archive; IRSA \footnote{http://irsa.ipac.caltech.edu/applications/Gator/}) we find  the accuracy of our final PRF-fitted coordinates to be generally better than $0$\arcsec.5, although in this work we allow for a more conservative matching radius of $2$\arcsec\ when bandmerging catalogs at different wavelengths and/or instruments. 
The internal background image used in APEX for photometry was produced choosing the option `SExtractor background method' for consistency with the previous part of this analysis.

As recommended by the IRAC handbook we chose aperture photometry for our main photometric analysis, using in this case the most accurate centroids returned by APEX. The use of this technique also ensured consistency with the most recent \spitzer\ studies of this cloud \citep{polychroni2010}. 

Aperture corrections and zero points were obtained from the IRAC/MIPS Instrument Handbooks\footnote{http://ssc.spitzer.caltech.edu/irac/iracinstrumenthandbook/},\footnote{http://ssc.spitzer.caltech.edu/mips/mipsinstrumenthandbook/}. Additional corrections (e.g., pixel phase and array correction) were applied when required. An aperture of 2 pixels ($2$\arcsec.4) with a sky annulus between 2 and 6 pixels ($7$\arcsec.3) for IRAC, and an aperture of $7$\arcsec.0 with a sky annulus between $7-13$\arcsec\ for MIPS, were found to yield results most closely agreeing with the magnitudes provided by \citet{ruch2007}, who used the version of {\sc daophot} \citep{stetson1987} modified by the Galactic Legacy Infrared Mid-Plane Survey Extraordinaire ({\sc glimpse}). Contrary to standard aperture photometry, the {\sc glimpse} technique is particularly useful for analysis in crowded fields and regions with variable background \footnote{see http://www.astro.wisc.edu/glimpse/photometry\textunderscore v1.0.pdf; http://www.astro.wisc.edu/glimpse/}, and we therefore checked the accuracy of our results by comparing the photometry for those YSO candidates in our list with counterparts in the source list provided by these authors.  
The root mean square (rms) difference between the MIPS\,$24$\micron\, photometry obtained in this work for the YSO list and that from \citet{ruch2007}  was found to be $\sim0.6\,$mag. Comparison of IRAC photometry yields an rms difference of $<0.2$\,mag in all four channels, which is consistent with the estimated $3\sigma$\,errors and equivalent to the minimum signal-to-noise (S/N) required for our final catalog (S/N$=5$; see below). 

The same procedures applied to the IRAC long exposure mosaics were also performed on the short exposure images. While the final catalog is based on the long exposure maps, we used the short exposure mosaics to obtain replacement photometry for those sources with bad pixels within their apertures or observed/suspected to be affected by saturation in the long exposure mosaics.

%%%%%%%%%%%%%%%%%%%%%%%%%%%%%%%%%%%%%%%%%%%%%%%%%%%%%%%%%%%%%%%%%%%%%%%%%%%%%%%%%%%%%%%%%%%%%%
\section{Stellar Classification} \label{sec:class}
\subsection{General YSO Classification: Methodology and Techniques}
\subsubsection{Protostars \& Optically Thick Disks}
We aimed to provide the most reliable sample of young stellar objects (YSOs) in this GMC. For the main classification in our analysis we chose the `revised', updated criteria in Appendix A of \citet{gutermuth2009}. The color and magnitude scheme developed by these authors includes a series of sequential steps (phases) to identify, clean, and classify the candidates: Phase 1) Removal of contaminants such as star-forming galaxies with strong polycyclic aromatic hydrocarbon (PAH) emission, AGN, unresolved knots of shock emission, and PAH-emission-contaminated apertures. This step also includes the first separation of YSOs by means of the four IRAC bands. Phase 2) A search for additional YSOs based on 2MASS photometry\footnote{One of the equations of Phase 2 (\citealp{gutermuth2008}; \citealp{gutermuth2009}) should read: E$_{[3.6]-[4.5]}$/E$_{H-K}$=(E$_{H-K}$/E$_{K-[4.5]}$)$^{-1}$-(E$_{H-K}$/E$_{K-[3.6]}$)$^{-1}$}. Phase 3) Identification and re-classification of previously identified sources with suitable MIPS $24$\,\micron\ photometry. When re-classifying photospheric sources into `transition disk' objects (those with significant $24$\,\micron\ emission; included within the Class II category) we required sources to have been classified as photospheric in both previous phases. To deredden the magnitudes we used the extinction maps and methodology described in \citet{RF2009}. The visual extinction map was transformed to a median A$_{\mathrm{H}}$ map using the extinction law from \citet{mathis1990}. We changed extinction in the 2MASS bands to extinction in the IRAC channels using the numbers from \citet{flaherty2007}.

The above classification was complemented and cross-checked with: i) the `red source' classification scheme from \citet{rob2008}, which should include all Class 0/I and several Class II sources; and ii) the `stage' phase from \citet{rob2006} (Stage 0/I: $\dot{M}_{\mathrm{env}}/M_{\star}>10^{-6}$\,yr$^{-1}$; Stage II: $\dot{M}_{\mathrm{env}}/M_{\star}<10^{-6}$\,yr$^{-1}$ and M$_{\mathrm{disk}}/$M$_{\star}>10^{-6}$; Stage III:  $\dot{M}_{\mathrm{env}}/M_{\star}<10^{-6}$\,yr$^{-1}$  and M$_{\mathrm{disk}}/$M$_{\star}<10^{-6}$). We used this last scheme to compare the above observational classification with an alternative method based on intrinsic physical properties (e.g., mass accretion rate and disk mass). This last analysis was carried out by studying the position of each YSO in the color-color diagrams (CCDs) with respect to the limits marking the areas where most of the sources of one particular `stage' are predicted to fall \citep{rob2006}.

\subsubsection{Pre-Main Sequence Population With Optically Thin Disks}
Separation and classification of pre-main sequence stars (PMS) with optically-thin disks is a particularly complicated process due to their similarity (in infrared color) with more evolved reddened main sequence and giant stars (photospheric-dominated). These transition objects are however essential to fully understand the different stages in star formation. In an attempt to estimate the population of sources with weak infrared excess and other PMS stars that may have been missed with the above color classification, we first excluded those objects in our source list already classified as Class 0/I and II. We then used the 2MASS catalog and a process similar to that used in \citet{kerton2008}, who attempted to separate the YSO population guided by a sample of known low-mass T-Tauri and intermediate-mass Herbig Ae/Be (HAeBe) stars. In this work (see Section \ref{sec:discussion}) we show that this method can only be applied successfully once the `younger' YSOs have been identified using additional data (e.g., IRAC). 

\begin{figure}[ht]
\centering
\includegraphics[scale=0.6,angle=0]{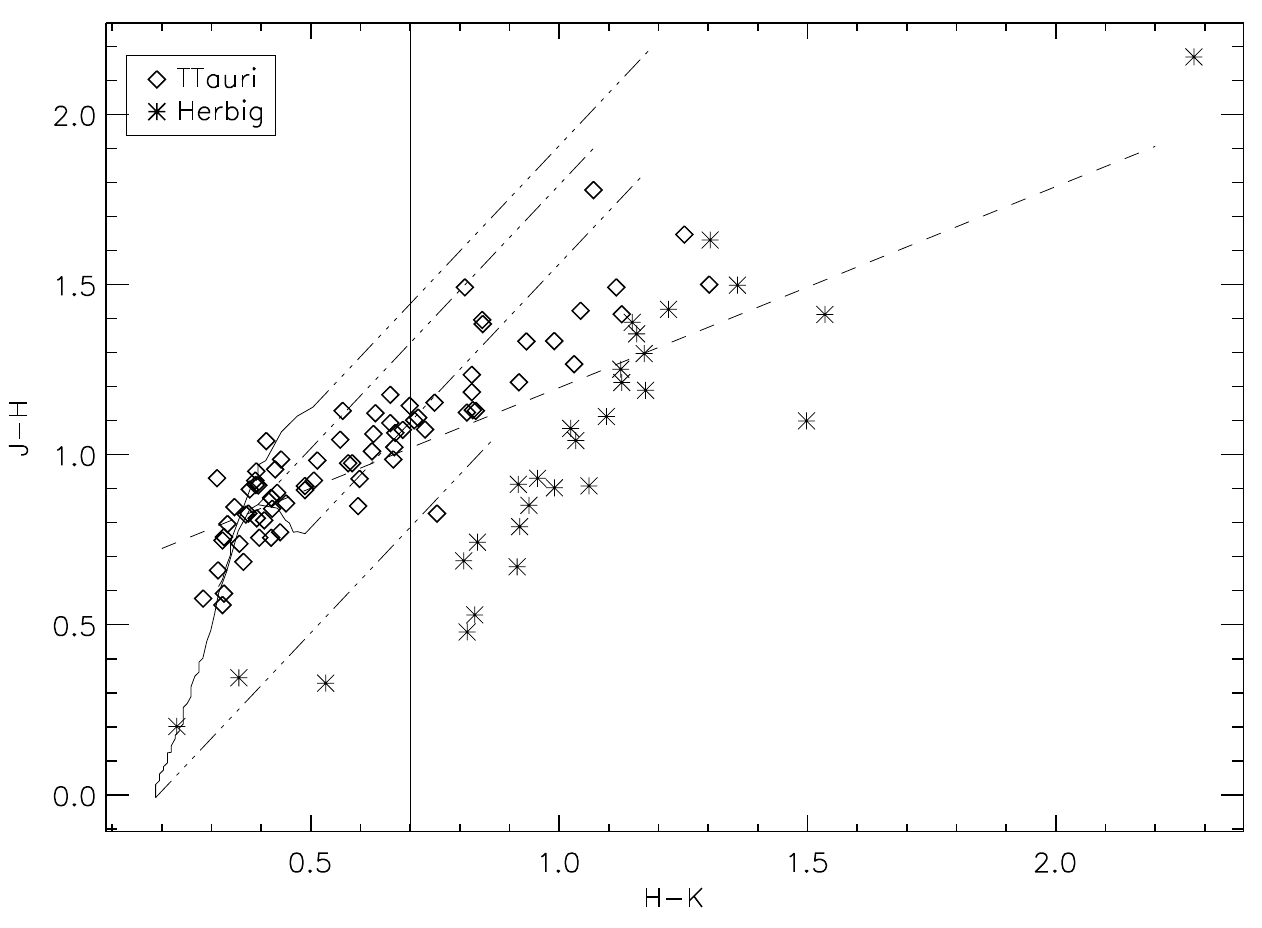}
\caption{Color-Color Diagram showing the T-Tauri and HAeBe samples shifted to d$=2$\,kpc. Solid lines mark the main sequence and giant branch from \citet{koornneef1983}, also shifted to a distance of $2$\,kpc. Dash-dotted lines are reddening vectors for an additional A$_{\mathrm{V}}=10$ for an O$6$\,V, M$8$\,V, M$2$\,V, and M$6$\,III star using the extinction law from \citet{mathis1990}. Dashed line marks the locus of T-Tauri stars from \citet{meyer1997} at the same distance. Vertical solid line marks the bluest [H-K] color accepted for PMS classification.}
\label{fig:ccd1}
\end{figure}

\begin{figure}[ht]
\centering
\includegraphics[scale=0.6,angle=0]{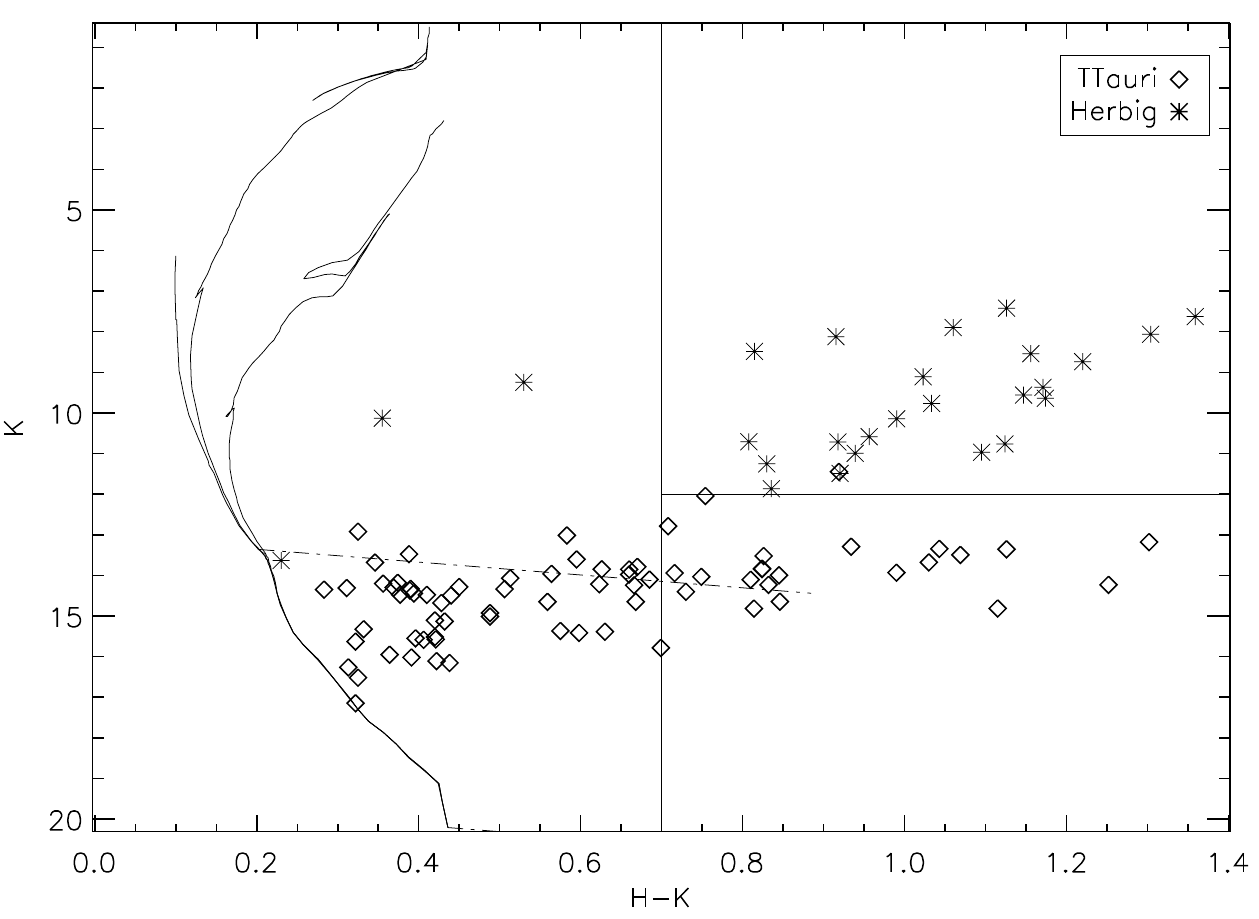}
\caption{Color-Magnitude Diagram for the T-Tauri and HAeBe samples at d$=2$\,kpc. Solid lines are solar metallicity isochrones (\citealp{marigo2008}; \citealp{girardi2010}) log(t\,yr$^{-1}$)=7, 8 and 9. Dashed-dotted line is the reddening vector for an $\sim$A0 star at this distance with A$_{\mathrm{V}}=10$. Horizontal solid line marks the magnitude limit separating T-Tauri and HAeBe candidates. Vertical solid line like in Fig.\ref{fig:ccd1}.}
\label{fig:cmd1}
\end{figure}

To investigate the possibility of missed candidates from the PMS population we first chose a sample of T-Tauri (\citealp{kenyon1995}) and HAeBe stars with known distances (\citealp{finkenzeller1984}; \citealp{the1994}; \citealp{mendigutia2011} and references therein). All sources where checked with {\sc simbad}\footnote{http://simbad.u-strasbg.fr/simbad/}, keeping variable, emission and pre-main sequence stars and rejecting those classified as double/multiple systems, low mass stars and brown dwarfs. Infrared photometry was obtained by matching the sample with the 2MASS Point Source Catalog through the {\sc gator} interface. Infrared photometric systems were converted to the 2MASS system using the transformations from \citet{carpenter2001}. All magnitudes were shifted to a distance of $2$\,kpc for W3 with the inclusion of interstellar extinction by means of the A$_{\mathrm{V}}-$distance relation from \citet{indebetouw2005}. Figures \ref{fig:ccd1} and  \ref{fig:cmd1} show the color-color diagram (CCD) and color-magnitude diagram (CMD) for the T-Tauri and HAeBe samples shifted to the distance of W3. 
The CCD shows the T-Tauri locus from \citet{meyer1997} and the main sequence and giant branch from \citet{koornneef1983} including interstellar reddening. Reddening vectors for an A$_{\mathrm{V}}=10$ have been included for an O$6$\,V, M$8$\,V, M$2$\,V, and an M$6$\,III star. The CMD shows solar metallicity isochrones (\citealp{marigo2008}; \citealp{girardi2010}) at log(t\,yr$^{-1})=$7, 8, and 9 for the same distance. The dashed-dotted line is the reddening vector for an $\sim$A$0$ star with A$_{\mathrm{V}}=10$ at d$=2$\,kpc, applied using the reddening law and the A$_{\mathrm{V}}/$A$_{\mathrm{J}}$ conversion from \citet{mathis1990}. The values used for extinction conversions between 2MASS bands are consistent with those from \citet{indebetouw2005} up to one decimal place taking into account uncertainties, which is a negligible difference compared to the expected uncertainty in the transformation from optical to infrared extinction.

Figure \ref{fig:ccd1} shows T-Tauri stars lying mainly above the T-Tauri locus. Many are within the reddening band formed by the reddening vectors \citep{mathis1990} of an O$6$ and $\sim$M$2$ main sequence stars (with a tail extending into the HAeBe region, to the right of the reddening band, and following the direction of the T-Tauri locus). The wide distribution implies variable amounts of extinctions toward these sources and variable disk emission. A large proportion of T-Tauri stars are indistinguishable from main sequence stars with just interstellar reddening or are consistent with weak-emission T-Tauri stars \citep{meyer1997}. The maximum A$_{\mathrm{V}}$ in the maps from \citet{RF2009} is about $\sim9.5-10$ in the region containing W3 Main/(OH), and $\sim7$ for those comprising KR 140 and AFGL 333. Thus there will be considerable extinction of objects in the W3 field, and so color identification of this type of T-Tauri star without the aid of spectroscopic data will be severely contaminated.

In consequence, we chose a more conservative approach and selected our T-Tauri sample by requiring these PMS stars to be reddened enough to lie above the T-Tauri locus, with colors satisfying  H-K$>0.7$ (similar to that used in \citet{kerton2008} for KR 140). 
 The above limits minimize the contamination from foreground or mildly reddened weak-emission T-Tauri stars (undistinguishable from main sequence), and early type stars. While the color cuts could still allow for non-negligible contamination from late main sequence and giant stars with moderate reddening, the magnitude selection criterion below reduces this contamination to mainly that caused by highly extinguished (A$_\mathrm{V}\ge7)$ stars of spectral type of $\sim$A or later (Figure \ref{fig:cmd1}).

HAeBe stars lie preferentially to the right of the reddening band. All suitable candidates should therefore be located in this region and satisfy the condition H-K$>0.7$. This limit aims to minimize contamination from reddened early type stars and luminous T-Tauri stars.
To separate candidates in regions of the CCD populated by both types of PMS (i.e., outside the reddening band) we used the information from the CMD (Figure \ref{fig:cmd1}), in which T-Tauri and HAeBe can be easily separated. All T-Tauri stars have K magnitudes $\gtrsim12$. HAeBe stars tend to be brighter, approaching this limit only for `late' stages reaching the main sequence, which we already discard with our imposed limit in the CCD to minimize contamination from reddened early type stars.
The combined color plus magnitude condition minimizes contamination of the T-Tauri sample from reddened giant stars. In addition, the color constraint imposed on HAeBe stars, which are mainly localized and `isolated' outside the reddening band of typical stars, already minimizes the contamination from reddened main sequence and giant stars.

We note that this relatively simple scheme for PMS classification may only be applied \textit{after} the Class 0/I and Class II populations have been identified using IRAC data, as there is considerable overlap in the CCD and CMD of T-Tauri candidates with \spitzer\, Class II/I/0 sources (see Section \ref{sec:discussion}).

\begin{table}[t!]
\caption{Selection Criteria for T-Tauri and HAeBe stars for d$=2$\,kpc}
\label{table:class3}
\centering
\begin{tabular}{l c l}
\hline
\hline
&T-Tauri&\\
\hline
&[H$-$K]$\geq 0.7$&\\
&[J$-$H]$\geq 0.59$([H$-$K]$-0.187)+0.72$&\\
&[J$-$H]$\leq 1.55$([H$-$K]$-0.39)+0.85$&\\
&K$>12$&\\
\hline
&HAeBe&\\
\hline
&[H$-$K]$\geq 0.7$&\\
&[J$-$H]$\leq 1.55$([H$-$K]$-0.187)-0.008$&\\
&K$<12$&\\
\hline
\end{tabular}
\end{table}

Our final PMS selection scheme has been summarized in (Table \ref{table:class3}). The color and magnitude selection criteria were applied to all sources in our initial IRAC list satisfying i) the cleaning/reliability conditions in the \spitzer\ channels; ii) matched to a 2MASS source with quality flag better than `D' in all 2MASS bands; and iii) classified as (mainly) photospheric or without a successful classification using the YSO scheme from \citet{gutermuth2009} (i.e., no Class 0/I, Class II, or contaminant).  

A search for additional PMS was also carried out by extending our analysis to 2MASS sources \textit{in} the area covered by the \spitzer\ survey but without a suitable IRAC counterpart (i.e., satisfying our initial cleaning and reliability conditions) in the short wavelength \spitzer\ channels. The lack of a detection in \spitzer\, minimizes the possibility of confusion with actual embedded protostars, which should have been previously identified with the color/magnitude criteria from \citet{gutermuth2009}.

\subsection{The \spitzer\ Catalog} \label{sec:catalog}
Here we present the final products and source lists derived from the analysis carried out in the previous section.
Results and statistics from our YSO detection and classification procedures are shown in Tables \ref{table:catyso} and \ref{table:class3a}. For the purpose of this paper, we define as `YSO' those Class 0/I and Class II candidates selected using the color/magnitude scheme from \citet{gutermuth2009}. The sample of PMS stars, that is those additional candidate young stellar objects selected using 2MASS photometry, will include stars with optically thick disks, e.g., classical T Tauri stars and HAeBe stars (Class II) missed by the IRAC color/magnitude classification, and optically thin disks, e.g., weak-lined T Tauri stars (Class III sources). An analysis of this sample will be relevant to investigating the `oldest' young stellar population in W3.

\subsubsection{Statistics, Completeness \& Reliability}
\begin{figure*}[ht]
\centering
\includegraphics[scale=0.85,angle=0]{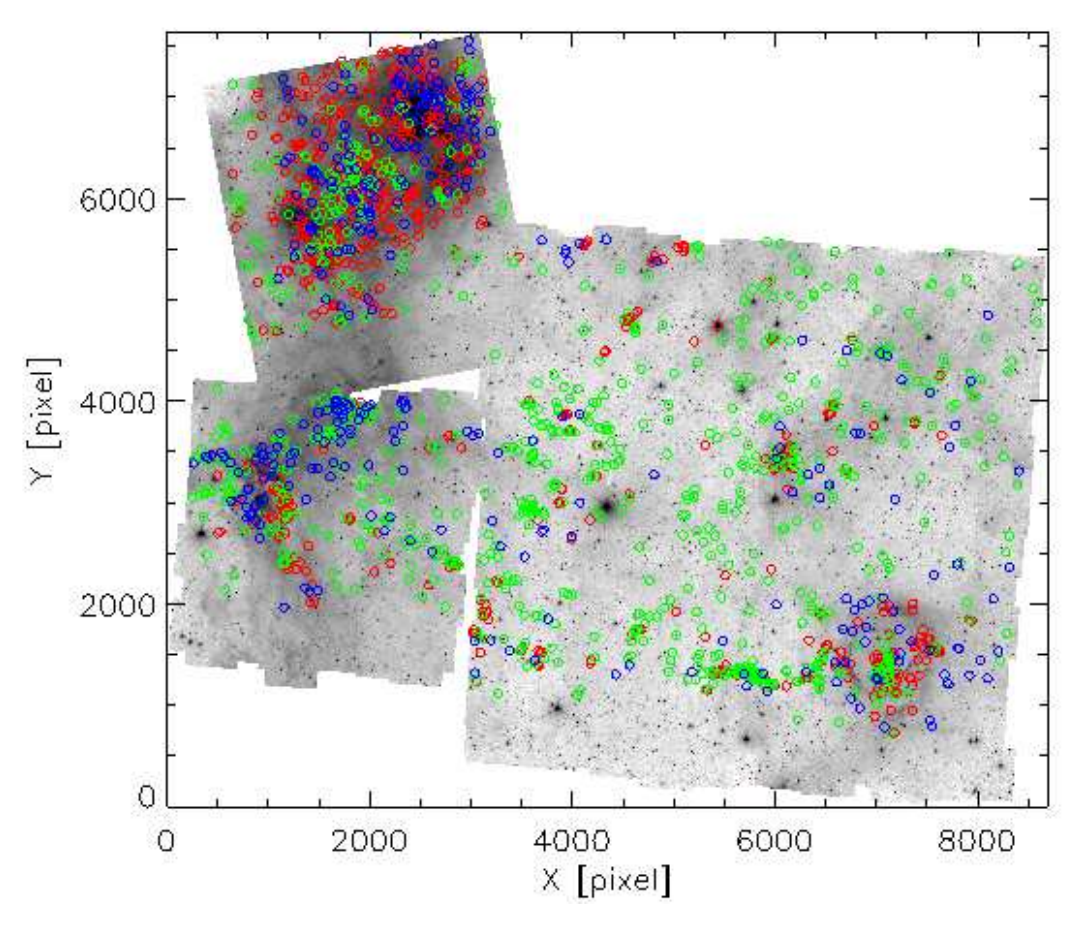}
\caption{Greyscale \spitzer\ channel 1 mosaic with Class 0/I (red), Class II (green) and PMS (blue) candidates. Sample includes all sources (and all flags) from Catalog 1, as well as PMS candidates with no IRAC counterparts (see text).}
\label{fig:cat1_mosaic}
\end{figure*}

The YSO search using the color and magnitude classification from \citet{gutermuth2009} yielded a total of $616$, $706$ (two of which were also observed in neighboring AORs), and $246$ YSOs in the regions surveyed in all four IRAC channels in each individual AOR: W3 Main/(OH), KR 140/KR 140-N, and AFGL 333 \spitzer\ regions, respectively (Fig. \ref{fig:cat1_mosaic}; Table \ref{table:class3a}). The full list of YSOs (Table \ref{table:catyso}) and 2MASS-based PMS candidates (Table \ref{table:catyso}a) appear in the electronic version of this article.

\begin{table*}[ht]
\caption{YSOs in each subregion of W3: Sample list$^a$}
\label{table:catyso}
\centering
\begin{tabular}{l l l l l l}
\hline
\hline
RA&Dec&Class&Class&Flag$^b$&Flag\\
h m s (J2000)&\degree\ \arcmin\ \arcsec\ (J2000)&Catalog 1&Catalog 2$^c$&Catalog 1&Catalog 2\\
\hline
...&&&&&\\
34 29 06.92&61 33 44.75&classII*&nomatch&$0$&$-1$\\
34 29 56.38&61 22 31.69&classII&classII&$1$&$1$\\
34 32 05.86&61 25 22.47&classII*&nomatch&$0$&$-1$\\
...&&&&&\\
\hline
\multicolumn{6}{l}{{$^a$Catalog is published in its entirety in the electronic version of this article.}}\\
\multicolumn{6}{l}{{$^b$ Reliability flag (see text):}}\\
\multicolumn{6}{l}{{1: Candidates satisfying cleaning/bandmerging requirements and with individual}}\\
\multicolumn{6}{l}{{detections in each band.}}\\
\multicolumn{6}{l}{{0: Candidates satisfying cleaning/bandmerging requirements but with IRAC}}\\
\multicolumn{6}{l}{{Channel 3 and 4, and MIPS 24\,\micron\ detection (centroid) and photometry}}\\
\multicolumn{6}{l}{{based on a successful detection in Channels 1/2.}}\\
\multicolumn{6}{l}{{$^c$ Same as Catalog 1, but without MIPS 24\,\micron-based re-classification}}\\
\multicolumn{6}{l}{{unless a successful detection was found in our original MIPS source list; i.e., MIPS}}\\
\multicolumn{6}{l}{{centroid (and photometry) not based exclusively on an IRAC detection.}}\\
\end{tabular}
\end{table*}

\begin{table*}[ht]
\caption{Photometry for YSOs in each subregion of W3$^a$}
\label{table:photometry}
\centering
\begin{tabular}{l l l l l l l l l l}
\hline
\hline
3.6\,\micron&Error&4.5\,\micron&Error&5.8\,\micron&Error&8.0\,\micron&Error&24\,\micron&Error\\
\hline
...&&&&&&&&&\\
12.08&0.01&12.05&0.01&11.93&0.03&12.03&0.09&9.34 &0.19\\
12.46&0.01&12.06&0.01&11.81&0.03&11.29&0.08&7.90 &0.36\\
12.98&0.01&12.98&0.01&12.86&0.05&12.95&0.12 &9.03 &0.14\\
...&&&&&&&&&\\
\hline
\multicolumn{10}{l}{{$^a$Photometry (magnitudes) for sample in Table \ref{table:catyso}.}}\\
\multicolumn{10}{l}{{Catalog is published in its entirety in the electronic version of this article.}}\\
\end{tabular}
\end{table*}

\begin{figure*}[ht]
\centering
\includegraphics[scale=0.85,angle=0]{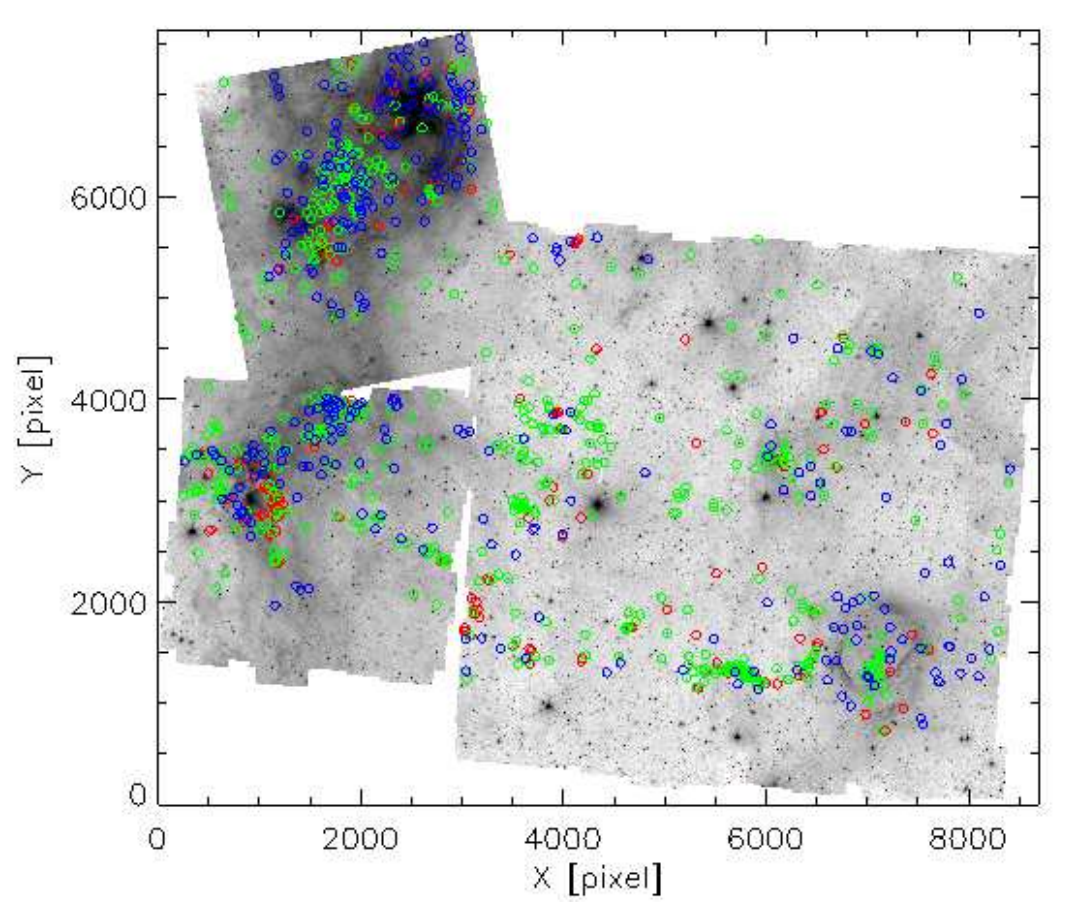}
\caption{Same as Fig.\ref{fig:cat1_mosaic} but for Catalog 2, without PMS candidates with no IRAC counterpart.}
\label{fig:cat2_mosaic}
\end{figure*}

For the purpose of the present analysis we chose reliability over completeness, and so this sample requires S/N$\geq5$ for any IRAC/MIPS photometry used in any particular method. When using 2MASS photometry in Phase 2 of the color/magnitude classification scheme, those sources with the closest distance to our IRAC sources (within $2$\arcsec) were chosen as suitable counterparts as long as they had a quality flag `A', `B', `C' or `D' in H and K-bands, and at least valid photometry (`D') in the J-band (if available). 

In order to remove as many artifacts and false detections as possible, we first bandmerged the IRAC channels using channel pairs. A candidate source could still be included in the initial list without suitable photometry in all four channels if it either appeared in channels 1 and 2, or channels 3 and 4. This was done as the first step to minimize the need for visual inspection of samples containing tens of thousands of sources, while at the same time attempting to minimize the loss of relatively `cold' sources without a suitable detection at shorter wavelengths.
In addition, we also imposed our own `internal' cleaning conditions based on the properties provided by APEX and SExtractor for those sources already published as reliable detections in the catalog of \citet{ruch2007} (e.g., successful PRF fitting, $\chi^2$ of fitting, ellipticity and successful deblending). We note, however, that these internal cleaning parameters are physically meaningless and were only used for `relative' classification of sources within a particular sample, with the only purpose being to reject as many artifacts (e.g., PSF residuals and spikes around bright sources) and false detections as possible. Despite this procedure, these conditions were still conservative, and visual inspection and manual rejection was still required and performed in the last stages of the catalog production process. All sources successfully classified as YSO candidates and satisfying all the cleaning conditions and bandmerging requirements have an entry$=1$ in the `flag' column in the source catalog (Table \ref{table:catyso}). This defines the `reliable' subset of the final list (Fig. \ref{fig:cat1_mosaic}).

In an attempt to improve the `completeness' of the sample we also used the detections in IRAC channels 1 and 2 (with the best sensitivity) as a base for a new source list. Photometry was performed on fixed centroids at longer wavelengths (including MIPS), and all the cleaning/selection conditions and the Gutermuth classification scheme were again applied to each source. Those additional detections satisfying all the catalog requirements were included in the final catalog with a flag entry$=0$. We note that in the scheme of \citet{gutermuth2009}, MIPS photometry was used mainly to reclassify as YSOs those sources initially rejected as galaxies, AGN, or fake excess sources in the first steps of the scheme.

Photometry derived from the MIPS $24$\,\micron\ maps was generally less reliable. We therefore produced a Catalog 2 (Fig. \ref{fig:cat2_mosaic}) based on the same procedure as Catalog 1, but allowing for re-classification of IRAC/2MASS sources only if there was a successful MIPS counterpart in our original (independently obtained) MIPS source list (i.e., the standard fixed-centroid photometry is not used in MIPS phase 3). Catalog 2 is therefore more conservative, because although the SExtractor extraction was visually observed to detect all significant sources, a large fraction of these detections did not satisfy the cleaning conditions after performing APEX PRF fitting on the MIPS mosaics, on account of the variable and complicated background at longer wavelengths. Both catalogs (Catalog 1 and Catalog 2) yield very similar source lists for Class 0/I and Class II sources, differing mainly on the number of Class 0/I$^*$ (highly embedded YSOs) and Class II$^*$ transition objects. Defined in this work as the ($^*$) population, both the highly embedded and transition objects rely on MIPS photometry for identification and classification. A summary of the number of candidates found of each class in each field is given in Table \ref{table:class3a}. As expected, highly embedded and transition objects are particularly abundant in Catalog 1 due to the use of MIPS photometry based on IRAC centroids, with Class 0/I$^*$ sources forming up to $\sim65\%$ of the population in W3 Main/(OH) ($\sim5.5\%$ Class II$^*$). In Catalog 2 these types of objects constitute less than $\sim3.5\%$ of the YSOs in each field. 

We created a sample resulting from bandmerging just channels 3 and 4 and ran it through the cleaning and classification procedures described above. This experiment produced no new sources. This shows that no significant sample of `cold' sources (without detections in IRAC channels 1 and 2) should have been missed by the initial bandmerging-by-pairs procedure (within the limitations of this technique). We therefore conclude that the majority of sources potentially missed in our (flag $=1$) analysis would have come from the samples in IRAC channel 1/channel 2 that do not have counterparts at longer wavelengths due to the diffuse emission and sensitivity loss (although we note that Catalogs 1 and 2 for flag $=1$ are very similar; Table \ref{table:class3a}). 

Unless mentioned otherwise, in the following sections we will use Catalog 1 (all flags) as the primary sample in our analysis. As we explain in Section \ref{sec:groupclass}, this source list is expected  to be a more reliable indicator of the YSO properties.  
We find that the use (or omission) of the ($^*$) population is mostly relevant in those highly populated regions with high extinction, mainly IC 1795 and KR 140. We cross-correlated our final YSO list with the list of infrared sources associated with the cluster IC 1795 presented in \citet{rocca2011}, and we find 76 YSOs in our sample which are consistent with being cluster members. Of these, 30 YSOs belong to the ($^*$) population in Catalog 1 (not classified as YSOs in Catalog 2), and yet all are also classified as YSOs (Class II) in \citet{rocca2011}. When applying the color classification from \citet{megeath2004} we find that only 4 out of the 76 sources would not have been classified as YSOs, which supports our decision and the need to keep the ($^*$) population in our analysis. Results based on Catalog 2 are only mentioned briefly when required.

We finally note that while the flag $=0$ and IRAC short wavelength-based catalogs intend to improve the completeness of the final sample  (and each source was visually inspected in channel 1), these detections are still tentative and should be treated with caution.

\begin{table}[ht]
\caption{YSOs in the \spitzer\ Survey}
\label{table:class3a}
\centering
\begin{tabular}{l l l l l}
\hline
\hline
\multicolumn{5}{c}{{W3 Main/(OH)}}\\
\hline
&Catalog 1&Catalog 1&Catalog 2&Catalog 2\\
Flag&all$^a$&1&all&1\\
\hline
Class 0/I&$39$&$32$&$39$&$32$\\
Class 0/I*&$405$&$7$&$7$&$7$\\
Class II&$138$&$85$&$132$&$85$\\
Class II*&$34$&$3$&$3$&$3$\\
HAeBe$^b$&$1$&$0$&$2$&$0$\\
T-Tauri$^b$&$110$&$50$&$126$&$58$\\
\hline
\multicolumn{5}{c}{{KR 140}}\\
\hline
Class 0/I&$88$&$80$&$84$&$78$\\
Class 0/I*&$130$&$3$&$3$&$3$\\
Class II&$271$&$252$&$271$&$252$\\
Class II*&$215$&$3$&$3$&$3$\\
HAeBe&$0$&$0$&$0$&$0$\\
T-Tauri&$94$&$56$&$96$&$57$\\
\hline
\multicolumn{5}{c}{{AFGL 333}}\\
\hline
Class 0/I&$57$&$51$&$57$&$51$\\
Class 0/I*&$25$&$0$&$0$&$0$\\
Class II&$140$&$129$&$140$&$129$\\
Class II*&$24$&$2$&$2$&$2$\\
HAeBe&$0$&$0$&$0$&$0$\\
T-Tauri&$86$&$43$&$86$&$43$\\
\hline
\multicolumn{5}{l}{{$^a$ Flag $=0$ \& Flag $=1$ combined.}}\\
\multicolumn{5}{l}{{$^b$ With \spitzer\ counterparts.}}\\
\end{tabular}
\end{table}

We compared Catalog 1 and Catalog 2 with the detections from \citet{ruch2007} for W3 Main/(OH), the region analyzed in their study. These authors detected a total of $295$ sources in the four IRAC channels, $21$ ($\sim7\%$) of which were classified as Class I sources, and $94$ ($\sim32\%$) as Class II. All $295$ sources were in our initial catalog resulting from the SExtractor source detection process. Our analysis identifies a similar number of Class I and Class II candidates in this sample, with a total of 117 sources classified as YSOs in Catalog 1 (93 in Catalog 2). Of the remaining sources in their list not classified as YSOs in Catalog 1, $130$ are stars, $0$ galaxies, $3$ shock/knots of emission or sources with PAH contaminated apertures, and $45$ do not satisfy the cleaning and reliability conditions. Sources in their sample not classified as YSOs in Catalog 2 consist of $149$ stars, $0$ galaxies, $3$ shock/contaminated aperture sources, and $50$ sources not satisfying cleaning/reliability conditions. In all cases, the percentage of Class II sources is $\sim3$ times that of Class 0/I sources. Table \ref{table:ruch} shows the results after applying our cleaning conditions and the Gutermuth scheme.

\begin{table}[ht]
\caption{Breakdown of candidate YSOs with counterparts in the catalog of \citet{ruch2007}}
\label{table:ruch}
\centering
\begin{tabular}{l l l l l}
\hline
\hline
Class&Catalog 1&Catalog 1&Catalog 2&Catalog 2\\
Flag&all&1&all&1\\
\hline
Total YSO&$117$&$77$&$93$&$77$\\
Class0/I&$21$&$20$&$21$&$20$\\
Class0/I*&$8$&$0$&$0$&$0$\\
ClassII&$72$&$57$&$72$&$57$\\
ClassII*&$16$&$0$&$0$&$0$\\
\hline
\end{tabular}
\end{table}

Table \ref{table:class3a} includes the number of PMS stars found based on 2MASS color and magnitude information, with counterparts in the \spitzer\ images. We found 51 more T-Tauri stars and 2 HAeBe stars when searching for 2MASS sources (from the 2MASS PSC) without suitable IRAC counterparts (i.e., IRAC sources satisfying our initial cleaning conditions), but located in the common (4-channels) areas surveyed by \spitzer. Many of these sources are located in the bright, diffuse region surrounding W3 Main, (OH) and AFGL 333 in the IRAC images, which did not allow for proper identification and extraction of candidate sources due to confusion. The imposed lower limit for the K magnitude of our HAeBe stars is within the 2MASS $10\sigma$ completeness limit for this band ($14.3$\,mag). This sample is also complete in the H band ($15.0$\,mag), and J band ($15.9$\,mag). The faintest 2MASS magnitudes for the T-Tauri sample lie beyond the 2MASS completeness limits by $\sim1$\,mag in K and H, and $\sim2$\,mag in J band, and therefore our list will be incomplete at the faint end of the population.

\begin{table*}[ht]
\caption{IRAC completeness limits for Catalog 1}
\label{table:complete}
\centering
\begin{tabular}{l c c c c}
\hline
\hline
AOR&Channel 1&Channel 2&Channel 3&Channel 4\\
&All/Flag$=1$&All/Flag$=1$&All/Flag$=1$&All/Flag$=1$\\
\hline
W3 Main/W3(OH)&$14.1/12.4$&$12.9/11.8$&$12.7/10.6$&$11.3/10.3$\\
KR 140&$13.1/13.7$&$13.1/13.0$&$13.2/12.8$&$12.2/11.9$\\
AFGL 333&$12.9/12.8$&$12.1/11.8$&$11.9/11.6$&$10.4/10.9$\\
\hline
\end{tabular}
\end{table*}

\begin{figure*}[ht]
\centering
\includegraphics[scale=0.7,angle=0]{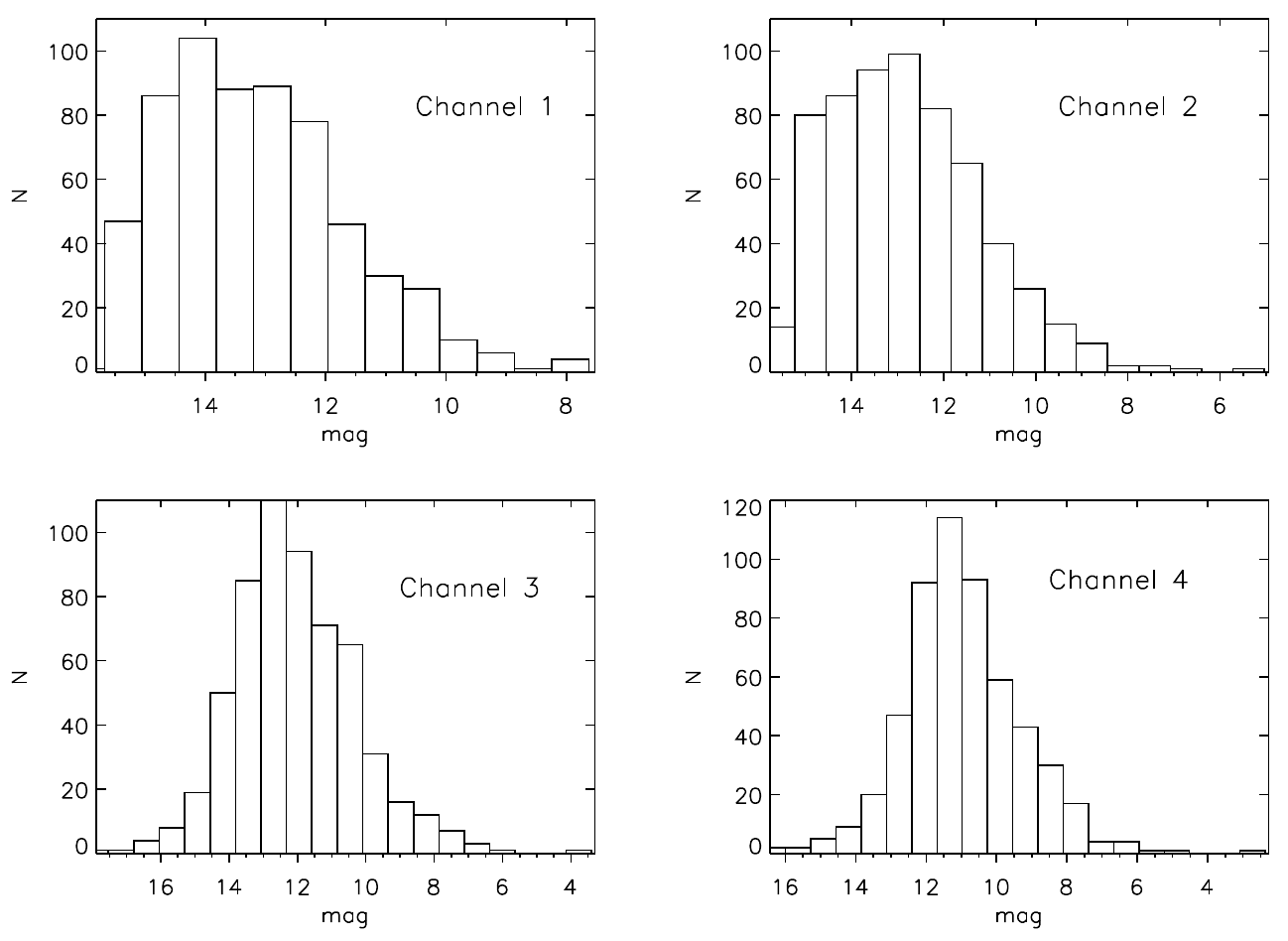}
\caption{Number of YSO candidates as a function of magnitude and channel for the region comprising W3 Main and W3 (OH). Channels 1-4 are shown in order from left to right, top to bottom. Completeness limits for each region (Table \ref{table:complete}) have been derived from the turnover points of the distributions.}
\label{fig:completeness}
\end{figure*}

The use of several steps of cleaning and reliability thresholds for our final sample, combined with different sensitivity limits in different regions for a given field, complicate the determination of a reliable completeness limit for our catalog. In order to provide an estimate of the magnitude at which a catalog is 100\% complete, which due to our cleaning and selection steps differs from a fainter limit at which some sources can be detected, we examined the number of detections as a function of magnitude for each field at each IRAC wavelength. Figure \ref{fig:completeness} shows the histograms for W3 Main and W3 (OH) for all YSOs in Catalog 1, the most complete sample derived in this work. Our estimates for 100\% completeness, given by the `turnover' points in the YSO distributions for each subregion of W3, are included in Table \ref{table:complete}. The effects of strong, large scale emission in dense regions with high stellar activity is evident from the limits derived for the HDL. 

Our choice of reliability over completeness excludes a large population of faint sources from our final list, and therefore the completeness limits for the different regions reveal a particularly conservative population for IRAC channels 1 and 2. The flag $=0$ sample (using the short wavelength channels as a template for photometry at longer wavelengths) improves completeness at these channels significantly, although we note that our estimated completeness limits are still conservative compared to samples derived from similar \spitzer\ studies of W3 \citep{polychroni2010}. Clearly, the completeness magnitudes quoted for our work are upper (faint) limits, especially for the long wavelength channels. The use of flag $=1$ data or the sample from Catalog 2 would be more conservative (and therefore less complete). The latter candidate source list, while more reliable than Catalog 1, is expected to suffer a more severe loss of highly embedded sources. This is due particularly to the rejection of poorly fitted sources (failed PRF fitting) with APEX, and our strict requirement of suitable fitting and photometry in the MIPS band (with severe sensitivity loss and confusion due to strong diffuse emission). We expect to compensate for this issue with the Herschel data currently being analyzed, which will be able to constrain the dust emission much more accurately for those embedded (younger) sources.

With regard to Class types, Class II sources are generally weaker at the longer IRAC wavelengths than more embedded Class 0/I objects, and are therefore more likely to be missed because of lower sensitivity and bright PAH emission (e.g., \citealp{chavarria2008}).

Despite our attempt to compensate for saturation in the IRAC sample by using the short exposure mosaics, some sources will still be missed in the brightest and most active regions such as W3 Main and W3 (OH) due to confusion and location within the bright infrared nebulosity. The catalogs would also exclude clearly extended sources and stellar groups. 

Due to our attempt to provide a reliable sample for our PMS sources, we again note that weak-lined and low extinction T-Tauri and HAeBe stars have been excluded from our catalog in order to avoid major contamination from reddened main sequence stars. We cross-checked our final list of candidate T-Tauri and HAeBe stars with the {\sc simbad} database, and found no matches with the exception of a couple of sources, classified in the database as `infrared sources' or `star in cluster'. Ultimately, spectroscopic surveys of stellar candidates are the most reliable method to find and confirm PMS.

\subsubsection{Contamination}
The \spitzer\ based sample was subjected to strict cleaning conditions and cross-checked with other classification schemes to ensure the highest consistency and reliability of our sample of YSO candidates. However, it is likely that some contamination from extraneous objects like galaxies/AGN, planetary nebulae (PNe) and AGB stars is present in the final sample.

A measure of the contamination by PNe and AGB stars was obtained by applying the conditions from \citet{rob2008} to the reliable photometry derived from the previous YSO classification. For successful classification as PN, detections needed to satisfy at least two of the four color conditions to ensure reliability in their position in the CCDs. While the selection scheme still allows for mutual contamination of YSO/AGB stars in both samples, this technique can still provide a useful measure of the general contamination of our YSO sample. The PN contamination using this method is predicted to be $\leq1.5\%$ for W3 Main/(OH), and $\leq1\%$ and $\leq2\%$ for KR 140 and AFGL 333, respectively. The contamination from AGB stars is expected to be $<3.0\%$ and $<0.5\%$ for W3 Main/(OH) and AFGL 333. No candidate AGB stars are found in KR 140 using this scheme. 

In their analysis, \citet{gutermuth2009} adapted their classification to work with the larger distances present in their survey of up to $\sim1$\,kpc. The use of this classification at the distance of W3 ($\sim2$\,kpc) is expected to shift the proportion of classified YSOs toward the brightest sources, because of the loss of dimmer sources rejected through the magnitude limits imposed on the sample and a possible contamination of the galaxy sample with YSOs.
To check this effect, we ran the codes on the YSO sample in W5 from the work of \citet{koenig2008}, applying the new classification conditions to the photometry provided by these authors.
W5 belongs to a massive complex (neighboring W3 and W4) and is considered to be at the same distance as these GMCs. These authors found a significant proportion of YSOs being misclassified as non YSOs, which was evident, for instance, in the `clustered' properties of the `galaxies'.

Although the code used in the present work is a `revised' version of that used by \citet{koenig2008}, we successfully classified $\sim95.5\%$ of their sources as YSOs of the same type (excluding the very few sources not satisfying the requirement of magnitude error $<0.2$\,mag). About two thirds of the $4.5\%$ sources where we disagreed with the classification from \citet{koenig2008} were classified as AGN, and $0\%$ as PAH galaxies. Increasing the magnitude limits for AGN classification by $0.5$\,mag to account for the larger distance of W3 (e.g., \citealp{megeath2009}), we find no differences in the percentage of sources classified as AGN. Although we cannot perform this test on their rejected (non-YSO) sample (as that data was not included in their publication), we expect this code to detect and recover successfully the main YSO population in W3, without the need for major additional modifications. 

This is in agreement with the following analysis of the distribution of those sources classified as galaxies or AGNs. Using the `Distance to Nearest Neighbor' technique \citep{clark1954} we measured the mean observed distance between galaxy candidates relative to the mean distance that would be expected for a random distribution for a population with the same characteristics. The ratio between these two quantities (R) would be equal to one for a perfectly random distribution. We obtain R$=0.94$, $0.98$, and $0.88$, with significance levels of $\sim25\%$, $\sim37\%$ and $\sim5\%$ for W3 Main/(OH), KR 140 and AFGL 333, respectively. This supports the random spatial distribution of our galaxy candidates and low YSO contamination based on the lack of significant clustering. Coordinates for the list of galaxy candidates is also available online (Table \ref{table:catyso}b).

%%%%%%%%%%%%%%%%%%%%%%%%%%%%%%%%%%%%%%%%%%%%%%%%%%%%%%%%%%%%%%%%%%%%%%%%%%%%%%%%%%%%%%%%%%%%%%
\begin{figure*}[ht]
\centering
\includegraphics[scale=0.74,angle=0]{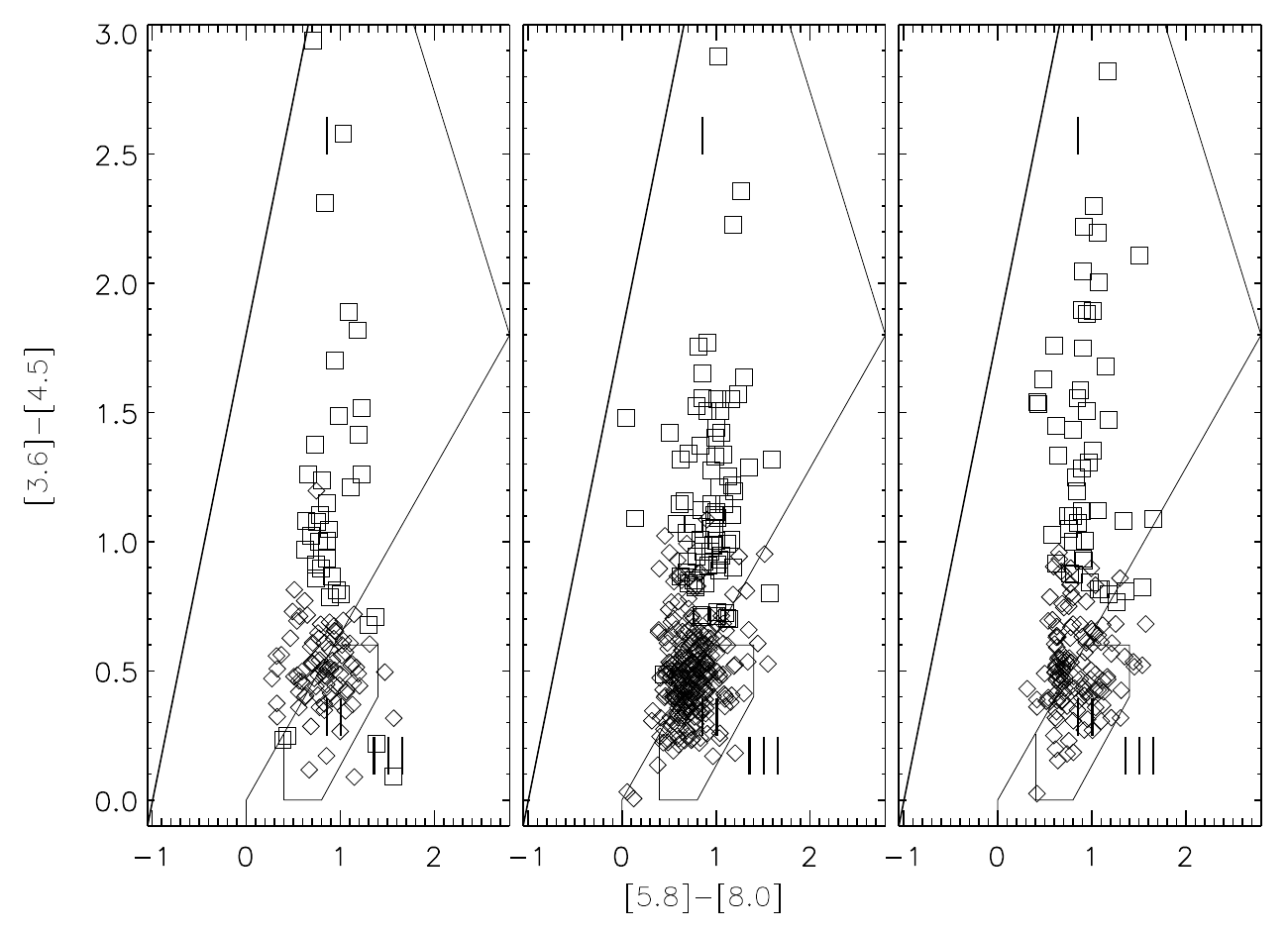}
\caption{IRAC CCD of Class0/I+(*) (squares) and ClassII+(*) (diamonds) YSOs from Catalog 1 in W3 Main/(OH) (left), KR 140 (middle) and AFGL 333 (right). Black solid lines mark the areas where the majority of Stage I, II and III sources from \citet{rob2006} are found.}
\label{fig:stage1}
\end{figure*}

\begin{table*}[ht]
\caption{General 2MASS Properties for `Class 0/I' and `Class II' Populations}
\label{table:stageprops2}
\centering
\begin{tabular}{l l l l l l l l}
\hline
\hline
Class&Mean [H-K]&$\sigma$&Mean [J-H]&$\sigma$&Mean K&$\sigma$&Objects\\
\hline
0/I&1.5&0.6&2.0&0.8&13.6&1.1&68\\
II&0.9&0.4&1.4&0.5&13.6&1.1&412\\
\hline
&&T-Test Stat.&Significance&F-Test Stat.&Significance&&\\
\hline
0/I vs. T-Tauri&[H-K]&11.5&8E-22&6.1&6E-13&&\\
&[J-H]&10.0&3E-18&9.3&6E-18&&\\
&[K]&-5.0&2E-6&1.3&0.3&&\\
0/I vs. II&[H-K]&10.8&2E-24&2.3&5E-07&&\\
&[J-H]&7.9&3E-14&2.6&1E-08&&\\
&[K]&-0.8&0.4&1.1&0.6&&\\
0/I vs. HAeBe&[H-K]&3.7&4E-04&2.8&4E-03&&\\
&[J-H]&6.2&1E-08&3.1&2E-03&&\\
&[K]&13.2&3E-23&2.2&0.01&&\\
II vs. T-Tauri&[H-K]&5.5&6E-08&2.6&2E-06&&\\
&[J-H]&7.1&5E-12&3.6&1E-09&&\\
&[K]&-5.8&1E-8&1.2&0.5&&\\
II vs. HAeBe&[H-K]&-2.2&0.03&1.2&0.6&&\\
&[J-H]&4.5&1.0E-05&1.2&0.6&&\\
&[K]&18.4&0.00&2.4&8E-04&&\\
\hline
\end{tabular}
\end{table*}

\section{YSO Analysis: Observed and Intrinsic Properties}\label{sec:discussion}
In this work we aim to produce a reliable list of YSO candidates, which will be used in our upcoming papers to investigate the early stages of massive star formation in W3. An estimate of the mass of each of our YSO candidates in the different classes is a crucial component of our analysis. Spectral energy distributions (SEDs) have been extensively used to estimate this parameter (e.g., \citealp{rob2006}). However, the highly embedded state of some of these sources demands proper modelling of the dust envelope and disk, which will be the focus of our future Herschel-based analysis. 

Based on the classification scheme proposed by \citet{gutermuth2009} we have produced a list of YSO candidates in the regions of W3 Main/(OH), KR 140, and AFGL 333, without any preliminary bias on selection according to clump/core association. This classification has been compared to and supplemented by alternative classification schemes, as described below.

We classified sources as `red' according to the nomenclature from \citet{rob2008}, which should include all Class I, Flat, and a large number of Class II sources as defined by \citet{lada1987} and \citet{greene1994}. Only three sources in KR 140 classified as Class 0/I in Catalog 1 (and two in Catalog 2) were not initially classified as red sources. These sources showed a slight flattening of the SED at longer wavelengths which is responsible for this misclassification, but all have been visually checked and their rising SEDs are consistent with embedded sources.

In this section we characterize the behavior of the \spitzer\ YSOs in the 2MASS and IRAC color-magnitude space. With this analysis we investigate the presence of possible identifying characteristics for the different classes, as well as near/mid infrared properties that may help in the identification of low mass and intermediate/high mass YSOs in our sample (e.g., T-Tauri and HAeBe PMS stars) based on the information used in this work.

\subsection{YSO Stages and the IRAC CCD}

The `stage' classification from \citet{rob2006} is particularly useful when combined with the `Class' scheme in order to avoid contradictions between observed (color, magnitude) and inferred  (e.g., $\dot{M}_{\mathrm{env}}$; M$_{\mathrm{disk}}$) properties. Figure \ref{fig:stage1} shows our YSO sample from Catalog 1 (flag $=1$) and the regions in the CCD that most sources of different stages are predicted to occupy (these and all the other figures in this analysis do not include error bars for clarity). We obtain a similar figure when using the candidate list from Catalog 2. While this scheme is useful to separate sources with and without circumstellar material, and a significant proportion of Stage I sources can be separated from the remaining population, some Stage I objects may however still exist in the regions occupied by Stage II (disk domain) and Stage III sources.   

For all catalogs, we observed that more than $\sim95\%$ of Class 0/I YSOs (flag $=1$) and $\sim90\%$ (flag $=0$) are classified as Stage I sources. Flag $=1$ Class II (including Class II$^*$) YSOs are more evenly distributed ($\sim50\%$) between Stage I and Stage II, although \textit{none} of these sources is as red in [3.6]-[4.5] as Class 0/I (including Class 0/I$^*$) candidates (Figure \ref{fig:stage1}). We observe $<1\%$ of Class II sources above [3.6]-[4.5]$>1$, and none above $1.2$. We consider this to be the limit separating the area exclusive to Class 0/I Stage I sources, and the area (bluer colors) where Class 0/I and Class II Stage I are mixed, perhaps indicative of a transition from envelope to optically thick disks. 

Very few sources lie in the Stage III area, which confirms the robustness of our sample and ability to separate embedded sources and optically thick disks from those consistent with photospheric colors and optically thin disks.

\subsection{The 2MASS CCD and CMD}
We next analyzed the behaviour of the `stage' vs `class' classification in the 2MASS CCD and CMD for those IRAC sources (Catalog 1, quality flag $=1$) with counterparts in the 2MASS PSC. As mentioned above, both Catalog 1 and 2 yield almost identical samples for Class 0/I and Class II sources, and therefore the conclusions below from this analysis are independent of the catalog used.

\subsubsection{Analysis of observed (Class) properties}

In Section \ref{sec:class} we established a scheme to separate the populations of T-Tauri and HAeBe stars based on magnitude and color information. HAeBe stars occupy a distinct region in the CCD, but K magnitude data is still essential to separate the two samples in the region near the T-Tauri locus populated by both types. 

%\clearpage
\begin{figure}[htp]
\centering
\includegraphics[scale=0.6,angle=0]{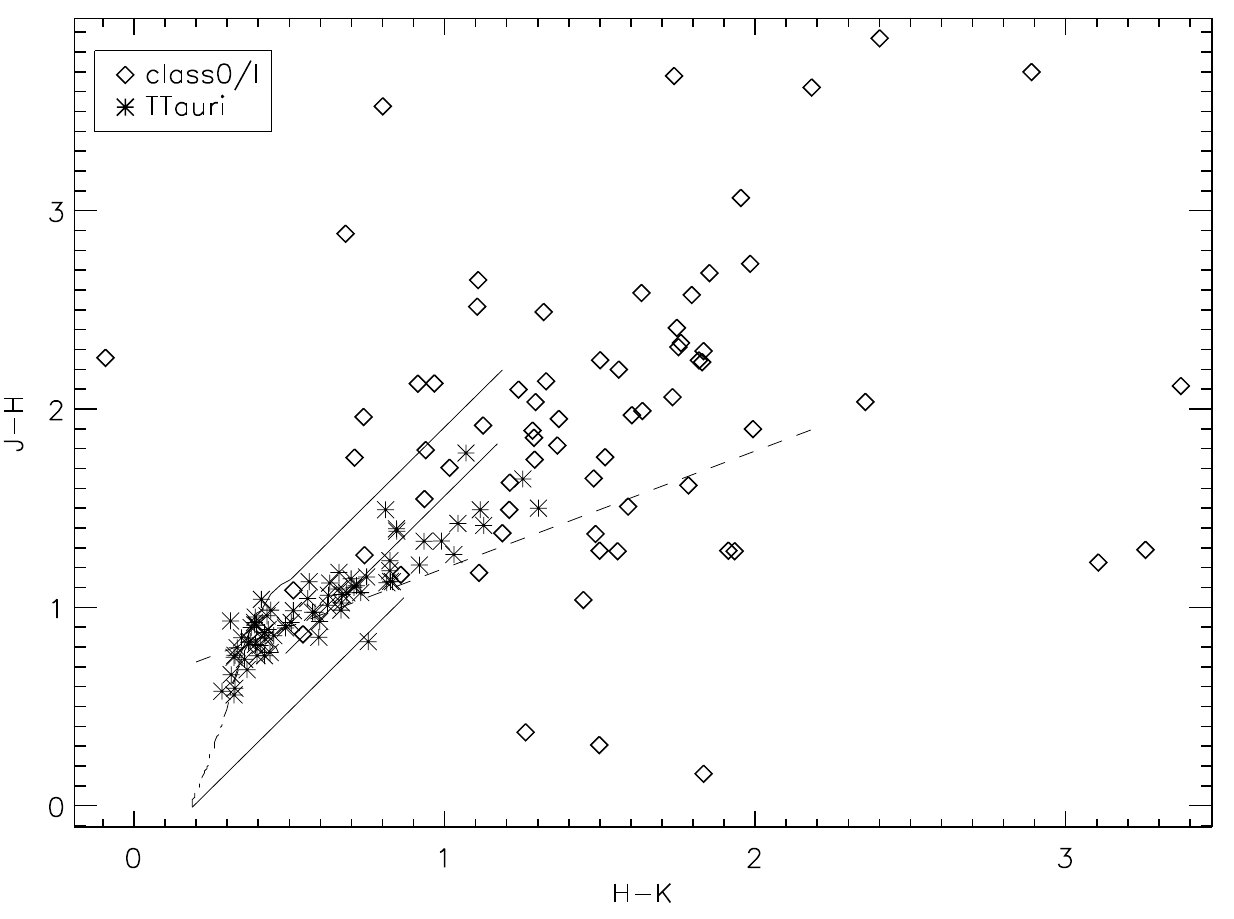}
\caption{2MASS CCD showing Class 0/I IRAC sources with 2MASS counterparts. Lines and T-Tauri data as in Fig. \ref{fig:ccd1}.}
\label{fig:class_ccd}
\end{figure}

\begin{figure}[htp]
\centering
\includegraphics[scale=0.6,angle=0]{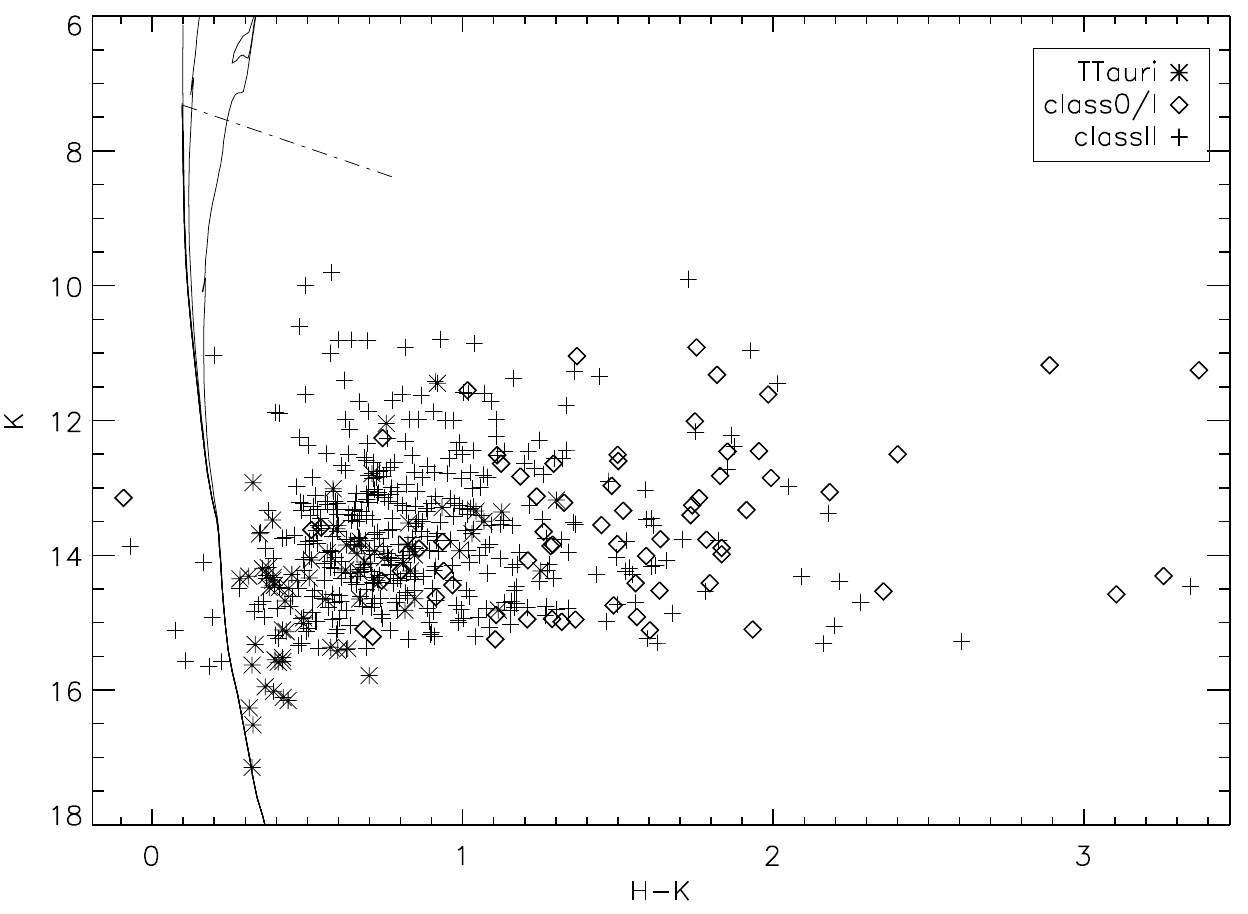}
\caption{2MASS CMD showing Class 0/I and Class II IRAC sources with 2MASS counterparts. Lines and T-Tauri data as in Fig. \ref{fig:cmd1}.}
\label{fig:class_cmd}
\end{figure}

\begin{figure}[htp]
\centering
\includegraphics[scale=0.6,angle=0]{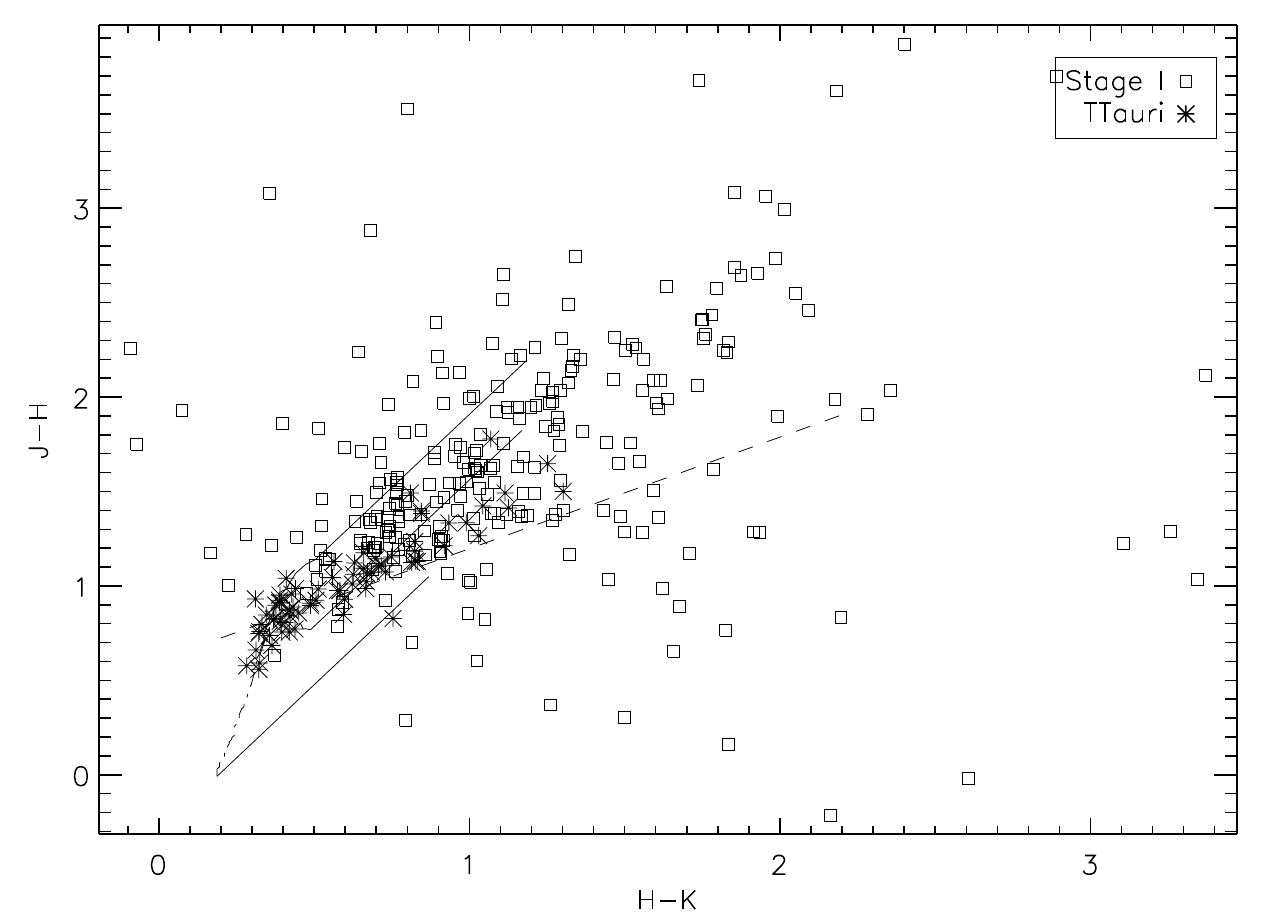}
\caption{2MASS CCD showing Stage I IRAC sources with 2MASS counterparts. Lines and T-Tauri data as in Fig. \ref{fig:ccd1}.}
\label{fig:stage1_ccd_overall}
\end{figure}

\begin{figure}[htp!]
\centering
\includegraphics[scale=0.6,angle=0]{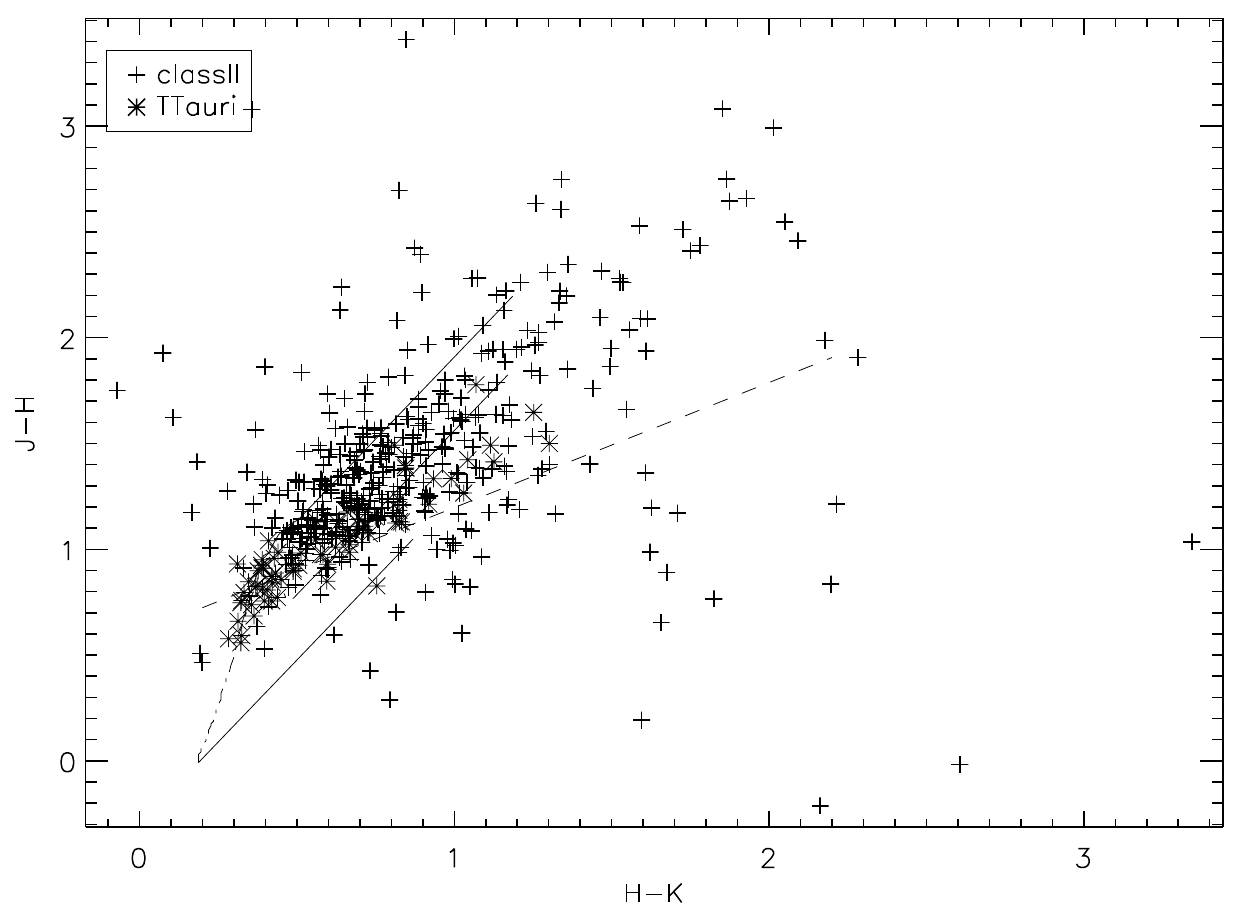}
\caption{Same as Fig. \ref{fig:class_ccd}, but for Class II sources.}
\label{fig:class_ccd2}
\end{figure}

\begin{figure}[htp]
\centering
\includegraphics[scale=0.6,angle=0]{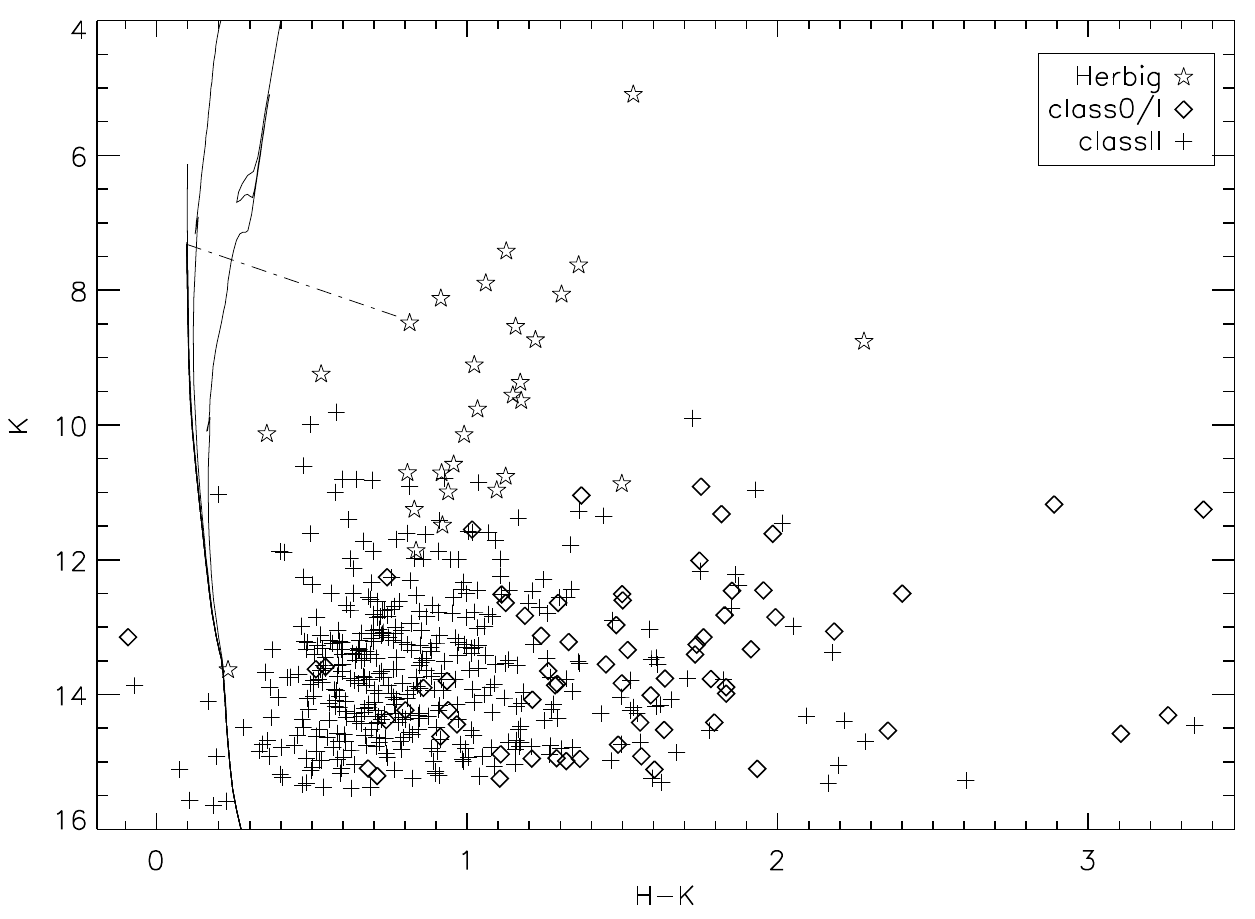}
\caption{Same as Fig. \ref{fig:class_cmd}, but with HAeBe stars.}
\label{fig:class_cmd_herbig}
\end{figure}

\begin{figure}[htp]
\centering
\includegraphics[scale=0.6,angle=0]{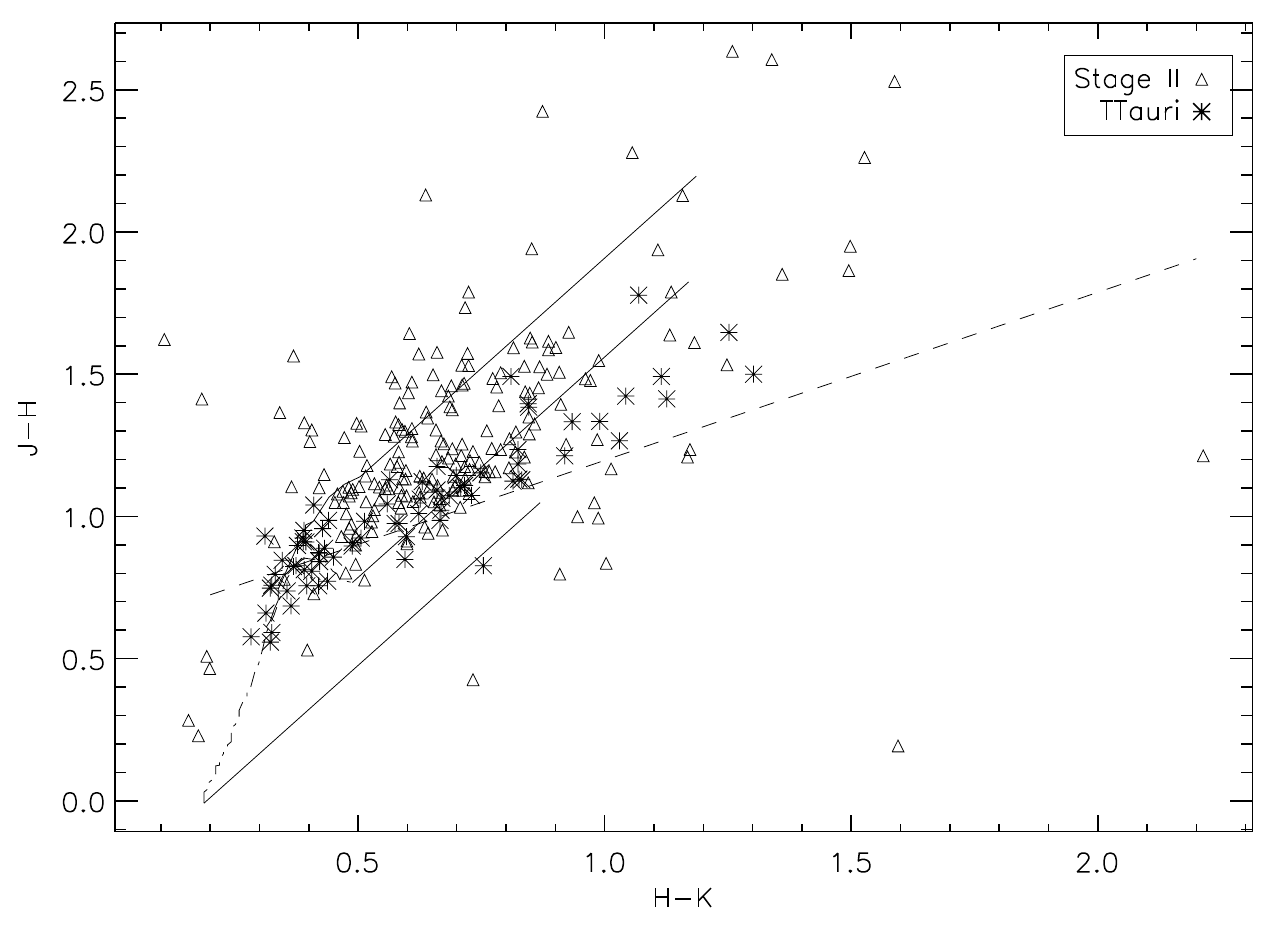}
\caption{Same as Fig. \ref{fig:stage1_ccd_overall}, but for Stage II sources.}
\label{fig:stage2_ccd_overall}
\end{figure}

We carried out T-Test and F-Test analyses comparing the color and magnitude distributions of various classes. Results are reported in Table \ref{table:stageprops2}. Statistically, and as in previous cases, the Class 0/I, Class II and T-Tauri samples are indistinguishable in the 2MASS CMD diagram, and therefore selection of low mass PMS stars is not possible in the infrared without prior knowledge of the protostar population. In color, the Class0/I sample is intrinsically redder, with the main population of Class II sources lying more intermediate between Class 0/I (and closer) to the T-Tauri sample (Figures \ref{fig:class_ccd}, \ref{fig:class_ccd2}, and \ref{fig:class_cmd}). $\sim90\%$ of Class 0/I sources lie between $0.7\leq$[H-K]$<2.5$, compared to Class II sources, located within $0.4\leq$[H-K]$<1.7$. While both have similar maximum [H-K] value of $\sim3.4$, $\sim80\%$ and $\sim50\%$ of Class 0/I sources lie above [H-K]$=1.0$ and $1.5$ respectively, compared to $\sim30\%$ and $\sim9\%$ for the Class II population.

We find a completely different scenario when focusing on the intermediate mass HAeBe population. 
As shown in Figure \ref{fig:cmd1}, Herbig stars are not only redder, but brighter than T-Tauri stars, with both populations clearly separated in the CMD. Class 0/I candidates \textit{are not} consistent with the colors \textit{or} magnitudes of HAeBe stars (Class 0/I being redder and dimmer). Class II sources are consistent in color with Herbig stars (i.e., intermediate between T-Tauri and Class 0/I), but are again dimmer than typical HAeBe stars (Figure \ref{fig:class_cmd_herbig}). While a spectroscopic analysis is still required to confirm our HAeBe candidates as such, we find that, contrary to T-Tauri stars, Herbig stars may still be selected without prior knowledge of the embedded population.

\subsubsection{Comparative Analysis of Intrinsically Different Populations (Stages)}

\begin{figure}[htp!]
\centering
\includegraphics[scale=0.6,angle=0]{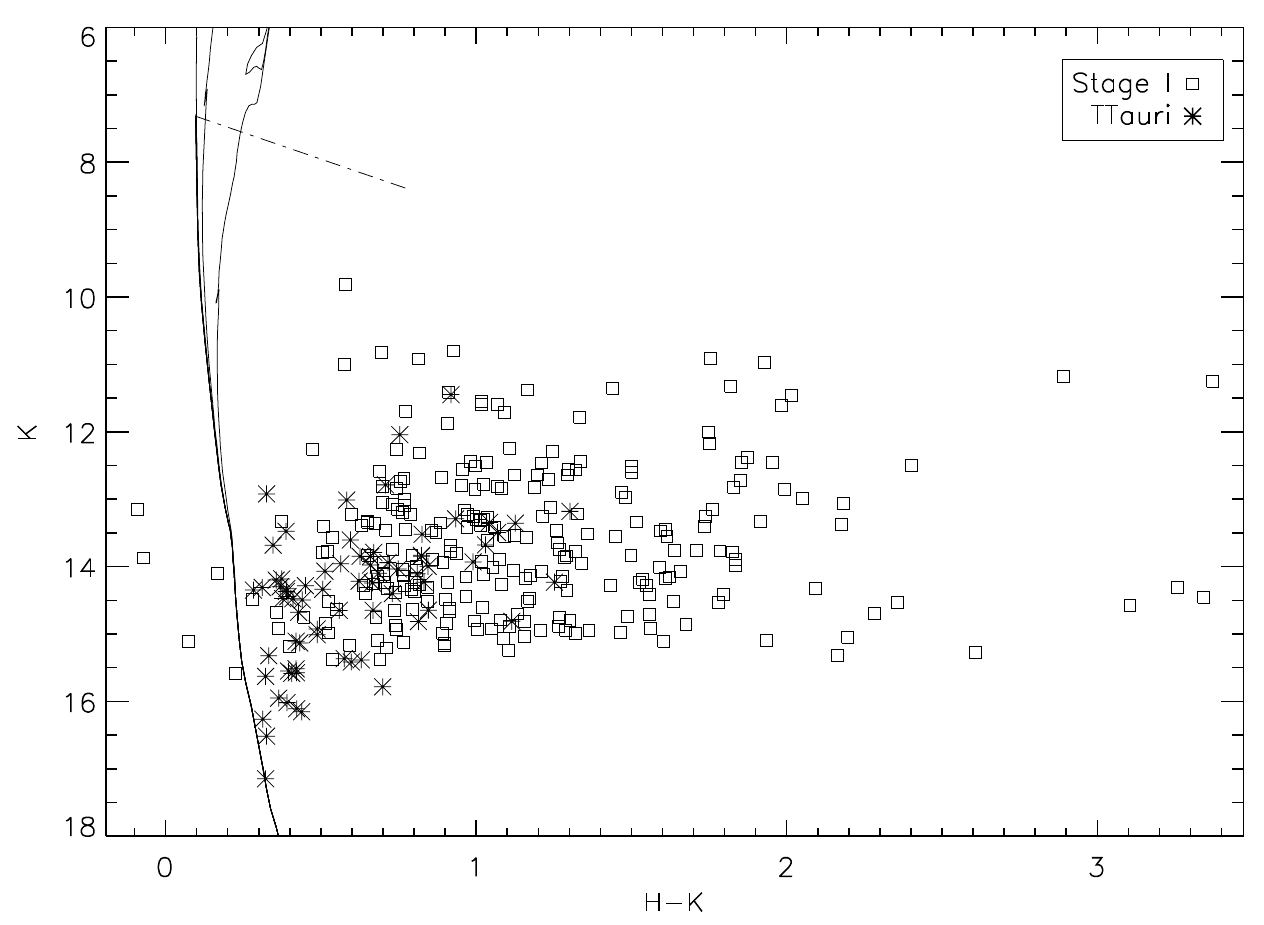}
\caption{2MASS CMD showing Stage I IRAC sources with 2MASS counterparts. Lines and T-Tauri data as in Fig. \ref{fig:cmd1}.}
\label{fig:stage1_cmd_overall}
\end{figure}

\begin{figure}[htp!]
\centering
\includegraphics[scale=0.6,angle=0]{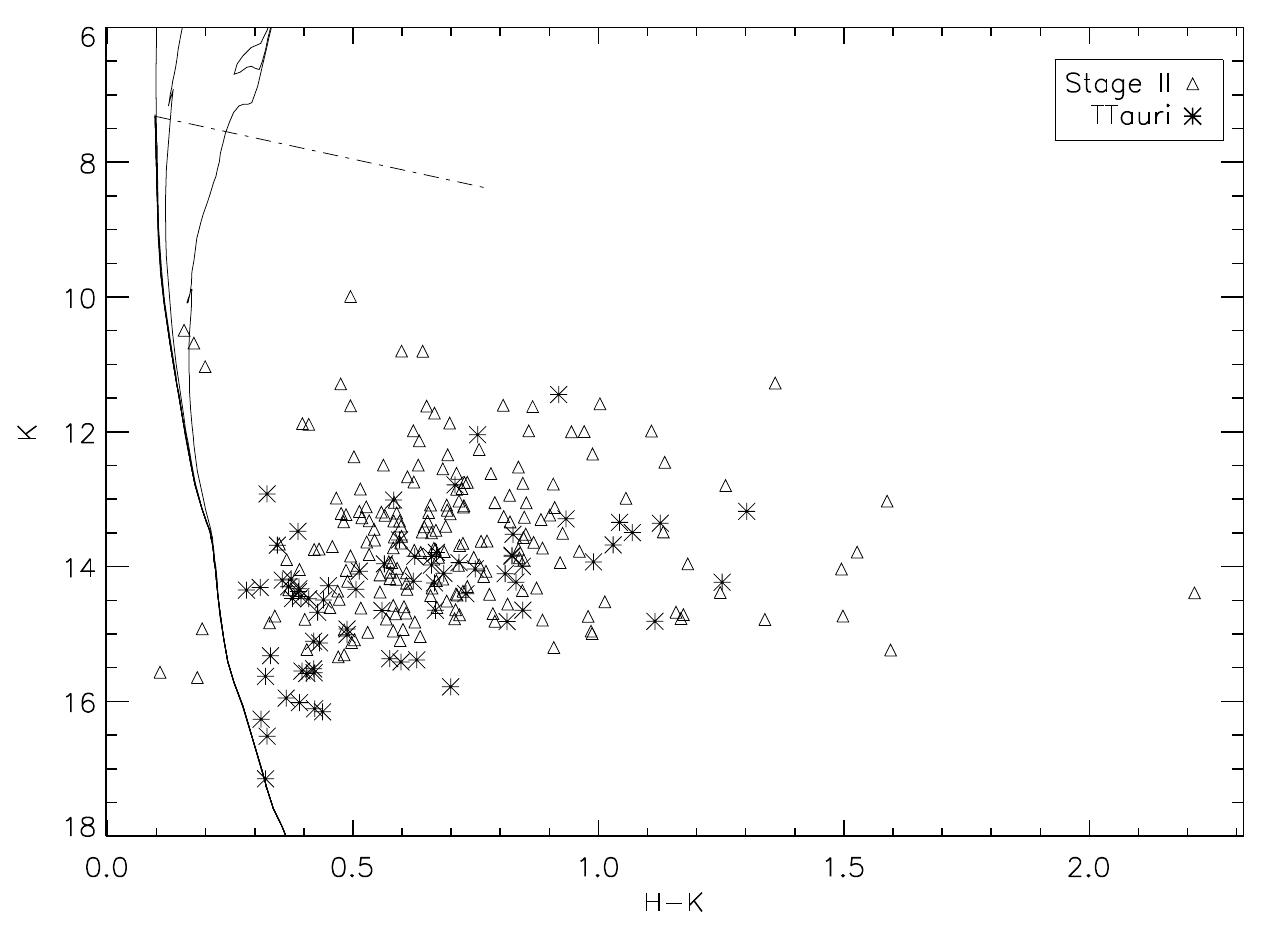}
\caption{Same as Fig. \ref{fig:stage1_cmd_overall}, but for Stage II sources.}
\label{fig:stage2_cmd_overall}
\end{figure}

\begin{figure}[htp!]
\centering
\includegraphics[scale=0.6,angle=0]{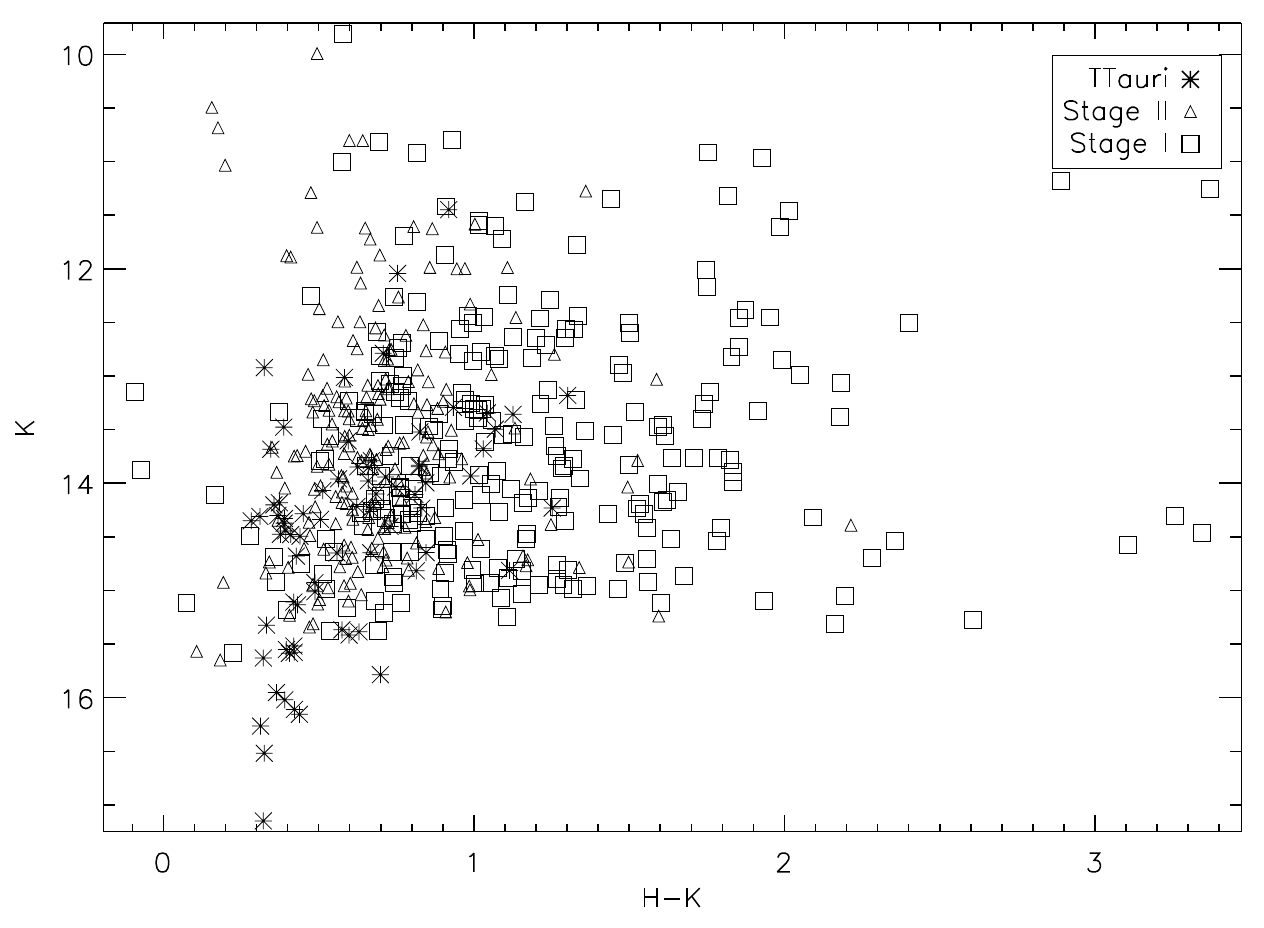}
\caption{Same as Fig. \ref{fig:stage1_cmd_overall}, but for both Stage I and Stage II sources.}
\label{fig:stages_all_cmd}
\end{figure}

Using the 2MASS data we find that the populations in Stage I and Stage II (as defined in \citealp{rob2006}), when treated as a whole, are indistinguishable with respect to observed K magnitude (e.g., Figs.\ref{fig:stage1_cmd_overall}, \ref{fig:stage2_cmd_overall}, and \ref{fig:stages_all_cmd}). However, while Stage I sources are found in the color region occupied by Stage II, members of the former population consistently reach redder colors: $\sim90\%$ of Stage I sources have [H-K]$>0.6$, $\sim85\%$ have [H-K]$>0.7$, $\sim55\%$ have [H-K]$>1.0$, and $\sim25\%$ have [H-K]$>1.5$ (maximum [H-K]$=\sim3.5$). The corresponding statistics for Stage II are $\sim65\%$, $\sim40\%$, $10\%$, and $2\%$ (maximum [H-K]$=2.2$). 

F-test significance levels between Stage I-Stage II and the T-Tauri population show both groups are statistically consistent with having been drawn from the same parent population in K magnitude as the T-Tauri sample (sig. levels: $0.41$ and $0.58$, respectively), with Stage II \textit{also} consistent in color (sig. level: $0.39$) with the PMS population (dominating in the reddening band of typical main-sequence and giant stars).
Stage I sources show a larger scatter in the CCD (e.g., Figures \ref{fig:stage1_ccd_overall}, \ref{fig:stage2_ccd_overall}), and significance levels obtained from Student's t-test and F-test between the two populations agree with both having different parent distributions based on color information (Table \ref{table:stageprops}).
 
We find no clear boundary in the CCD or CMD separating sources with intrinsic, physically different characteristics, which could easily be explained by different inclination angles. About $90\%$ of Stage 0/I sources lie in the range $0.5\leq$[H-K]$<2.0$, $\sim90\%$ of Stage II between $0.4\leq$[H-K]$<1.2$, leaving the region in [H-K]$\geq1.2$ as the area dominated by (younger) sources with $\dot{M}_{\mathrm{env}}/M_{\star}>10^{-6}$\,yr$^{-1}$ (e.g., Figure \ref{fig:stages_all_cmd}).

\begin{table*}[ht]
\caption{General 2MASS Properties: `Stage I' vs `Stage II' Population}
\label{table:stageprops}
\centering
\begin{tabular}{l l l l l l l l}
\hline
\hline
Stage&Mean [H-K]&$\sigma$&Mean [J-H]&$\sigma$&Mean K&$\sigma$&Objects\\
\hline
1&1.2&0.6&1.7&0.6&13.6&1.1&254\\
2&0.7&0.3&1.3&0.4&13.6&1.1&203\\
\hline
&&T-Test Stat.&Significance&F-Test Stat.&Significance&&\\
\hline
&[H-K]&10.6&1E-23&4.1&4E-23&&\\
&[J-H]&8.0&1E-14&2.9&5E-14&&\\
&[K]&0.4&0.7&1.1&0.7&&\\
\hline
%\multicolumn{8}{{$^*$ $1\sigma$ Standard Deviation.}}\\
\end{tabular}
\end{table*}

\section{Spatial Distribution and Clustering: Group Classification and Characterization}\label{sec:groupclass}
There is strong evidence that star formation in GMCs occurs primarily in clusters. The strong link between `clustered star-formation' and massive star formation implies that in order to investigate the nature and processes involved in the formation of the most massive stars it is crucial to investigate the characteristics of young stellar clusters and their pre-stellar progenitors. This analysis, described below, was based on our YSO surface density maps (e.g., \citealp{chavarria2008}) and on the so-called Minimum Spanning Tree Algorithm (MST).

\subsection{Minimum Spanning Tree Analysis}
In order to identify and characterize regions of YSO clustering we first implemented the nearest-neighbor and MST techniques \citep{gower1969}. In the MST algorithm there is an optimized break length, D$_{\mathrm{break}}$ (the branch length whose removal maximizes the number of groups, N$_{\mathrm{g}}$, with members in each group separated by distances shorter than this length) for the specified N$_{\mathrm{YSO}}$, the minimum number of YSOs for the group to be considered a `cluster' (N$_{\mathrm{YSO}}$). To facilitate comparison with previous analyses of adjacent regions at the same distance (e.g., W5; \citealp{koenig2008}) we chose N$_{\mathrm{YSO}}=10$. We also obtained a new set of results for N$_{\mathrm{YSO}}=5$ to investigate the clustering properties at smaller scales. D$_{\mathrm{break}}$ was estimated from the peak of the distribution of the number of groups satisfying the YSO requirement (N$_{\mathrm{g}}$) as a function of d$_{\mathrm{break}}$ \citep{battinelli1991}; the minimum d$_{\mathrm{break}}$, as well as the incremental step, were chosen to be $0.05$\,pc, similar to the resolution of MIPS $24$\,\micron\ at $2$\,kpc.  

D$_{\mathrm{break}}$ and N$_{\mathrm{g}}$ are expected to be affected by incompleteness and resolution (e.g., \citealp{bastian2007}). Nevertheless, this technique is particularly useful when comparing the relative degree of clustering for different types of YSOs in different regions with similar data, as in the case of the different subregions in W3. Since we cannot confirm physical associations in these groups, cluster membership cannot be determined without additional data. Therefore, for the following analysis we define the `sub-branches' resulting for D$_{\mathrm{break}}$ as `stellar groups', to avoid confusion from true clusters and associations.
Global results for the \spitzer\ survey are presented in Table \ref{table:clusters}. When a range of break lengths is found to have the same maximum  N$_{\mathrm{g}}$ we include the range. MST parameters for the entire YSO population (analyzed as a whole) and for each subregion (each AOR analyzed individually) are shown in Tables \ref{table:clusters} and \ref{table:clusters_regions}, respectively. Tables also include results for the YSO population as a whole and divided into classes. Each Class is also separated according to N$_{\mathrm{YSO}}$ in order to investigate a possible hierarchical structure within larger groups.

\begin{table*}[ht]
\caption{Identified Stellar Groups in W3}
\label{table:clusters}
\centering
\begin{tabular}{l l l | c c c | c c c}
\hline
\hline
\multicolumn{9}{c}{{All Survey}}\\
&&&\multicolumn{3}{c}{{Flag$=0$}}&\multicolumn{3}{c}{{Flag$=1$}}\\
\hline
Catalog&Class&N$_{\mathrm{YSO}}$&D$_{\mathrm{break}}$&N$_{\mathrm{g}}$&$\%$ Assoc.&D$_{\mathrm{break}}$&N$_{\mathrm{g}}$&$\%$ Assoc.\\
\hline
1&All$^a$&$10$&$0.60$&$27$&$56\%$&$1.2$&$13$&$76\%$\\
2&All&$10$&$0.60-1.2$&$15$&$40-78\%$&$1.2$&$13$&$76\%$\\
1&Class0/I&$10$&$3.1-3.4$&$6$&$82-83\%$&$3.1-3.4$&$6$&$80-82\%$\\
1&ClassII&$10$&$0.85$&$14$&$55\%$&$0.85$&$14$&$56\%$\\
1&PMS&$10$&$0.70-1.2$&$6$&$30-61\%$&&&\\
1&Class0/I/II&$10$&$0.60-1.2$&$15$&$40-78\%$&$0.80$&$13$&$54\%$\\
1&All&$5$&$0.45$&$60$&$53\%$&$0.55$&$33$&$54\%$\\
2&All&$5$&$0.55-0.60$&$37$&$55-60\%$&$0.95$&$33$&$54\%$\\
1&Class0/I&$5$&$2.3-2.8$&$10$&$80-88\%$&$2.3-2.8$&$10$&$80-87\%$\\
1&ClassII&$5$&$0.60$&$29$&$53\%$&$0.60$&$24$&$51\%$\\
1&PMS&$5$&$0.65$&$14$&$37\%$&&&\\
1&Class0/I/II&$5$&$0.55-0.60$&$39$&$57-62\%$&$0.55-0.65$&$32$&$54-63\%$\\
\hline
\multicolumn{9}{l}{{$^a$ Includes Class0/I, Class0/I$^*$, ClassII, and ClassII$^*$. Other classes exclude}}\\
\multicolumn{9}{l}{{ ($^*$) candidates unless specifically mentioned. }}\\
\end{tabular}
\end{table*}

Using the group information provided by the MST algorithm and the technique from \citet{battinelli1991} we first explored whether there might be a characteristic scale in the entire \spitzer\ survey (Table \ref{table:clusters}).

Our list of YSO candidates (Class 0/I, Class II, Class 0/I* and Class II* candidates: `All' class in Table {\ref{table:clusters}) from Catalog 1 yields D$_{\mathrm{break}}=0.6$  for a minimum group membership of N$_{\mathrm{YSO}}=10$, resulting in $56\%$ of the YSO population associated with a group. Flag $=1$-only sources and Catalog 2 (both containing less than half the sources in Catalog 1) are more consistent with a larger length and grouped fraction of $\sim1.1$\,pc and $73.5\%$, respectively. It is clear that incompleteness will affect the optimal break length, as missing sources will affect the YSO grouping and result in spatially larger groups in order to satisfy the minimum group membership requirement. As shown in Table \ref{table:clusters_regions}, this appears to be an issue mainly when considering the (*) population, which constitutes the most uncertain sample and the main difference between Catalog 1 and Catalog 2. Exclusion of this sample affects the Class 0/I candidate sample in particular, because when the highly embedded population is omitted the few remaining typical Class 0/I sources are more widely distributed, resulting in spatially larger groups for a given N$_{\mathrm{YSO}}$.
The information from Catalog 1 (all flags) is therefore expected to be a more (statistically) significant indicator of the properties of the overall population, and so this is the primary catalog used in the following analysis. We also find that all the results derived for N$_{\mathrm{YSO}}=5$ are comparable with the results derived from Catalog 1 and N$_{\mathrm{YSO}}=10$, with a typical D$_{\mathrm{break}}\sim0.54$\,pc containing $55\%$ of the YSO population. This is in good agreement with the results from \citet{koenig2008}, who found optimal break lengths in the neighboring region of W5 (for a sample of the same characteristics) of 0.54\,pc (YSO fraction: $44\%$). 

\begin{table*}[ht]
\caption{Identified Stellar Groups in Individual Subregions of W3}
\label{table:clusters_regions}
\centering
\begin{tabular}{l c c c c c c}
\hline
\hline
\multicolumn{7}{c}{{W3 Main/(OH)}}\\
\\[0.1pt]
\hline
&\multicolumn{3}{c}{{N$_{\mathrm{YSO}}=10$}}&\multicolumn{3}{c}{{N$_{\mathrm{YSO}}=5$}}\\
Class&D$_{\mathrm{break}}$&N$_{\mathrm{g}}$&$\%$ Assoc.&D$_{\mathrm{break}}$&N$_{\mathrm{g}}$&$\%$ Assoc.\\
\hline
All$^a$&$0.45$&$11$&$48\%$&$0.45$&$23$&$62\%$\\
Class0/I/II&$0.75$&$5$&$63\%$&$0.55$&$11$&$51\%$\\
Class0/I&$2.1-3.6$&$2$&$74-97\%$&$2.0$&$4$&$77\%$\\
Class0/I+Class0/I$^*$&$0.45-0.55$&$6$&$35-49\%$&$0.55$&$22$&$72\%$\\
ClassII&$0.65-0.85$&$2$&$35-49\%$&$0.55$&$8$&$44\%$\\
ClassII+ClassII$^*$&$0.65-0.85$&$2$&$30-43\%$&$0.55$&$9$&$40\%$\\
PMS&$0.75-1.2$&$2$&$38-85\%$&$0.65$&$8$&$45\%$\\
\hline
\\[0.1pt]
\multicolumn{7}{c}{{KR 140}}
\\[0.1pt]
\hline
All&$1.1$&$14$&$67\%$&$0.85-0.95$&$30$&$72-77\%$\\
Class0/I/II&$0.65-1.6$&$7$&$42-76\%$&$0.65$&$18$&$61\%$\\
Class0/I&$3.1-3.4$&$3$&$66\%$&$2.3-3.0$&$6$&$68-83\%$\\
Class0/I+Class0/I$^*$&$2.2-2.9$&$5$&$78-90\%$&$0.75$&$13$&$62\%$\\
ClassII&$0.75-0.85$&$8$&$54-55\%$&$1.1$&$12$&$69\%$\\
ClassII+ClassII$^*$&$1.2$&$10$&$61\%$&$0.95$&$24$&$65\%$\\
PMS&$3.9-4.7$&$3$&$75-93\%$&$2.3-2.5$&$7$&$73-77\%$\\
\hline
\\[0.1pt]
\multicolumn{7}{c}{{AFGL 333}}
\\[0.1pt]
\hline
All&$0.55$&$6$&$44\%$&$0.55$&$12$&$59\%$\\
Class0/I/II&$0.55-1.3$&$4$&$41-93\%$&$0.55$&$11$&$66\%$\\
Class0/I&$0.25-0.35$&$2$&$40-49\%$&$0.15-0.35$&$2$&$19-49\%$\\
Class0/I+Class0/I$^*$&$0.25-2.7$&$2$&$28-94\%$&$1.3-1.5$&$4$&$78\%$\\
ClassII&$0.85-1.3$&$4$&$60-90\%$&$0.55$&$9$&$49\%$\\
ClassII+ClassII$^*$&$0.85-1.3$&$4$&$52-85\%$&$0.55$&$10$&$46\%$\\
PMS&$0.65-1.1$&$3$&$43-60\%$&$0.65-1.1$&$6$&$60-82\%$\\
\hline
\multicolumn{7}{l}{{$^a$ Class 0/I, Class 0/I$^*$, Class II, and Class II$^*$}}\\
\end{tabular}
\end{table*}

\subsection{Determination of Group Intrinsic Properties}
Results above indicate that the global YSO population of W3 as a whole shows a tendency to group with a scale D$_{\mathrm{break}}$ comparable to or less than half a parsec. While larger than typical core sizes associated with the formation of individual stars ($\sim0.1$\,pc; e.g., \citealp{mckee2007}; \citealp{motte2007}), these scales are consistent with clump-like objects ($\sim0.5$\,pc; e.g., \citealp{zinnecker2007}) considered to be the likely birth place of stellar clusters. This led \citet{koenig2008} to conclude that these scales might well be typical of high mass star forming regions. However, the relevance of this result and its underlying relation to the actual physical processes in star formation remains ambiguous. 

It is important to quantify how representative this value is of the inter-YSO separations and how relevant this grouping is to the original birth configuration and conditions of the eventual stellar members.
We examined the distribution of D$_{\mathrm{near}}$, the distance of a YSO to its nearest neighbor, both for the entire sample (Table \ref{table:parameters1}) and for members within a specific group (Table \ref{table:parameters2}). Typically, D$_{\mathrm{near}}$ is considerably smaller than half a parsec, and therefore the optimal D$_{\mathrm{break}}$ is more indicative of `inter-cluster' separation than YSO separation, and therefore more relevant for cluster formation than that of individual stars. We note that embedded clusters of massive stars like the one forming IRS5 (W3 Main) or the Trapezium in the Orion Nebula have \textit{maximum} projected stellar separations of the order of $0.02-0.05$\,pc \citep{megeath2005}. These approach the resolution limits of IRAC and MIPS $24$\,\micron, respectively. Therefore, a given \spitzer\ `YSO' may in fact contain more than one protostar. The clear link between massive stars and clusters, however, make the present study a required step for understanding the physics behind high-mass star formation and the differences with respect to that of low mass stars. 

With this goal in mind, we analyzed each subregion of W3 in more detail to investigate the underlying properties of the stellar groups found by the MST algorithm, as well as possible local differences on the intrinsic characteristics of the stellar population. Tables \ref{table:parameters1} and \ref{table:parameters2} include the parameters derived from the YSO candidate list in Catalog 1. The former includes, for each subregion, the total area surveyed, the extinction range, the total mass in the region derived from the extinction maps (e.g., \citealp{heiderman2010}), the number of YSOs (Table \ref{table:class3a}), the total surface density, the YSO surface density (e.g., \citealp{chavarria2008}), and the star formation efficiency (SFE) of the region assuming that the YSOs are solar-mass stars. The latter assumption is not expected to be accurate, especially considering the possibility of some \spitzer\ YSOs actually being more than one object. However, we used this parameter as a measure of the \textit{relative} properties of the different stellar groups, just like when considering the `ages' of the regions in W3 (Section \ref{sec:history}). In both tables uncertainties for the mass and surface density have been derived from the statistical uncertainties in the extinction calculations. These uncertainties do not include effects such as variations in the extinction law or background fitting uncertainties during the creation of the maps. The final errors in these maps do not account for the larger differences observed with respect to other extinction estimates in the literature, and which depend on knowledge of the dust emissivity and temperature \citep{RF2009}. Changes in these last parameters can, by themselves, affect the estimated extinction values by a factor of $\sim2$, and therefore the uncertainties derived from this work will be underestimated.
We note that SFE is the amount of mass in YSOs \textit{at present} compared to the total mass, and therefore not necessarily the ultimate conversion efficiency.

Table \ref{table:parameters2} summarizes the average properties of the identified groups for N$_{\mathrm{YSO}}=10$ using the break length derived from the global analysis of the entire \spitzer\, survey (D$_{\mathrm{break}}=0.6$\,pc; Table \ref{table:clusters}). Several methods have been used to characterize the size and shape of identified groups, from circular to convex-hull techniques (e.g., \citealp{bastian2007}; \citealp{gutermuth2009}). To characterize the size and elongation of the group we chose to use an elliptical area. The center was defined as the average position of the YSOs within the group. The semimajor axis is the vector from the chosen center to the farthest YSO. The semiminor axis is the minimum size required to keep all the YSOs within the ellipse (Table \ref{table:parameters2}). Parameters for individual groups have been included in the electronic version of this article (Table \ref{table:parameters2}a).

\clearpage
\begin{sidewaystable}[p!]
%\begin{table*}
\caption{Average Global Parameters in Subregions of W3}
\label{table:parameters1}
\centering
\begin{tabular}{l | c c c c c c c c c c c c}
\hline
\multicolumn{12}{c}{{Global AOR Data$^a$}}\\
\hline
Region&D$_{\mathrm{Near}}$&D$_{\mathrm{Near}}$&Area&A$_{\mathrm{V}}$&A$_{\mathrm{V}}$&Mass$_{\mathrm{gas}}$&$\Sigma_{\mathrm{gas}}$&n$_{\mathrm{YSO}}$&$\Sigma_{\mathrm{YSO}}$&$\Sigma_{\mathrm{YSO}}$&SFE\\
&Min-Max&Mean&&Min-Max&Mean&&&&Min-Max&Mean&\\
&[pc]&[pc]&[pc$^2$]&[mag]&[mag]&[$10^4$\,M$_{\odot}$]&[M$_{\odot}$\,pc$^{-2}$]&&[pc$^{-2}$]&[pc$^{-2}$]&\\
\hline
\hline
All-Survey&$0.01-2.9$&$0.33\pm0.01$&$1316$&$1.2-9.9$&$3.5$&$6.2\pm0.005$&$59.44\pm0.04$&1566$^b$&$0.05-569.13$&$1.73\pm0.001$&$0.02$\\
W3Main/(OH)&$0.01-1.8$&$0.26\pm0.01$&$231.5$&$1.2-9.9$&$4.0$&$1.4\pm0.002$&$61.31\pm0.08$&616&$0.12-558.25$&$3.41\pm0.005$&$0.04$\\
KR140&$0.02-2.9$&$0.38\pm0.02$&$853$&$1.4-6.6$&$3.0$&$3.8\pm0.004$&$45.38\pm0.04$&706&$0.05-569.13$&$1.25\pm0.002$&$0.02$\\
AFGL333&$0.03-2.2$&$0.34\pm0.02$&$231.5$&$1.7-7.7$&$3.4$&$1.0\pm0.002$&$53.25\pm0.10$&246&$0.10-336.68$&$1.77\pm0.004$&$0.02$\\
\hline
\multicolumn{12}{l}{{$^a$ Average parameters for the entire AOR.}}\\
\multicolumn{12}{l}{{$^b$ Excludes repeats: two YSOs appearing in two fields due to AOR overlap.}}\\
\end{tabular}
\end{sidewaystable}
%\end{table*}

\begin{sidewaystable}[p!]
%\begin{table*}
\caption{Average Parameters of Groups in Subregions of W3$^a$}
\label{table:parameters2}
\centering
\begin{tabular}{l c c c c c c c c c c c c c}
\hline
Data&D$_{\mathrm{Near}}$&D$_{\mathrm{Near}}$&Area&A$_{\mathrm{V}}$&A$_{\mathrm{V}}$&Mass$_{\mathrm{gas}}$&$\Sigma_{{\mathrm{gas}}}$&n$_{\mathrm{YSO}}$&$\Sigma_{\mathrm{YSO}}$&$\Sigma_{\mathrm{YSO}}$&SFE&a$^b$&a$/$b\\
&Min-Max&Mean&&Min-Max&Mean&&&&Min-Max&Mean&&&\\
&[pc]&[pc]&[pc$^2$]&[mag]&[mag]&[$10^2$\,M$_{\odot}$]&[M$_{\odot}$\,pc$^{-2}$]&&[pc$^{-2}$]&[pc$^{-2}$]&&[pc]&\\
\hline
\hline
\\[0.1pt]
\multicolumn{14}{c}{{Class0/I($^*$)+ClassII($^*$); D$_{\mathrm{break}}=0.60$\,pc; N$_{\mathrm{YSO}}=10$}}\\
\\[0.1pt]
\hline
W3Main/(OH)&$0.05-0.51$&$0.21\pm0.05$&$14.7$&$2.9-7.4$&$4.8$&$11.0\pm0.02$&$71.9\pm0.2$&449&$1.61-178.21$&$7.75\pm0.06$&$0.07$&1.97&1.1\\
KR140&$0.04-0.48$&$0.16\pm0.04$&$5.05$&$2.7-4.2$&$3.4$&$2.9\pm0.009$&$51.4\pm0.3$&294&$1.62-141.95$&$8.91\pm0.03$&$0.14$&1.39&1.65\\
AFGL333&$0.07-0.44$&$0.21\pm0.05$&$2.43$&$3.0-5.2$&$4.1$&$1.6\pm0.007$&$61.7\pm0.4$&134&$3.02-73.90$&$12.59\pm0.12$&$0.15$&1.02&1.64\\
\hline
\\[0.1pt]
\multicolumn{14}{c}{{Class0/I/II; D$_{\mathrm{break}}=0.60$\,pc; N$_{\mathrm{YSO}}=10$}}\\
\\[0.1pt]
\hline
W3Main/(OH)&$0.04-0.54$&$0.19\pm0.09$&$2.84$&$3.6-6.0$&$4.6$&$2.1\pm0.01$&$69.5\pm0.5$&45&$2.44-64.50$&$8.64\pm0.06$&$0.10$&1.06&1.49\\
KR140&$0.04-0.54$&$0.19\pm0.06$&$4.51$&$3.1-4.8$&$4.1$&$2.8\pm0.01$&$61.0\pm0.3$&141&$1.44-155.20$&$7.88\pm0.03$&$0.09$&1.37&1.44\\
AFGL333&$0.08-0.46$&$0.23\pm0.05$&$2.93$&$2.9-5.8$&$4.3$&$2.0\pm0.009$&$64.7\pm0.3$&106&$1.70-32.71$&$6.43\pm0.02$&$0.10$&1.12&1.51\\
\hline
\multicolumn{14}{l}{{$^a$ Parameters for individual groups (Table \ref{table:parameters2}a) available in the electronic version of this article.}}\\
\multicolumn{14}{l}{{$^b$ Ellipse semi-major axis.}}\\
\end{tabular}
%\end{table*}
\end{sidewaystable}
\clearpage

%%%%%%%%%%%%%%%%%%%%%%%%%%%%%%%%

%%%%%%%%%%%%%%%%%%%%%%%%%%%%%%%%%%%%%%

\begin{figure*}[ht]
\centering
\includegraphics[scale=0.8,angle=270]{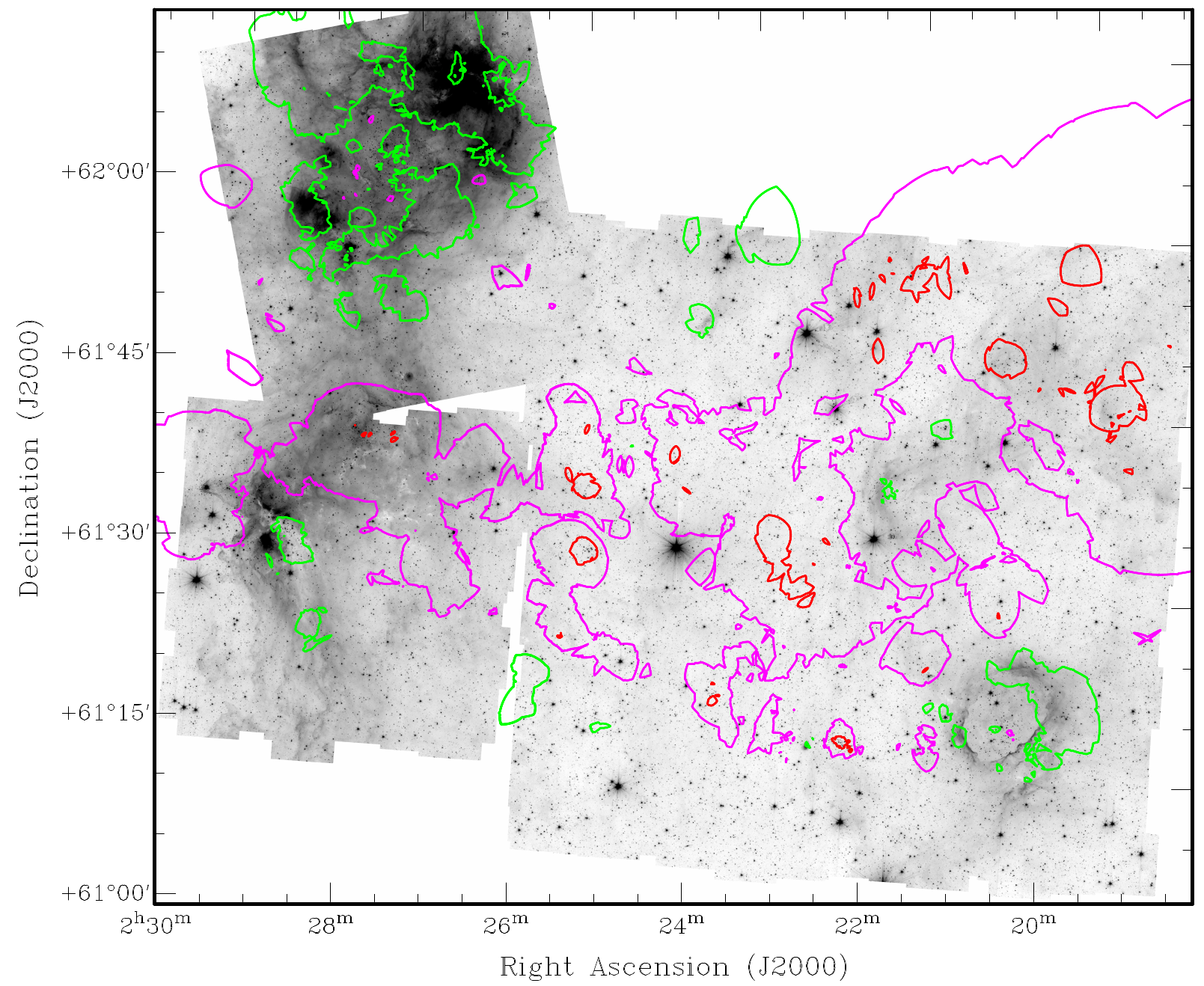}
\caption{Greyscale \spitzer\ channel 1 image of W3 with a superposition of `age' contours from the ratio of Class II/Class 0/I, including ($^*$) population. Only specific contours are shown for clarity: relatively old, $3\%$ of map peak value (red); intermediate, $0.5\%$ (magenta); and relatively young, $0.05\%$ (green).}
\label{fig:age_all}
\end{figure*}

\begin{figure*}[h!]
\centering
\includegraphics[scale=0.65,angle=270]{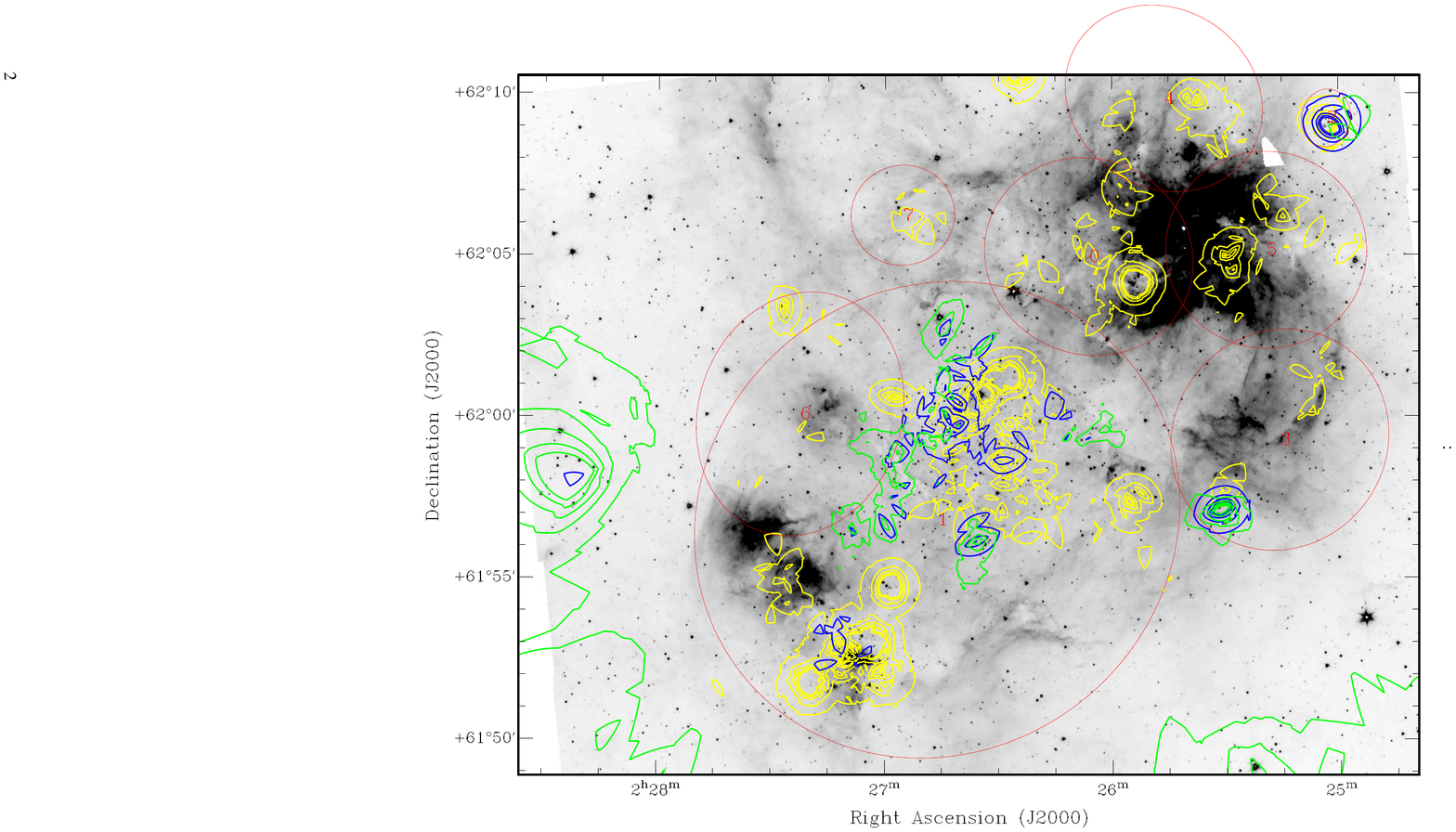}
\caption{Greyscale \spitzer\ channel 1 image of W3 Main/(OH) with identified groups for N$_{\mathrm{YSO}}=10$ and D$_{\mathrm{break}}=0.6$\,pc (red ellipses). Yellow contours are Class 0/I surface density contours between $1-5\%$ of peak value $\sim560$\,YSO\,pc$^{-2}$ in $1\%$\,steps. Blue contours are Class II contours between $5-25\%$ of peak value $\sim100$\,YSO\,pc$^{-2}$ in $5\%$\,steps.  YSO contours have been chosen to span a common YSO range for both classes of $\sim5-25$\,YSO\,pc$^{-2}$. Green contours are of the ratio of Class II/Class 0/I YSO surface density maps; these include transition and highly embedded candidates: $^*$ classification. These `age' contours are for 0.5-2.5$\%$ of the peak value of $\sim300$ in $0.5\%$ steps, a range chosen to highlight the youngest regions.}
\label{fig:main1}
\end{figure*}

\begin{figure*}[ht!]
\centering
\includegraphics[scale=0.65,angle=270]{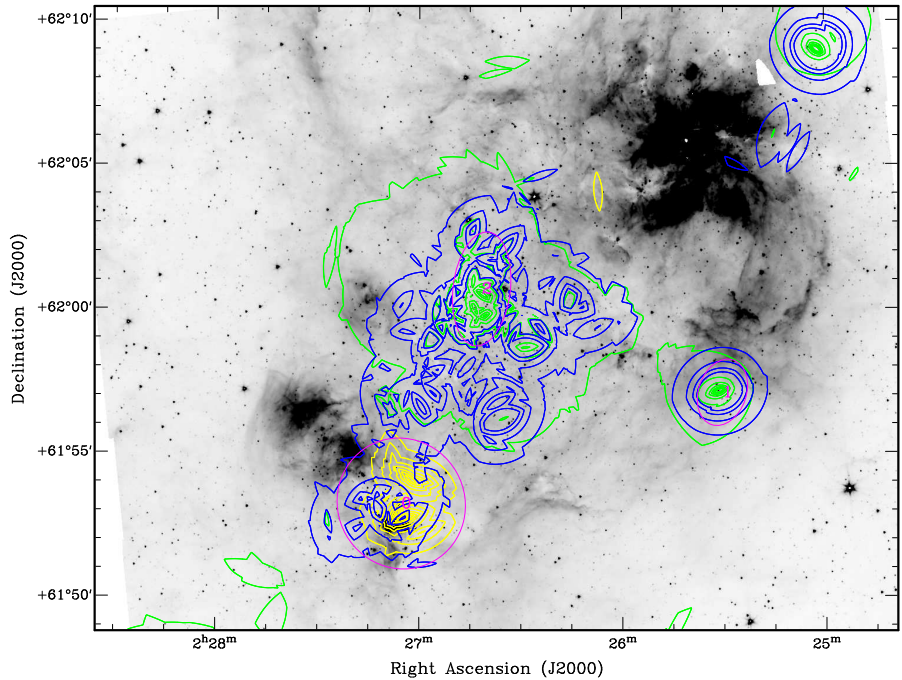}
\caption{Like Fig. \ref{fig:main1}, but excluding the less reliable ($^*$) population. Groups are marked as magenta ellipses. Class 0/I surface density contours (yellow) between $20-90\%$ of peak value $\sim8$\,YSO\,pc$^{-2}$ in $10\%$\,steps. Class II contours (blue) between $1.5-7\%$ of peak value $\sim100$\,YSO\,pc$^{-2}$ in $\sim1.4\%$\,steps. YSO contours have been chosen to span a common YSO range of $\sim1.5-7$\,YSO\,pc$^{-2}$. `Age' contours (green) are between 1-31$\%$ of peak value of $\sim500$ in $5\%$ steps. }
\label{fig:main2}
\end{figure*}

\begin{figure*}[ht!]
\centering
\includegraphics[scale=0.65,angle=270]{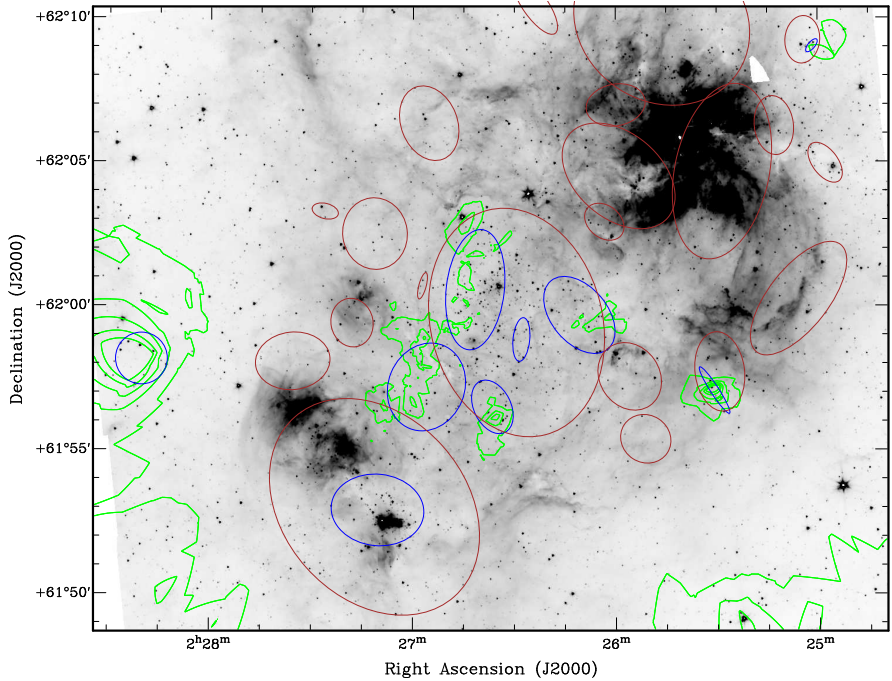}
\caption{Like Fig. \ref{fig:main1} but for N$_{\mathrm{YSO}}=5$ and D$_{\mathrm{break}}=0.55$\,pc, which is optimal in this region for both Class 0/I + Class 0/I$^*$ groups (brown ellipses), and Class II + Class II$^*$ groups (blue ellipses). Parameters are from Table \ref{table:clusters_regions}. Figure shows the location of the highest concentrations of YSOs according to Class.}
\label{fig:main1_5}
\end{figure*}

\begin{figure*}[ht!]
\centering
\includegraphics[scale=0.65,angle=270]{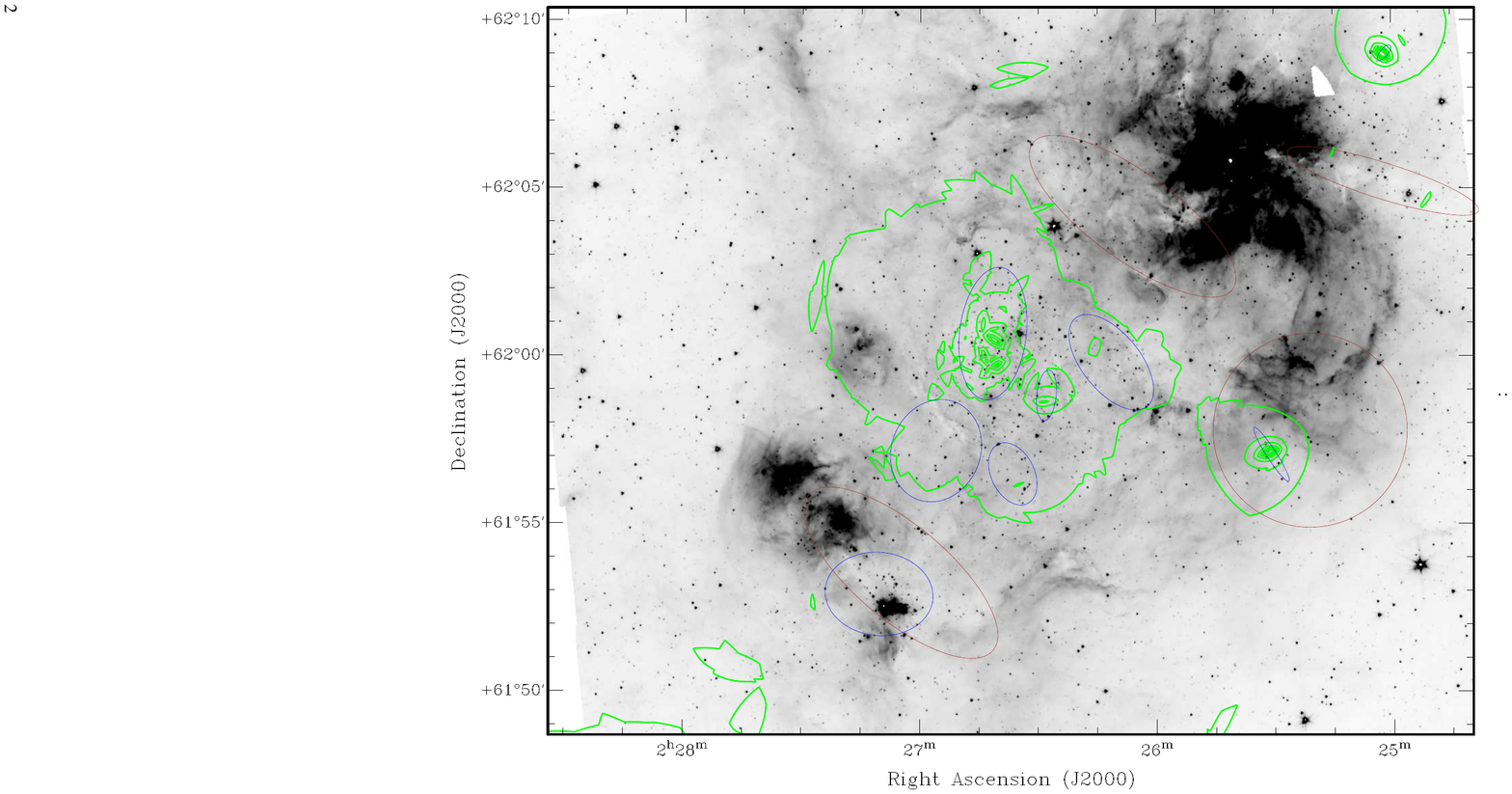}
\caption{Like Fig. \ref{fig:main1_5}, but excluding the ($^*$) population. Now D$_{\mathrm{break}}=1.95$\,pc, which is optimal for Class 0/I (brown ellipses), and D$_{\mathrm{break}}=0.55$, which is optimal for Class II (blue ellipses). Parameters are from Table \ref{table:clusters_regions}.}
\label{fig:main2_5}
\end{figure*}

%%%%%%%%%%%%%%%%%%%%%%%%%
\section{W3 in Perspective: Stellar Content, Cluster Properties, and Star Formation Activity} \label{sec:history}
 W3 is believed to have signatures of both triggered (HDL; e.g., \citealp{oey2005}) and isolated (KR140; e.g., \citealp{kerton2008}) massive star formation. The existence of different star formation processes in the same cloud makes W3 a prime location for further investigation. The identification and characterization of the embedded/young ($1-2$\,Myr) cluster population (including their relative age) can shed some light on cluster and massive star formation. 
With this goal in mind, in this section we use the spatial distribution of YSO classes, the properties of the identified groups, and the surface density/age maps to present a description of the star formation history and activity in W3. Characteristics of the groups (including the ($^*$) population) discussed below can be found in Table \ref{table:parameters2}a (online).

We carried out the main analysis on W3 by working on each subregion individually, while using common group properties (e.g.,  N$_{\mathrm{YSO}}$ and D$_{\mathrm{break}}$) for the entire survey. Clearly, the definition and properties of the populations that are `clustered' or `distributed' (objects outside the `cluster' boundaries, stars formed in isolation and members displaced from their original birthplaces) depend on the fundamental definition of `cluster' and the chosen boundary (see e.g., \citealp{bressert2010} for a compilation of recent cluster identification techniques). The application of a common definition and technique throughout an entire cloud like W3 (same distance/resolution) can, however, be used to determine the \textit{relative}  properties of each subregion, avoiding major systematic effects arising from distance dependent factors or an assumed definition. For W3, this approach is used to investigate the possibility of intrinsic differences between stellar groups in different environments and with different stellar activity (e.g., HDL vs. KR 140), and so illuminate possible intrinsic differences between the clustered and distributed populations (e.g., \citealp{allen2007}). Unless mentioned otherwise, the following analysis will be based on the YSO sample from Catalog 1 and groups with N$_{\mathrm{YSO}}=10$ and D$_{\mathrm{break}}=0.6$\,pc (`red' groups).

We made use of the YSO density distribution maps and carried out an individual analysis of the identified groups in each subfield in W3 (W3 Main/(OH), AFGL 333, and KR 140). `Age' maps were also created from the ratio of the YSO surface density images, for example the ratio of Class II to Class 0/I candidates. Peaks in this map represent the `oldest' of those regions \textit{containing YSOs}, whereas a low value in a region populated by YSOs indicates a relatively young region. The maps used in this work are best suited for the analysis of regions known to contain YSO groups, because the low map values in inactive or unpopulated regions depend on YSOs rather distant from the pixel in question. Figure \ref{fig:age_all} shows the age map for the entire W3 cloud. The `oldest' star forming regions (peaks) are located in the KR 140 field, although we note that many of these old regions are not always associated with identified groups (see below).

Finally, it is important to distinguish between the actual `age' of the system (from the start of star formation activity) and `apparent age'. In an undisturbed system initiated by a short-lived burst of star formation, our ratio method yields reliable ages. High density, central regions would contain the oldest population and shorter free-fall timescales (t$_{\mathrm{ff}}$\,$\alpha$\,$\rho^{-1/2}$). Such an idealized system would show a relatively ordered distribution of its population, and the age estimate would be an acceptable upper limit. On the other hand, our age estimates for triggered regions are expected to be biased toward younger ages, because there is strong evidence (e.g., IC 1795; see below) for star formation being an on-going process lasting for at least a few Myr, rather than being the result of an isolated burst of star formation. Thus, on average, the YSO groups in W3 Main/ (OH) might only appear to be the youngest using the simple age map contours (Fig. \ref{fig:age_all}).

\subsection{\textit{Star Formation in the HDL: W3 Main/(OH) \& AFGL 333}}
AFGL 333 and W3 Main/(OH) are located in the eastern HDL neighboring W4 (Fig. \ref{fig:intro}). The HDL is the most dense structure and essentially all the major activity of the GMC is found within its boundaries. In both fields we have identified signatures indicative of several modes of star formation contributing to the overall structure and young stellar population.

\subsubsection{The W3 Main/(OH) Field}
The cluster IC 1795 is located at the center of an eroded shell-like structure containing the W3 Main and W3 (OH) complexes. Both show the most active massive star formation, including deeply embedded clusters, \ion{H}{2} regions, and a trapezium-like system rivaling that in the Orion nebula (e.g., \citealp{megeath2005}). 

\begin{figure}[h!]
\centering
\includegraphics[scale=0.6,angle=0]{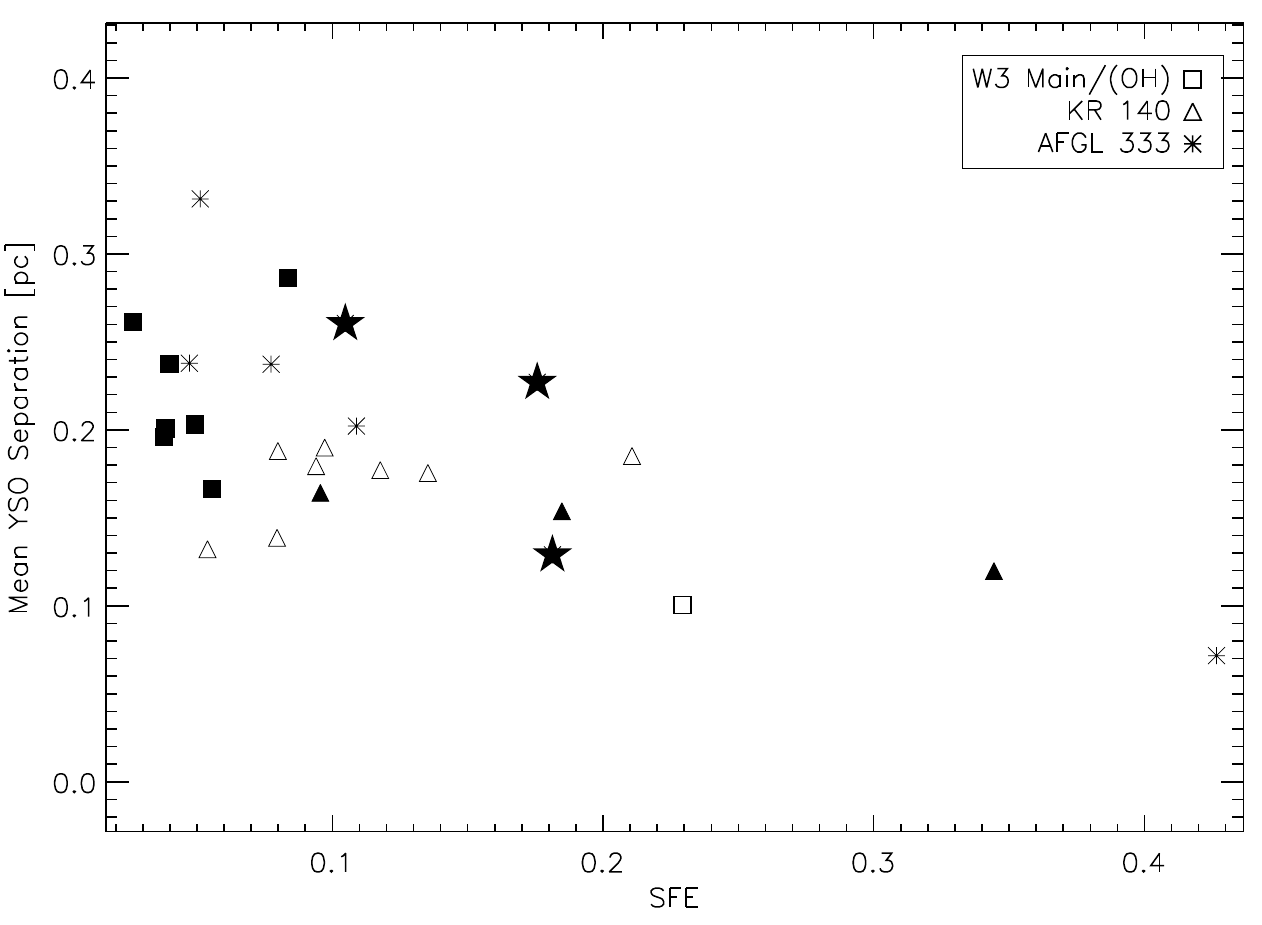}
\caption{Inter-YSO separations as a function of SFE for W3 Main/(OH), KR 140, and AFGL 333. YSOs (including highly embedded and transition candidates) are associated in groups with N$_{\mathrm{YSO}}=10$ and D$_{\mathrm{break}}=0.6$\,pc. Filled symbols mark those groups with the youngest ages (Class II/Class 0/I $< 1$).}
\label{fig:sfe_meandis}
\end{figure}

\begin{figure*}[h!]
\centering
\includegraphics[scale=0.65,angle=270]{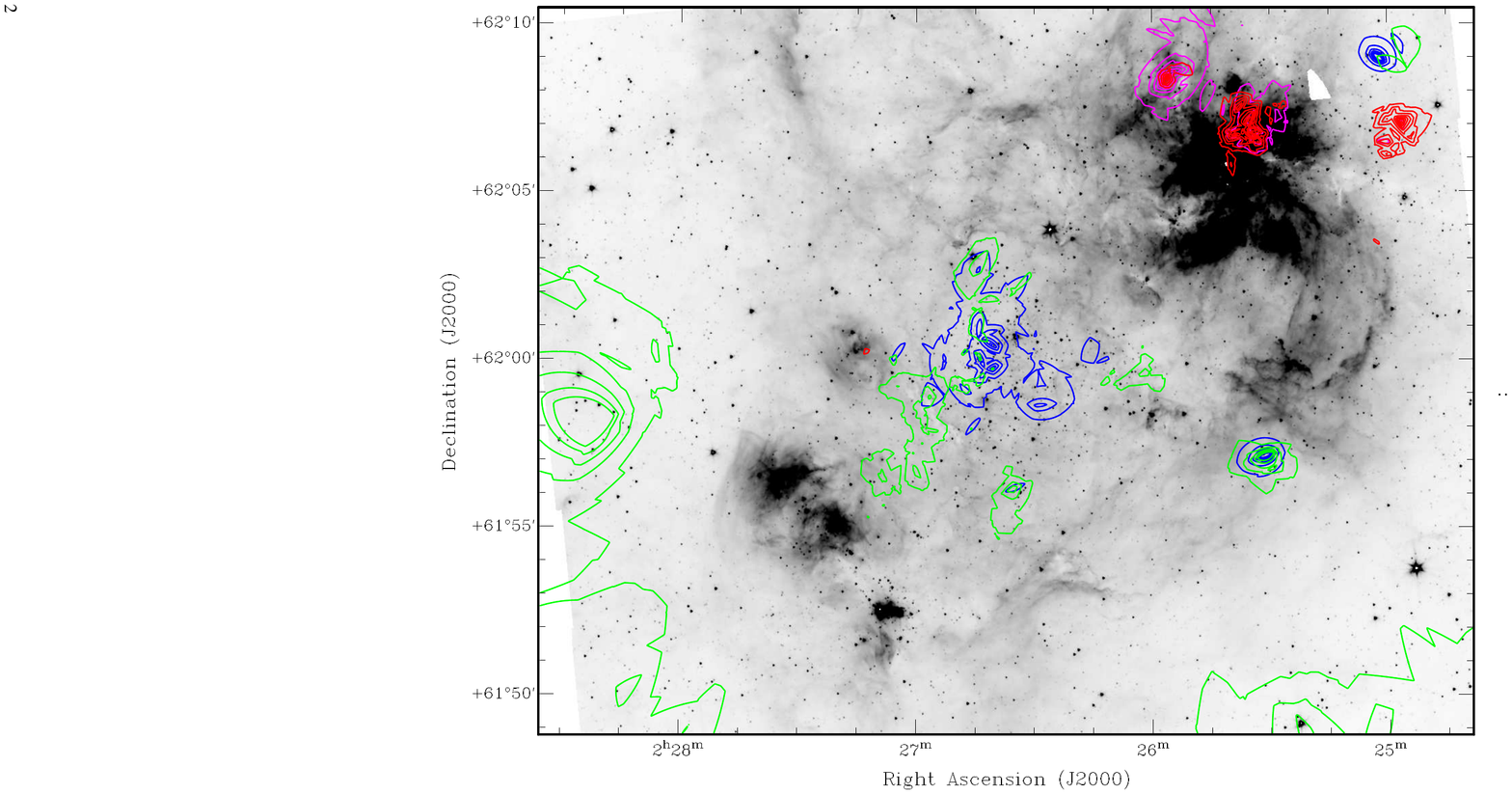}
\caption{Same as Fig. \ref{fig:main1} with various `age' contours superimposed. Blue contours are Class II/Class 0/I surface density contours between 5-30$\%$ of peak value of $\sim500$ in $5\%$ steps, excluding the ($^*$) population. Red and magenta contours are for PMS/[Class 0/I+Class II] between 3-10$\%$ of peak value of $\sim80$ in $1\%$ steps (red), and 10-60$\%$ in $10\%$ steps (magenta), with and without the ($^*$) population, respectively. Green contours like Fig. \ref{fig:main1}.}
\label{fig:age1}
\end{figure*}

Figure \ref{fig:main1} shows the W3 Main/(OH) field with the identified groups, the contours representing the surface density distributions for YSOs calculated on a grid identical to that of the \spitzer\ images, and including highly embedded and transition stage candidates (Class 0/I$^*$ and Class II$^*$: ($^*$) sample). This figure also includes `age' contours, obtained from the ratio of the surface density maps: [Class II+Class II$^*$]/[Class 0/I+Class 0/I$^*$].

Figure \ref{fig:main2} shows the groups found for the same N$_{\mathrm{YSO}}$ and break length as Figure \ref{fig:main1}, but excluding the less reliable ($^*$) population. For convenience these will be specifically referred to as `magenta' groups, to distinguish them from the `red' groups in Figure \ref{fig:main1}. Reduction of the number of candidate YSOs yielded smaller and fewer groups, as expected. In addition, the omission of highly embedded sources, combined with the confusion in the regions of strong IR emission, highly affected the numbers of YSOs detected in the innermost and most active regions, especially around W3 Main. We find two magenta groups within red Group 1, concident with IC 1795. The oldest (magenta Group 2; Fig. \ref{fig:main2}) coincides with the oldest part of IC 1795. Magenta Group 0 coincides with a high extinction region and contains numerous \ion{H}{2} regions and known clusters, including the well known maser sources W3 (OH) and W3 (H$_2$O). Although with N$_{\mathrm{m}}<25$, the characteristics of these two groups are consistent with those of mid-rich triggered groups in Figure \ref{fig:main1} (see below) linked to massive star formation, with surface densities greater than $70$\,M$_\odot$\,pc$^{-2}$, A$_{\mathrm{V-mean}}>5.0$, and mean inter-YSO separations $D<0.25$\,pc.

In Figures \ref{fig:main1_5} and \ref{fig:main2_5} we show the groups identified for each particular Class (with and without the ($^*$) population) with the relaxed requirement of N$_{\mathrm{YSO}}=5$.  The identified groups more closely trace particular structures observed in extinction (e.g., filaments), as well as smaller separate concentrations of Class 0/I and Class II candidates.

We suggest that, overall, clustered formation in compact clumps/cores can be distinguished from the star formation process associated with young groups that have low surface density and extinction and that are found in triggered regions (e.g., Groups 6 and 7; Fig. \ref{fig:main1}). Figure \ref{fig:sfe_meandis} shows the inter-YSO separations as a function of SFE. Groups in `induced' regions with the largest separations have been less efficient in forming young stars, which could be due to formation in a turbulent environment. This could be indicative of the `distributed' star formation mode taking over in those regions that lack the conditions, like high column density and low turbulence, required to form massive stars and their (richer) parent clusters. Indeed, a {\sc simbad} search reveals no \ion{H}{2} regions or clusters associated with any of these triggered groups, which also lack localized and significant radio continuum emission in the Canadian Galactic Plane Survey (CGPS; \citealp{taylor2003}) maps. Rich groups ($25 \le $ N$_{\mathrm{m}} < 50$) such as Group 0 and Group 5 have average inter-YSO separations $<0.25$\,pc and the highest surface densities, and both contain within their perimeters a large proportion of the major massive star activity of the W3 GMC. This supports the classical view that one can have clustered star formation caused by triggering when at \textit{early stages} high surface densities without disruptive turbulence are present, to ensure that a gravitationally (unstable) bound system is formed. 

The privileged location of IC 1795 (Group 1; Fig. \ref{fig:main1}), in the most central and dense parts of the HDL, was likely key in the formation of its rich population. Although it depends on the pre-existence of the HDL, its location, morphology and population characteristics are compatible with isolated/quiescent formation, and not necessarily the product of a triggering event as suggested in previous studies.   
A similar (quiescent) origin is also suggested for some individual isolated groups in the outer boundaries of the field, characterized by a relatively poor and yet closely spaced (contrary to the poor groups associated with triggered regions discussed above) old population in a relatively compact configuration (e.g., Group 2 in Fig. \ref{fig:main1}). Group 1 is the next `oldest' system after Group 2, and it is the most massive and richest group, with the smallest separation distances down to $0.02$\,pc. 

W3 Main and (OH) (Fig. \ref{fig:intro}) show the highest extinction and both are likely to have (at least part) of their stellar population induced by IC 1795, in agreement with the conclusions from \citet{oey2005} and \citet{polychroni2010}. A triggered scenario is supported by i) an elongated `older' Class II dominated region within Group 1 extending in the directions of W3 Main and W3 (OH) (see Fig. \ref{fig:main1}); ii) the cavity and shell like structure around IC 1795 hosting both systems; and iii) a tendency of Class II candidates to be located toward the most central regions, with younger groups of YSOs following the shell around Group 1 (in increasing age: Groups 4, 7, 0, 5, 3, 6; Fig. \ref{fig:main1}). Using the ratio of Class II/Class 0/I as a relative measurement of age, Group 6, the oldest after Group 2 and Group 1, happens to be also the nearest to IC 1795, while younger groups have larger distances. 

Secondary bursts of star formation, as well as a non-negligible star formation duration, are needed to reconcile the characteristics of the YSO population in IC 1795 with the age estimated by \citet{oey2005} ($3-5$\,Myr) using optical photometry and spectral analysis. We find a highly distributed population of Class II and Class 0/I sources in the vicinity of IC 1795. Star formation is estimated to occur in 1-2 dynamical crossing times \citep{elmegreen2000}, but larger age spreads are possible if multiple events occur within a certain system. Figures \ref{fig:main1_5} and \ref{fig:main2_5} show the groups obtained when separating the YSOs according to Class with a minimum group membership requirement of N$_{\mathrm{YSO}}=5$. The `class subgroups' observed within Group 1 and the triggered population could be indicative of such additional star formation events. This combined activity likely reinforced the effects of the central cluster when forming the cavity and dense surrounding shell hosting the most massive stars in W3.

Contrary to the suggestion of \citet{oey2005}, some of the stellar activity in W3 Main might also have originally initiated in quiescent mode. Figure \ref{fig:age1} shows an `older' (low and intermediate mass) PMS population toward the outer edge of this region, shown by the magenta/red contours. On the other hand there are young groups at the inner edge of W3 Main, closer to IC 1795. By calibrating the Class II/Class 0/I ratio of Group 1 to the age of IC 1795 we obtain an average age for these young groups of $\sim1.5-2.5$\,Myr. In this calculation we assumed the onset of star formation occurred in a single event, and so a larger proportion of Class II sources implies a more evolved population. We ignored effects such as secondary bursts of star formation in the region, and assumed that for groups with similar initial stellar characteristics there is a direct relation between the class ratio and the actual age of the system. A non-negligible period of star formation and internal triggering would mean that it would be more appropriate to consider our age estimates as lower limits. This age estimate for the young groups is however comparable to the estimated age for the oldest (diffuse) \ion{H}{2} regions and the PMS population known to dominate in this region ($\sim10^6$\,yrs; \citealp{feigelson2008}). Therefore, triggering (either indirectly by IC 1795, and/or by this pre-existing PMS population) could have been responsible for the observed young proto-OB population in the central regions of W3 Main, a secondary burst of massive star formation in a region with an already highly enhanced surface density. Such a scenario would nicely link IC 1795 to one of the possible models suggested by \citet{feigelson2008} to explain the origin of W3 Main (option 4 in their list), and could explain the anomalous age distribution (a central young cluster surrounded by an older population) described by these authors. 

\subsubsection{The AFGL 333 Field}
W3 Main and W3 (OH) dominate both in massive stars and in star formation activity. The other HDL field, containing AFGL 333, is characterized by a young stellar population in its most part associated with high extinction regions and filamentary structures, that have remained undetected in the 2MASS-based extinction maps. There are also clear similarities and some differences relative to W3 Main/(OH).

Figures \ref{fig:afgl1} and \ref{fig:afgl2} show the identified YSO groups (with and without the ($^*$) population) for the AFGL 333 field, including YSO surface density contours and age contours. Figures \ref{fig:afgl1_5} and \ref{fig:afgl2_5} show the groups identified for N$_{\mathrm{YSO}}=5$ and different classes (each with their corresponding D$_{\mathrm{break}}$). The oldest and youngest groups are clearly separated, with the youngest (brown ellipses) localized in the central regions, and the oldest (blue ellipses) toward the outer parts of AFGL 333 (northern parts and triggered regions neighboring W4).

\begin{figure*}[ht!]
\centering
\includegraphics[scale=0.65,angle=270]{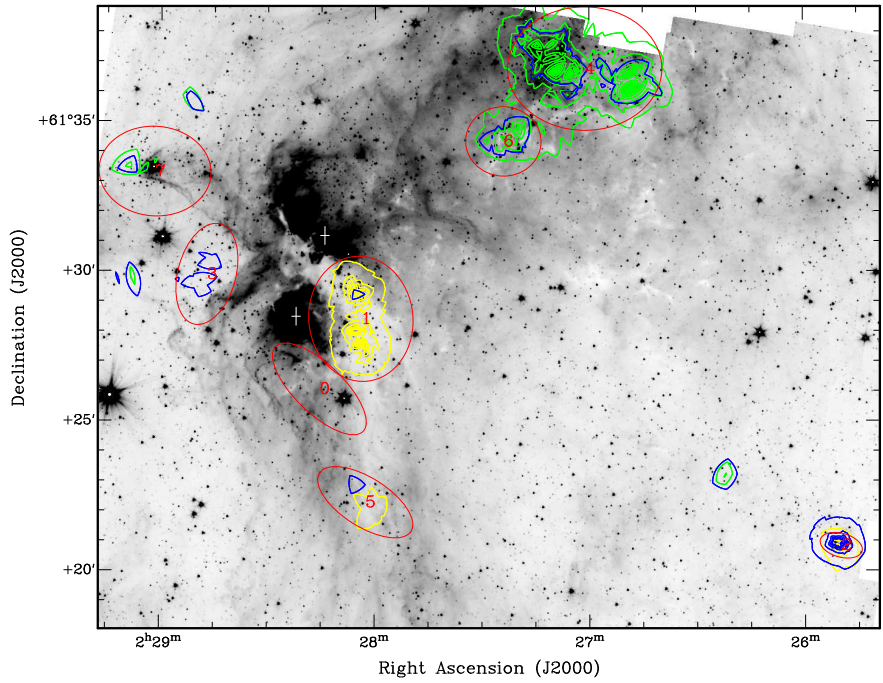}
\caption{Greyscale \spitzer\ channel 1 image of AFGL 333 with identified groups for N$_{\mathrm{YSO}}=10$ and D$_{\mathrm{break}}=0.6$\,pc (red ellipses). Yellow contours are Class 0/I surface density contours between $5-65\%$ of peak value $\sim120$\,YSO\,pc$^{-2}$ in $10\%$\,steps. Blue contours are Class II contours between $3.5-43.5\%$ of peak value $\sim170$\,YSO\,pc$^{-2}$ in $10\%$\,steps. Crosses are IRAS 02245+6115 (bottom) and IRAS 02244+6117 (top). YSO contours have been chosen to span a common YSO range for both classes of $\sim6-75$\,YSO\,pc$^{-2}$. Green contours are of the ratio of Class II/Class 0/I YSO surface density maps; these include transition and highly embedded candidates: $^*$ classification. These `age' contours are for 10-90$\%$ of peak value of $\sim300$ in $10\%$ steps. Groups like Group 1 are relatively young, while those like Group 4 are older.}
\label{fig:afgl1}
\end{figure*}

\begin{figure*}[ht!]
\centering
\includegraphics[scale=0.65,angle=270]{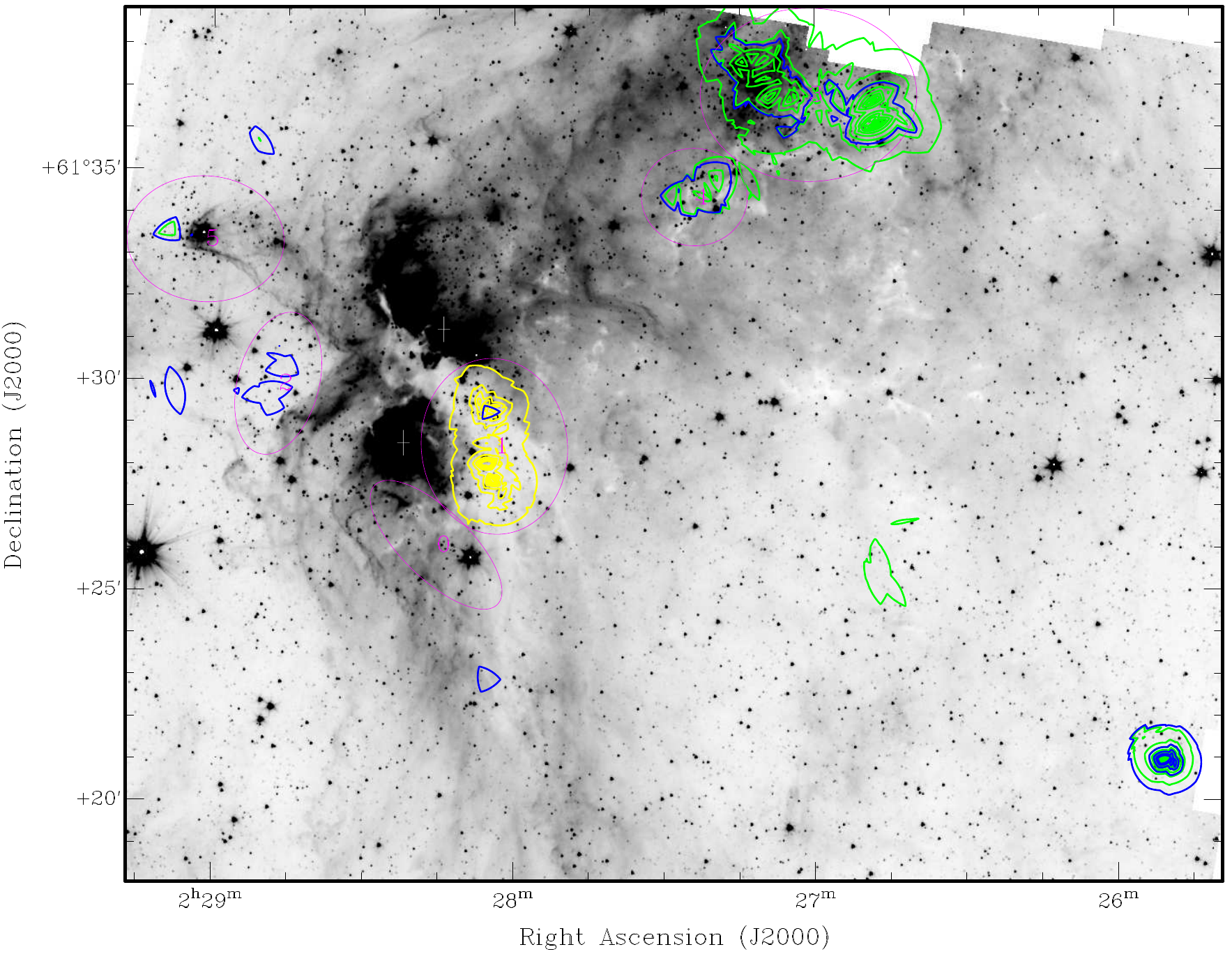}
\caption{Like Fig. \ref{fig:afgl1}, but excluding the less reliable ($^*$) population. Groups are marked as magenta ellipses. `Age' contours between 5-80$\%$ of peak value of $\sim1000$ in $5\%$ steps.}
\label{fig:afgl2}
\end{figure*}

\begin{figure*}[ht!]
\centering
\includegraphics[scale=0.65,angle=270]{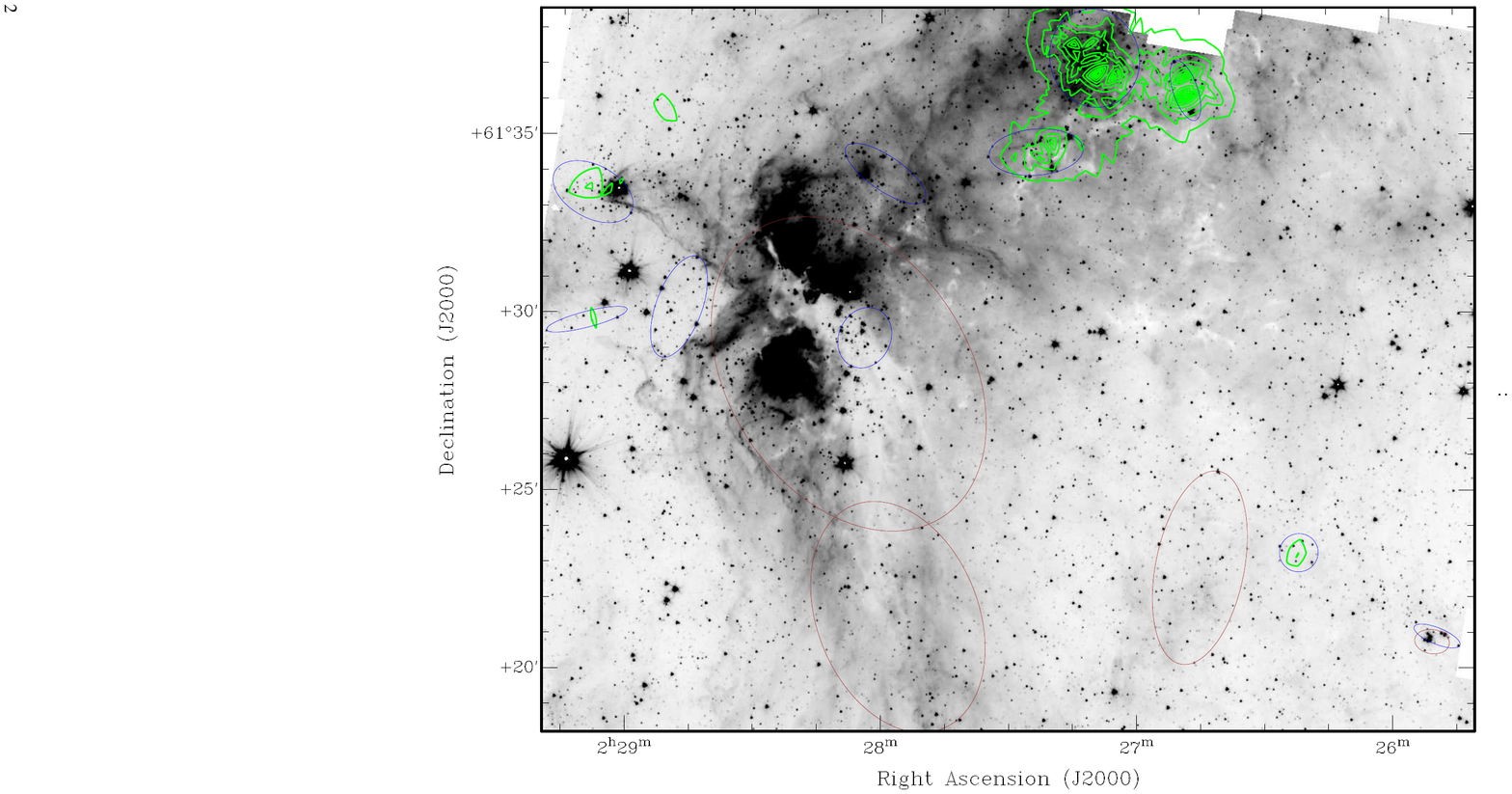}
\caption{Like Fig. \ref{fig:afgl1} but for N$_{\mathrm{YSO}}=5$, D$_{\mathrm{break}}=1.25$\,pc, and Class 0/I + Class 0/I$^*$ (brown ellipses), and N$_{\mathrm{YSO}}=5$, D$_{\mathrm{break}}=0.55$, and Class II + Class II$^*$ (blue ellipses). Parameters are from Table \ref{table:clusters_regions}.}
\label{fig:afgl1_5}
\end{figure*}

\begin{figure*}[ht!]
\centering
\includegraphics[scale=0.65,angle=270]{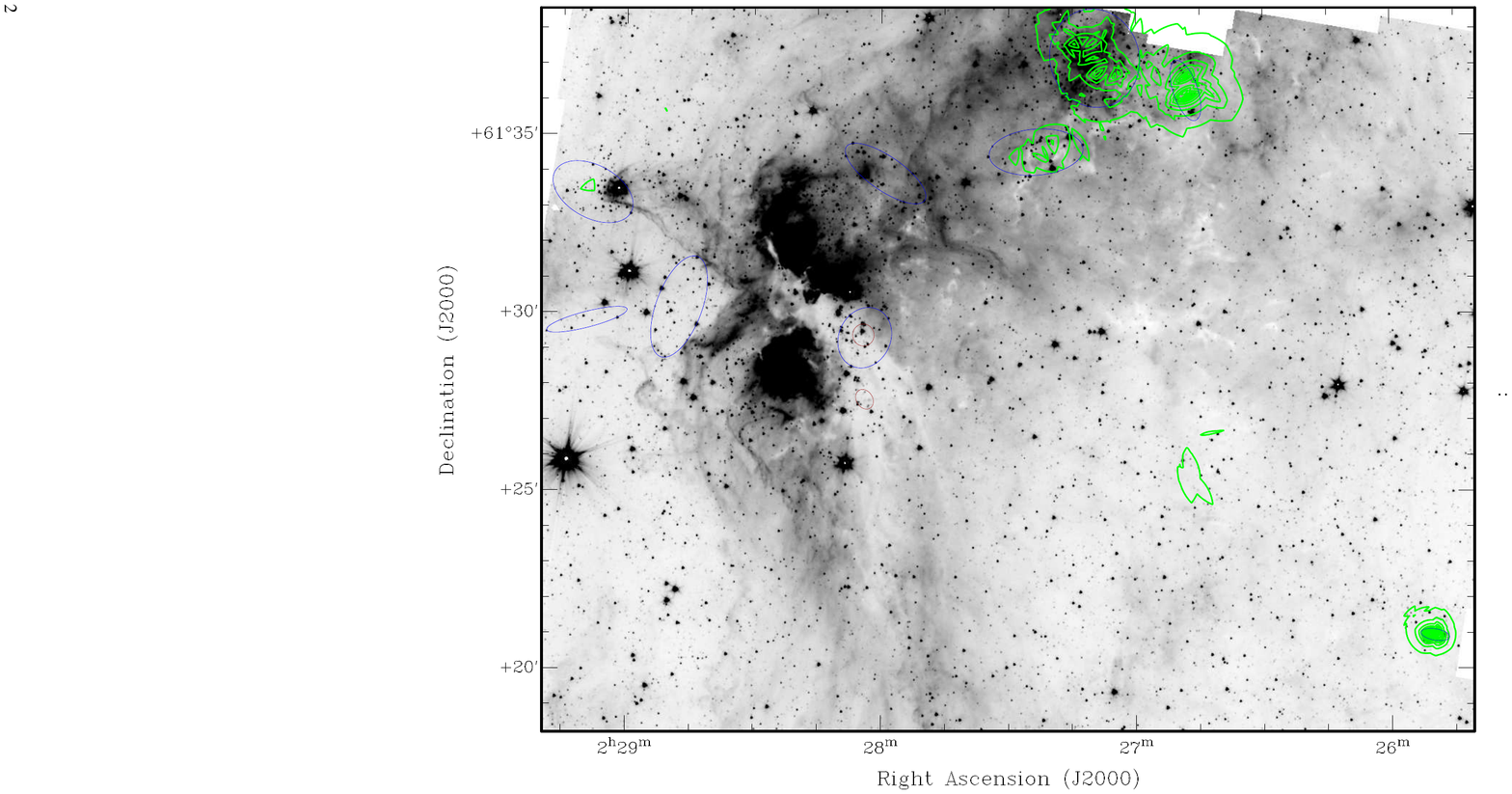}
\caption{Like Fig. \ref{fig:afgl2} but for N$_{\mathrm{YSO}}=5$, D$_{\mathrm{break}}=0.15$\,pc, and Class 0/I (brown ellipses), and N$_{\mathrm{YSO}}=5$, D$_{\mathrm{break}}=0.55$, and Class II (blue ellipses). Parameters are from Table \ref{table:clusters_regions}.}
\label{fig:afgl2_5}
\end{figure*}
Group 4 (Fig. \ref{fig:afgl1}) is the oldest and most massive group in the field, with the largest area, surface density, and highest extinction of $\sim$A$_{\mathrm{V-peak}}\sim7.5$). From our analysis we conclude that its population formed first, clearing a region and eroding a cavity-like structure much like IC 1795 (sizes $\sim6-7$\,pc) but in YSOs less populated. Within this cavity, the YSO population shows a more compact configuration in the eastern side of the group, and a more widely separated population toward the western side, where some high extinction filamentary structures are identifiable. There is also a population of PMS candidates toward the edge of this cavity, on the western side of Group 4 (opposite W4). Just as with IC 1795, the properties of this group (its location in the middle of the HDL and the geometrically `ordered' population, already eroding the surrounding material) are compatible with it being produced by quiescent formation, rather than having a triggered origin by the Perseus superbubble.

A significant difference relative to W3 Main/(OH) is the abundance of filaments with and without stellar activity.  Groups in this field  are all relatively poor, with only two groups (red Groups 1 and 4) with N$>25$ members (Fig. \ref{fig:afgl1}) and with the YSO population mainly localized to these filamentary structures. The morphology of such filaments, seen well in \spitzer\ channel 4 (Fig. \ref{fig:intro_ch4}), suggests a turbulent origin, a suggestion supported by their remarkable similarity and continuity with nearby (non-extinction) structures. A good example is Group 6 (Fig. \ref{fig:afgl1}), associated with a filamentary structure within the cavity caused by the stellar population in Group 4, and containing a relatively old YSO population (but still younger than the latter). Its associated population, while slightly offset from that filament either due to displacement since formation or clearing of the immediate surroundings, clearly traces the overall shape of the high extinction, indicating a birth association. While the filament in Group 6 is easily traced, only dispersed `remnants' of high extinction structures are visible at the western side of Group 4. This supports the `older' evolutionary stage for the latter, where the apparently dissociated population may be due to the dispersal of their common parental filaments.

A triggered origin is clearly identified for Groups 7 and 3 (N$_{\mathrm{m}} < 25$), both in the outer boundary of the HDL and associated within structures carved by the activity in W4.  The pillar associated with Group 7 coincides with a known cluster (IRAS-02252+6120; \citealp{bica2003}), containing both a Class II population, located toward the outer edges of the pillar, and Class 0/I sources, mainly in the innermost regions. Such a configuration has previously been observed in other pillar-like structures (e.g., \citealp{choud2010}) and suggested to be the result of radiative driven implosion (RDI) triggered star formation, in which an ionization/shock front driven by an expanding \ion{H}{2} region causes a neighboring overdensity to collapse, triggering the formation of stars.

Group 1 is the youngest and the richest group in Figure \ref{fig:afgl1}. It is located in the innermost regions of AFGL 333 and it is associated with the most dense and prominent filament in this field. If filaments are formed (or induced to collapse) by turbulence and compression by nearby star activity, then Group 1 might have been induced by the activity at opposite sides associated with IRAS 02245+6115 and IRAS 02244+6117, bottom and top crosses in Figure \ref{fig:afgl1}, respectively. The latter contains bright infrared sources (BIRS; \citealp{elmegreen1980}), and the former hosts a known cluster and massive star activity \citep{bica2003}. These form the brightest regions in the mid-infrared in AFGL 333 and both contain a population of PMS in their outermost parts. Thus two active older regions sandwich the central young stellar population of Group 1.   

Group 5 (Fig. \ref{fig:afgl1}), while associated with the same high extinction structure as Group 1, is however relatively member-poor. Its location away from the influence of the main (infrared-bright) star forming activity in AFGL 333 may explain the low membership, in contrast with the richness of Group 1. 

High extinction structures border Group 0 as well (especially noticeable in IRAC channel 4; Fig. \ref{fig:intro_ch4}). The associated filaments are located within an evacuated region, much like those in Group 6. However, as mentioned above, triggering in such regions of low surface density will likely result in isolated star formation (instead of clustered). 

The results from the above analysis remain unchanged when the $(^*)$ population is excluded, except for the omission of red Groups 2 and 5 (Fig. \ref{fig:afgl2}). 

We note that the features seen in extinction in the mid-infrared are better tracers of high column density material than the 2MASS-based extinction map, the latter missing some major high extinction areas (e.g., Group 1, Group 7; Fig. \ref{fig:afgl1}). Thus the mass (and surface density) estimates for associated groups are lower limits.  Indeed, Group 1 has the largest number of associated YSOs, and it is expected to have a large surface density comparable to or greater than groups of similar membership, $>80$\,M$_\odot$\,pc$^{-2}$. The extinction structure associated with this group is the brightest in the AFGL 333 field in the $850$\,\micron\ SCUBA map, which is an excellent tracer of column density. 
We find a similar situation in the KR 140 field (below). For small structures and filaments prominent in the submillimeter maps, including KR 140-N, north of the KR 140 \ion{H}{2} region near the center of the field, information is lost in the 2MASS-based extinction maps.
Our Herschel analysis (in preparation) will be able to provide accurate masses and properties for each of these structures based on temperature-calibrated dust emission.

\subsubsection{Modes and Sequential Star Formation}

It has been argued that the presence of massive star forming sites along the interface between W3 and W4 is evidence that the main activity in the HDL is triggered by the expansion of the W4 \ion{H}{2} region (e.g., \citealp{carpenter2000} and references therein). This is supported by the estimated ages of the structures (e.g., \citealp{oey2005}). We agree that the HDL structure itself was created by W4, with conditions (e.g., turbulence, surface density) favorable for star/cluster formation. However, whether the formation and ultimate collapse of clumps and cores in W3 were directly triggered is less clear. Based on the above evidence it seems also plausible to us that formation of the HDL was followed in sequence by a more quiescent evolution governed by local conditions.

 Local triggering does play a major role in enhancing (massive) star/cluster formation: i) internal triggering within evacuated regions (inner shells) generates secondary bursts of star formation and a distributed population either by forming or collapsing pre-existing small overdensities (clumps and filaments; IC 1795, AFGL-Group 4, AFGL-Group 0); ii) compression of high density regions generates major bursts of star formation, including massive stars and highly embedded clusters (e.g., inner part of W3-Main, W3 (OH), AFGL-Group 1, AFGL-Group 7).

Overall, the star formation activity and processes in the AFGL 333 and W3 Main/(OH) fields operate similarly, albeit less vigorously in the former. Environmental physical differences between the two regions, such as column density distribution and kinematics, will be investigated in upcoming papers. The preponderance of filaments and the characteristics of these and other star forming and starless structures in the HDL will be the subject of our Herschel paper currently in preparation.

%%%%%%%%%%%%%%%%%%%%%%%%%%%%%%%%%%%%%%%%%%%%%%%
\begin{figure*}[ht!]
\centering
\includegraphics[scale=0.65,angle=270]{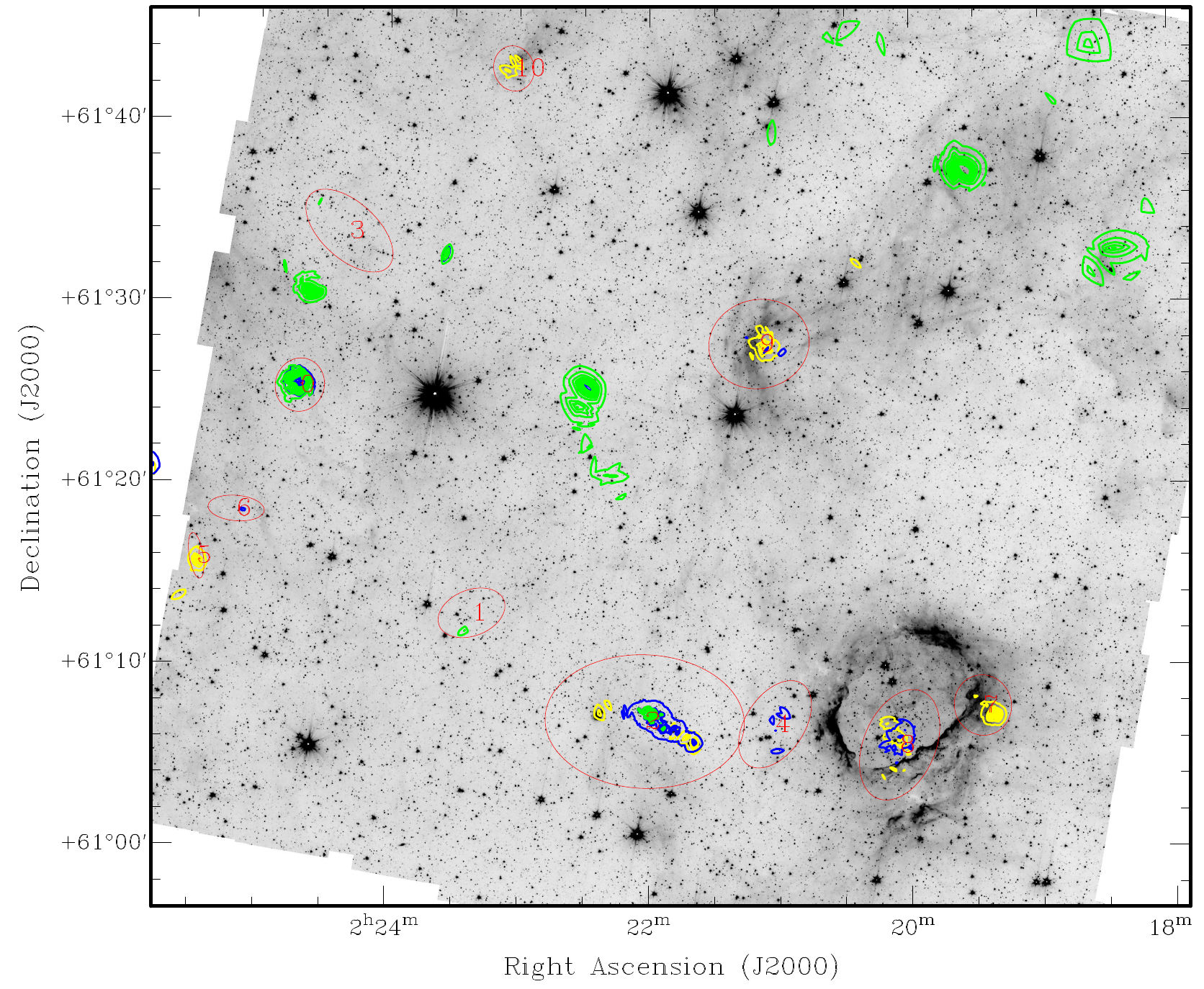}
\caption{Greyscale \spitzer\ channel 1 image of the KR 140 field with identified groups for N$_{\mathrm{YSO}}=10$ and D$_{\mathrm{break}}=0.6$\,pc (red ellipses). Yellow contours are Class 0/I surface density contours between $10-90\%$ of peak value $\sim100$\,YSO\,pc$^{-2}$ in $10\%$\,steps.  Blue contours are Class II contours between $4.0-34.0\%$ of peak value $\sim250$\,YSO\,pc$^{-2}$ in $10\%$\,steps. YSO contours have been chosen to span a common YSO range for both classes of $\sim10-90$\,YSO\,pc$^{-2}$. Green contours are of the ratio of Class II/Class 0/I YSO surface density maps; these include transition and highly embedded candidates: $^*$ classification. These `age' contours are for 4-20$\%$ of peak value of $\sim680$ in $2\%$ steps.}
\label{fig:kr1}
\end{figure*}

\begin{figure*}[ht!]
\centering
\includegraphics[scale=0.65,angle=270]{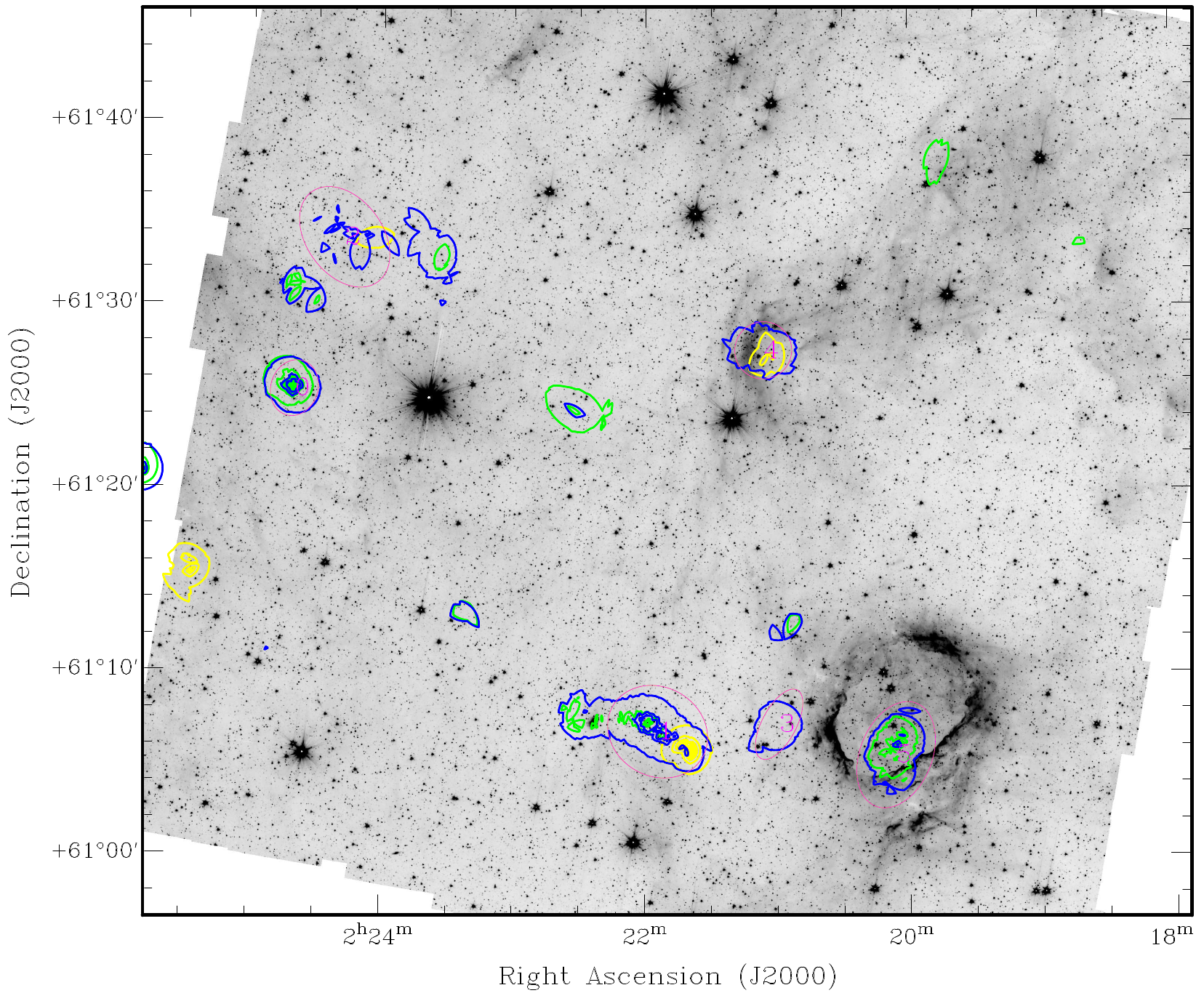}
\caption{Like Fig. \ref{fig:kr1}, but excluding the less reliable ($^*$) population. Groups are marked as magenta ellipses. Class 0/I surface density contours between $5-95\%$ of peak value $\sim50$\,YSO\,pc$^{-2}$ in $10\%$\,steps. Class II contours between $1.5-30.5\%$ of peak value $\sim175$\,YSO\,pc$^{-2}$ in $10\%$\,steps. YSO contours have been chosen to span a common YSO range for both classes of $\sim2.5-50$\,YSO\,pc$^{-2}$. Age contours are for 2-52$\%$ of peak value of $\sim1000$ in $5\%$ steps.}
\label{fig:kr2}
\end{figure*}

\begin{figure*}[ht!]
\centering
\includegraphics[scale=0.65,angle=270]{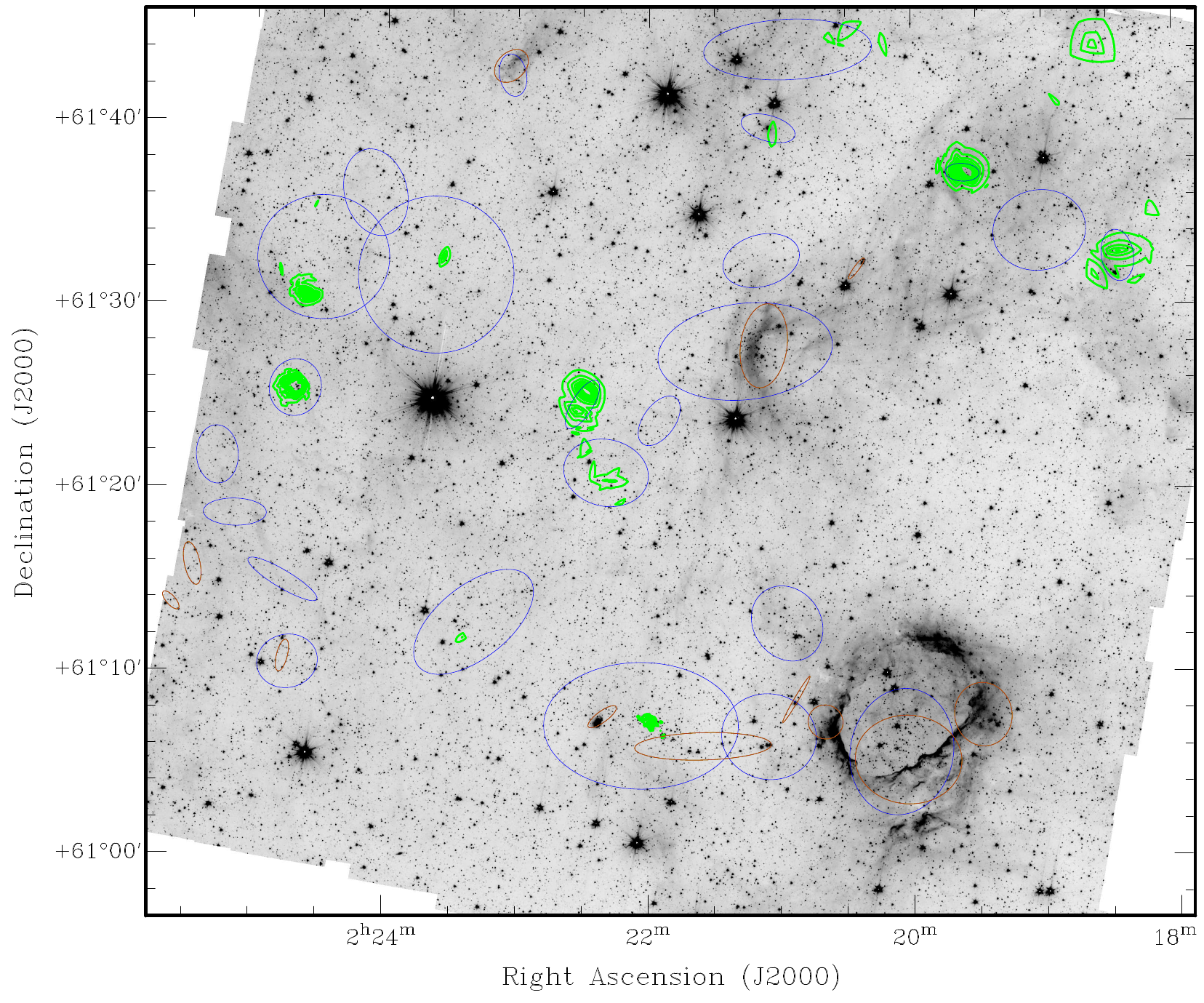}
\caption{Like Fig. \ref{fig:kr1} but for N$_{\mathrm{YSO}}=5$, D$_{\mathrm{break}}=0.75$\,pc, and Class 0/I + Class 0/I$^*$ (brown ellipses), and N$_{\mathrm{YSO}}=5$, D$_{\mathrm{break}}=0.95$, and Class II + Class II$^*$ (blue ellipses). Parameters are from Table \ref{table:clusters_regions}.}
\label{fig:kr1_5}
\end{figure*}

\begin{figure*}[ht!]
\centering
\includegraphics[scale=0.65,angle=270]{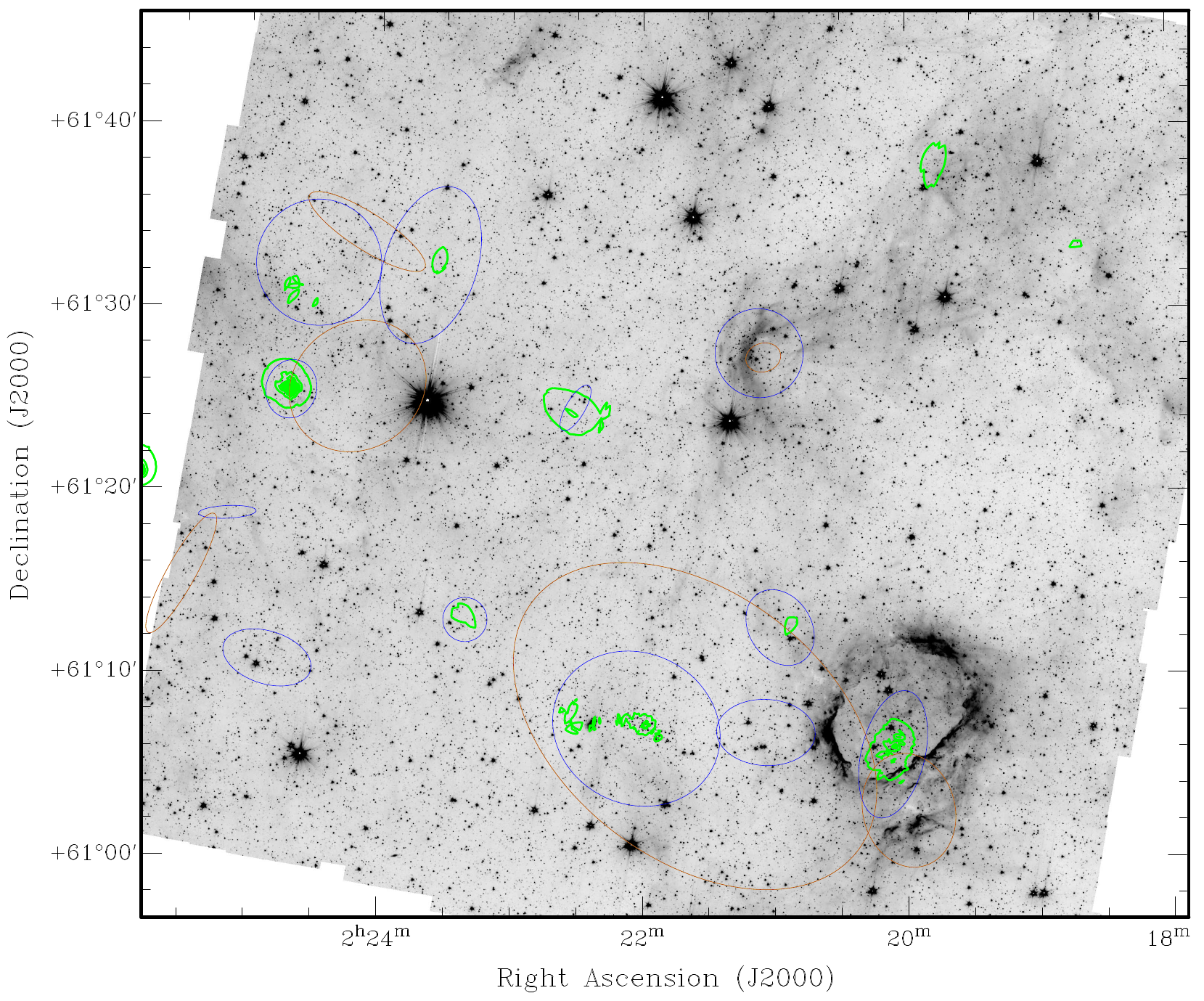}
\caption{Like Fig. \ref{fig:kr2} but for N$_{\mathrm{YSO}}=5$, D$_{\mathrm{break}}=2.25$\,pc, and Class 0/I (brown ellipses), and N$_{\mathrm{YSO}}=5$, D$_{\mathrm{break}}=1.05$, and Class II (blue ellipses). Parameters are from Table \ref{table:clusters_regions}.}
\label{fig:kr2_5}
\end{figure*}

\subsection{Star Formation in the Central and Western Region: KR 140-N and KR 140 \ion{H}{2} Region}

\begin{figure*}[ht!]
\centering
\includegraphics[scale=0.65,angle=270]{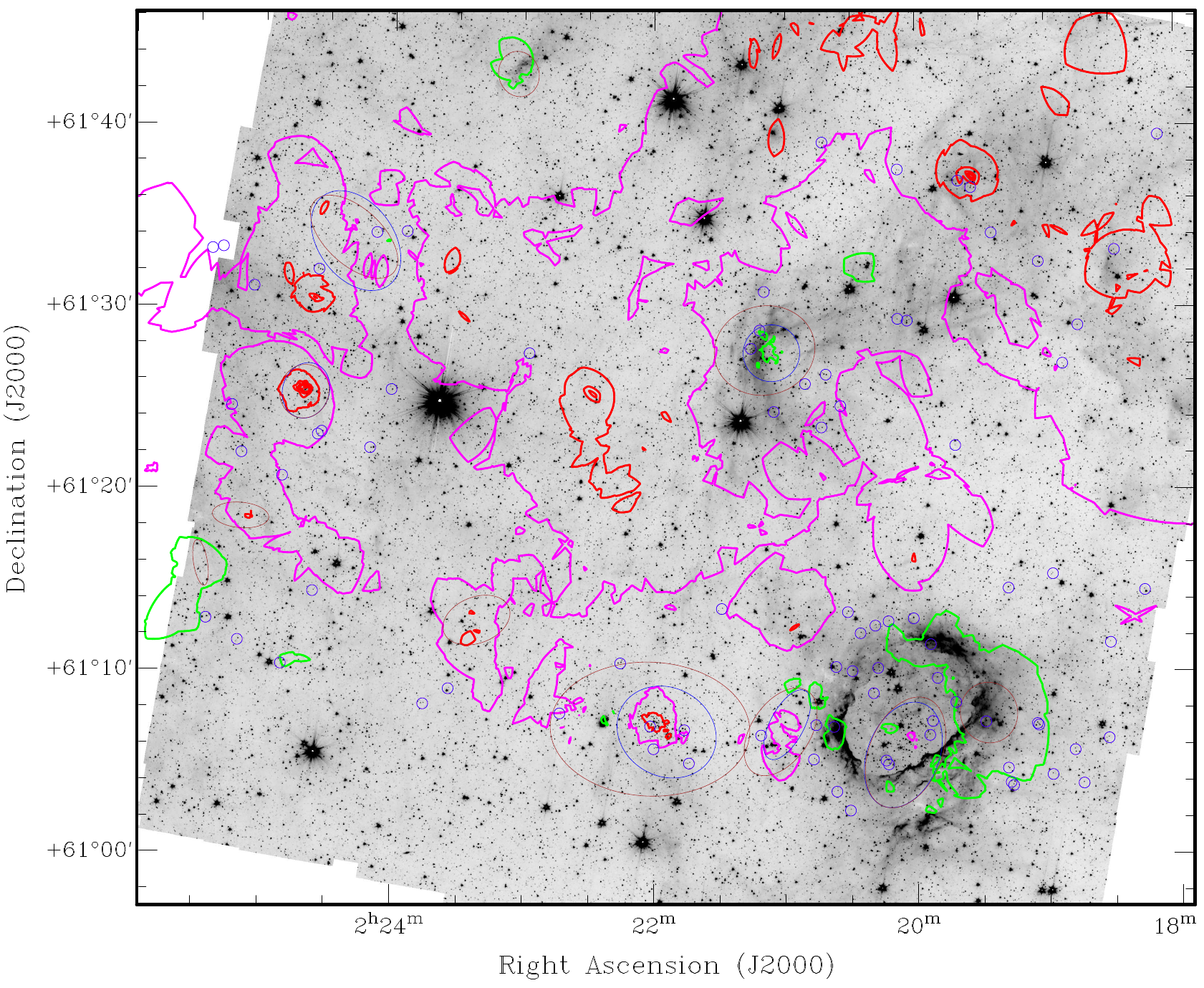}
\caption{Greyscale \spitzer\ channel 1 image of the KR 140 field. Contours like Fig.\ref{fig:age_all}, but with red contours between $3-33\%$ in $10\%$ steps (red). Small blue circles are 2MASS-based PMS candidates. Brown and blue ellipses like in Figs. \ref{fig:kr1} and \ref{fig:kr2}, respectively.}
\label{fig:age_kr}
\end{figure*}

The W3 molecular cloud is, overall, in an advanced state of evolution. Channel 4 \spitzer\ images reveal a highly turbulent and dynamic environment with a variety of structures such as bright rims, pillars with varying orientation, cavities, and filaments (Fig. \ref{fig:intro_ch4}). In this third field, groups have a wide range of ages and properties and can be separated mainly into i) groups associated with filaments; ii) groups associated with the active star forming regions: KR 140-N and KR 140 (Fig.\ref{fig:intro}).

Figures \ref{fig:kr1} and \ref{fig:kr2} show the YSO groups (with and without the ($^*$) population) identified in the central and western regions of the W3 GMC. The figures also show YSO surface density contours and age contours. Figures \ref{fig:kr1_5} and \ref{fig:kr2_5} are like Figures \ref{fig:kr1} and \ref{fig:kr2}, but for N$_{\mathrm{YSO}}=5$ and separating YSOs according to class. These figures show a tendency for Class II sources to be more widely distributed. While this effect could be enhanced by motion away from their birthsites, when Class II sources are found systematically located preferentially on one particular side of the parent structure, like in the case of Group 2 (Fig. \ref{fig:kr1}), a propagating triggering disturbance seems more likely responsible for the observed distribution.

The Class 0/I population is well localized, almost exclusively in the high extinction filaments and bright rims. The Class II population is found distributed throughout the entire cloud, following the infrared (PAH) emission structures (channel 4) and within the central `cavity' that separates the HDL and the western region comprising KR 140-N and KR 140. Independent star forming events likely produced the observed distributed population. Star formation events throughout the cloud also triggered secondary bursts of star formation resulting in the eroded and turbulent environment observed at longer wavelengths. An almost square lower column density region is noticeable in the extinction maps, but we find no evidence of any major event that could have caused an evacuation. An age sequence could be inferred for the star forming regions associated with the HDL. However, the structures and groups found in the central and western regions of the W3 GMC (Fig. \ref{fig:kr1}) appear more characteristic of sequential or even isolated events, perhaps more typical of the normal evolution and aging of a cloud  without a neighbor like W4, and/or perhaps related to the original formation of the GMC. 

Groups in the central region of the W3 cloud are member-poor (N$_{\mathrm{m}} < 25$; Groups 6, 5, 1, 10, 3, and 0, in order of increasing membership; Fig. \ref{fig:kr1}). Analysis of the environment of the filaments (Groups 6, 5, 1, 0, the latter also identified as magenta Group 0; Fig. \ref{fig:kr2}) suggests their YSO population could have been triggered by external events. The fact that most run parallel to the HDL might indicate some interaction (pillars and elephant trunks in the region point toward the upper (and older) part of AFGL 333), but might, on the other hand, just reflect the high density there, based on their range of ages. Group 10, located in a bright rim of emission in the PAH band in the northern parts, does suggest a triggered formation. The orientation of other nearby pillar structures points to a triggering source located in the most northern parts of the cloud and outside the \spitzer\, surveyed area.

The age of the systems in the central regions of W3 range from the youngest (Group 5 or 10) to the oldest (Groups 0 and 3) in the GMC. Group 0 is also the filament with lowest extinction. Group 3 exists within a cavity-like structure in the neighborhood of weak emission pillar-remnants that point toward the location of Group 10. This, combined with the lack of association with an extinction structure and its low density is suggestive of an old triggered population that formed from material associated with the now low density pillar structures.

The richest groups (N$_{\mathrm{m}} \ge 25$; Groups 4, 9, 8, and 2, in order of increasing membership; Fig. \ref{fig:kr1}) are associated with KR 140-N, KR 140 (e.g., \citealp{kerton2001}; \citealp{kerton2008}), and the major filament east of this \ion{H}{2} region. 

The structure and YSO population of KR 140-N (Group 9) support a triggered origin from RDI. The innermost region of this structure contains a large population of Class 0/I sources, with Class II candidates extending in front and behind the shocked region. A similar YSO distribution was observed in pillars associated with embedded clusters triggered by W4 (e.g., Group 7 in the AFGL 333 field; Fig. \ref{fig:afgl1}) that were also suggested to have been produced by RDI.  The cometary-like morphology also resembles that obtained by theoretical models of RDI of a cloud exposed to a high ionizing flux ($\Phi_{\mathrm{LyC}}=3\times10^{11}$\,cm$^{-2}$\,s$^{-1}$; \citealp{bisbas2011}). A significant tail of `blown' material extends toward the west, containing a largely distributed population of Class II and some Class 0/I sources, the latter generally associated with knots of bright PAH emission. Just as Group 7 in Figure \ref{fig:afgl1} is triggered by W4, the presence of Class II sources `ahead' of the front for KR 140-N suggests triggering by a source `external' to this structure and to the east. The location of the rim, at the edge of an evacuated region, also suggests an external (albeit so far unidentified) influence.

The (low density) YSO population of Group 7 likely associated with the shell of KR 140 is the youngest, indicating active star formation triggered by the ionizing star VES 735. We find a population of Class II and PMS sources extending toward the north of the \ion{H}{2} region (Fig. \ref{fig:age_kr}) that are likely to have originated because of the activity in the latter.
Group 8, a rich (N$_{\mathrm{m}} > 50$) group projected on the \ion{H}{2} region, contains a mixed population of Class 0/I and II sources (ratio $\sim1$) indicative of an extended period of star formation. This group, together with Group 2, are not only the richest in the field, but are also associated with the highest surface densities ($>60-70$\,M$_\odot$\,pc$^{-2}$) and extinctions (A$_{\mathrm{V-peak}}\sim5.5$). This extinction is in agreement with that estimated for the exciting O star of KR 140, VES 735, by \citet{kerton1999} from spectroscopy (A$_{\mathrm{V}}\sim5.4$) and optical photometry (A$_{\mathrm{V}}\sim5.7\pm0.2$), confirming the reliability of the extinction map at least within the resolution-related limitations. 

The filaments associated with Groups 2 and 4 are the most prominent of the region. Our analysis indicates that Group 2 is in a relatively highly evolved stage and was not triggered by KR 140, in agreement with previous analysis \citep{kerton2008}. Contrary to the population of Group 4, Class II sources in Group 2 are displaced from the filament toward the north, with a string of Class 0/I still deeply embedded in the innermost regions of the filament. The east-west orientation and arc-like shape of this structure, as well as the distribution of the Class II population and the material traced in the mid-infrared, suggest a possible trigger located in the direction of KR 140-N. Group 4 is younger, with characteristics and orientation that suggest a link to the activity in KR 140, at least in the northernmost parts.

All member-rich groups are also identified even after excluding the ($^*$) sample (Fig. \ref{fig:kr2}). When reducing the membership requirement to N$_{\mathrm{YSO}}>5$ we again reach similar conclusions (Figs. \ref{fig:kr1_5} and \ref{fig:kr2_5}).

Overall, Class 0/I candidates are confined to the innermost regions of the filaments (high ellipticity groups), and therefore trace these structures with high accuracy. We also detect a distribution of mainly Class II groups around the `cavity-like' region  bounded by Group 10 (Fig. \ref{fig:kr1}), KR 140-N, and Group 3. While such a configuration is reminiscent of a triggered origin, the distributed Class II population might actually be the result of percolating low level spontaneous star formation. A string of such Class II groups is observed crossing the entire field, extending from the easternmost regions to KR 140, and incorporating the filaments identified with Groups 2 and 4 (Fig. \ref{fig:kr1_5}), suggesting that the filaments in the southern central and western parts of W3 are the peak overdensities in a region of overall enhanced extinction.

VES 735 might have formed as part of a growing loose association of massive stars. \citet{kerton2001} indicate that the nearby sources IRAS 02171+6058 and IRAS 02174+6052 have luminosities consistent with lower mass embedded B stars. We find a Class 0/I YSO matching IRAS 02171+6058 and no significant radio continuum emission, which supports their hypothesis of an embedded B-type star. We do not detect a \spitzer\ counterpart for the other source. A more detailed analysis is required to investigate this possible association for VES 735, including the numerous bright infrared stars (BIRS) in this region\footnote{The YSO classification and association with clumps and cores will be investigated in our upcoming Herschel paper}.

In Figure \ref{fig:age_kr} we plot the `age' contours deduced from the ratio of Class II/Class 0/I with ($^*$) sources. Peaks in the red contours indicate the oldest age, while green represents the youngest. The distribution of ages throughout the field is evident. Group 7, associated with the shell, is the youngest. The age of VES 735 is estimated to be $\sim2$\,Myr, consistent with an \ion{H}{2} region of $\sim1-2$\,Myr (\citealp{kerton1999}; \citealp{ballantyne2000}), and so this age can be considered an upper limit for the YSO population associated with this group. Group 7 is followed in age by Group 5, and those associated with the other active (infrared-bright) star forming sites: Group 10, Group 9 (KR 140-N), and Group 8 (KR 140). Excluding the ($^*$) sources we find the youngest group to be that associated with the filament closest to KR 140 (Group 4 in Fig. \ref{fig:kr1}), followed by the population in KR 140-N (Group 9), the filament associated with Group 2, and Group 3 (central W3), with the \textit{oldest} groups being Group 8, projected on KR 140, and Group 0.

The regions undergoing star formation (Class 0/I) in the central and western part of W3 are at very localized spots in the cloud. Neighboring groups have an age spread suggestive of individual star formation events and evolution. Filaments host some of the oldest populations, while those with triggered morphology (Group 10, KR 140-N: Group 9) are younger, as expected if actually triggered by (and therefore dependent on) a \textit{previous} event. Star formation events must also continue over at least a few Myr, as suggested by the Class II population distributed throughout the cloud surrounding the active star forming sites.

Figure \ref{fig:age_kr} also shows the candidate PMS sources, which dominate near the KR 140 \ion{H}{2} region. Finding the oldest population to be associated with KR 140 itself (and therefore independent of any previous star formation), supports the `spontaneous' origin of its massive exciting star.

%%%%%%%%%%%%%%%%%%%%%%%%%%%%%%%%%%%%%%%%%%%%%%%%5

\section{Conclusion}
By means of \spitzer\ and 2MASS data we have identified and analyzed the young stellar population in the W3 GMC. These were classified according to the standard `CLASS' nomenclature for YSOs, and compared to other classification schemes based on intrinsic stellar properties, such as envelope accretion rate and disk mass. 
We find distinct regions in the CCD separating the intrinsically (and observationally) young population from the disk-dominated sample, with an intermediate region likely containing edge-on optically thick disks and weaker envelope candidates. 

Observationally, the low-mass PMS population cannot be identified unambiguously without first identifying the protostellar Class 0/I and Class II sources. Intermediate mass stars could in principle be identified \textit{without} the need for any previous protostar classification, although  it is likely that some HAeBe sources would then be missed through misclassification as Class II and T-Tauri objects.

The YSO population was divided into spatial groups according to the minimum spanning tree algorithm. We also created YSO surface density maps and `age' maps. These data were used to investigate the characteristics and history of the star formation in W3.

The distinctive HDL is no doubt influenced by the expansion of W4. The high density conditions there favored the formation of particularly rich bursts of star formation. The HDL contains the main star formation activity of W3 and has signatures of both spontaneous and triggered star formation from both external (e.g., W4) and internal events. Whether W4 was just key to creating the initial favorable \textit{conditions} for (massive) star and cluster formation, the very high surface density regions, clumps and cores which then collapse at a later stage, or whether it also \textit{triggered} the star formation in such clumps and cores is a more subtle question.

Our finding of a relatively older population in the western side of W3 Main and AFGL 333, opposite to shells formed by compression by IC 1795 and W4, respectively, suggest star formation could have started first in quiescent mode throughout the GMC, including these structures as well as the cluster IC 1795 itself. Subsequently,  triggering mechanisms by the intense activity (e.g., IC 1795) were responsible for compressing an already dense environment, greatly enhancing and/or inducing major bursts of massive star and cluster formation (e.g., W3 Main, W3 (OH)).  

The evolved state of W3 is also evident in the central and western parts of this cloud, an overall highly turbulent and eroded environment. This is believed to have been produced by numerous individual star formation events that were responsible for triggering secondary episodes of star forming activity. The central regions between AFGL 333 and KR 140 are particularly rich in filaments, seen in extinction in the mid-infrared, whose formation and star forming activity appear to be associated with this highly turbulent environment. Recent star formation is confined mainly to these very localized regions.

The overlapping spatial distributions of the YSO class populations in the most active areas of W3 indicate on-going periods of star formation in regions of massive star and cluster formation, as opposed to a single, major short-lived event. 

Based on number of YSO types (in groups) and ratios of class surface densities (in regions) as relative measurements of age, we find that the region comprising W3 Main and W3 (OH) is the youngest, followed by the central/west region of W3 and AFGL 333.  However, in AFGL 333 confusion does not allow us to detect groups in the currently most active areas, and therefore the age is representative of the YSO population associated with filaments. We cannot determine if the activity in AFGL 333 did start first or if the activity all across the HDL started at similar times triggered by W4, but of all the ages derived for the groups associated with isolated filaments we find those in this region to be the oldest. Indeed, the characteristics of the YSO population and individual group ages suggest that the activity immediately west of AFGL 333 could have induced some secondary triggered events in filamentary structures in the central regions of the cloud. 
 
On-going stellar activity in W3 Main, IC 1795, and KR 140 results in younger apparent ages for their associated groups. Nevertheless, the KR 140 region appears to contain the oldest population in the western part of W3. This age is less than that of IC 1795, and the absence of a suitable nearby triggering mechanism supports the spontaneous origin for VES 735 and stars in this possible association. We find that the age of KR 140-N must be intrinsically young. This structure shows not only a morphology consistent with triggering by RDI, but also a well defined distribution of Class II sources surrounding the Class 0/I population, similar to that associated with pillar structures. 

The age of IC 1795 has been estimated to be around $\sim3-5$\,Myr. Even when considering a supersonic velocity (for typical velocity in the ionized medium) of c$\sim10$\,km$^{-1}$,  IRS5 would require $\sim2.5$\,Myr, and $\sim3$\,Myr for interaction with KR 140-N. If IRS5 itself were actually triggered by IC 1795, then this makes the former an even more unlikely candidate for triggering activity in the western regions. If a source of triggering is indeed located in the HDL, then at least the western side of these structures must have been undergoing star formation activity well before the onset of the present major activity in the HDL. This is supported by findings of a relatively older population in the western side of these structures.

Results presented above reveal that despite having considered a range of relatively evolved stages of pre-stellar evolution, the W3 GMC is still a prime target for investigating different modes of star formation. We classified as a \textit{grouped} population those YSOs not part of a previously cataloged cluster, but belonging to a group identified though the MST analysis.

\textit{Primordial/quiescent \textit{grouped} formation} is identified by old, relatively isolated systems with the closest inter-YSO separations and compact configurations (not necessarily circular if parental material is filamentary), whose richness depends on the primordial surface density of the clump/core. An example is Group 2 in Figure \ref{fig:main1}. \textit{Triggering} may be a more efficient mechanism for creating high surface density structures favorable to high multiplicity and/or for inducing the collapse of such structures (e.g., by overcoming internal pressure and turbulent support). All current major clustered massive star activity in W3 is believed to have been `triggered' to some extent. 
\textit{Quiescent clustered formation} might nevertheless have occurred initially to produce IC 1795, whose preferential location in the HDL resulted in a rich population that subsequently triggered secondary bursts of star formation within an already overdense region. 
The outcome for the GMC would also depend ultimately on the physics, effectiveness, and limitations of the processes creating the precursors of groups and clusters in the parental cloud (e.g., filaments, clumps, and cores).

\textit{Triggered formation} has occurred throughout W3. 
In some cases, unstable structures with enough surface density and low turbulence are created (collect/collapse model). Elsewhere there is collapse induced (RDI) in the neighborhood of the triggering source. Examples for triggered \textit{grouped} formation are Group 7 in Figure \ref{fig:kr1}, a YSO population associated with a shell of material compressed by the expanding KR 140 \ion{H}{2} region, and Group 9 associated with a cometary like structure in KR 140-N. \textit{Clustered} triggered formation can be observed in pillar structures such as Group 7 facing W4 (Fig. \ref{fig:afgl1}). The present data cannot confirm whether pre-existing cold seeds had formed in the cloud prior to the triggering mechanisms, and this issue will be revisited with the available Herschel data (Rivera-Ingraham et al., in preparation). 

Regardless of the detailed sequence of processes, our work indicates that triggering is the main mechanism associated with those structures dense enough to host the current main cluster and massive star formation activity (i.e., W3 Main). Clusters formed within such structures are the richest. IRS5 in particular, unresolved in our data, has been suggested to be a Trapezium-like system of proto OB stars within an envelope $<0.02$\,pc in size; its protostellar density, of $\sim0.5-5\times10^{6}$\,pc$^{-3}$, makes it one of the most dense clusters known (\citealp{megeath2005}; \citealp{rodon2008}). This system is itself embedded in a massive cluster of low mass and PMS stars (\citealp{megeath1996}; \citealp{feigelson2008}) within the region of highest extinction in the entire W3 GMC (A$_{\mathrm{V}}=9.6\pm0.3$). 

We classified as \textit{distributed} those YSO candidates not in groups and not presently associated with extinction structures in the infrared.
Some of the distributed population has an origin associated with high extinction structures (e.g., filaments, pillars) cleared by the stellar activity (e.g., Fig. \ref{fig:afgl1}, western part of Group 4). A filament can follow a highly irregular and curved morphology, likely linked to large scale turbulence in the environment, and the YSOs born in the filament will looked highly distributed after the disappearance of their parent structure (e.g., Group 6, Fig. \ref{fig:afgl1}). The fact that the Class II population of Group 4 is surrounded by possible `remnants' of parent structures suggests lifetimes for the filaments and clumps of the order of a few Myr. Whether the star formation processes within filaments differ from those in clustered formation needs to be determined. Note that a population of filamentary origin can be confused with that truly originated in isolation. The latter can be observed in the `tail' of KR 140-N and in highly turbulent, low surface density regions in the vicinity of active star formation. An example can be found in the western side of W3, where `knots' of infrared emission trace isolated overdensities (clumps, cores), more relevant for the formation of individual objects. 

Herschel submillimeter imaging data (Rivera-Ingraham et al., in preparation) will be used to identify and characterize the population of cores and clumps in the W3 GMC. By examining how these differ in regions of quiescent and triggered activity or with and without an associated YSO population, this analysis will probe the processes involved in the earliest stages of massive star/cluster formation and the different modes of star formation. 

In the present analysis we observe a link between the star formation modes and their environment. Filamentary structures appear to trace the turbulence in the cloud. Since the environment is a key parameter in understanding the processes and properties of the stellar progenitors, molecular data (Polychroni et al. in preparation) will also be used to characterize the dynamics of the material associated with the Herschel sources, and link their differing properties back to their environment. 

\acknowledgements We thank the anonymous referee for very useful suggestions and improvements to this paper. This research was supported in part by the Natural Sciences and Engineering Research Council of Canada and the funds received by A. Rivera-Ingraham as Connaught Fellow at the University of Toronto.

%\bibliographystyle{apj}
%\bibliography{biblio}

\end{document}